\documentclass[a4paper,fleqn,usenatbib]{mnras}

\usepackage{mathptmx}
\usepackage[T1]{fontenc}
\usepackage{ae,aecompl}

\usepackage{graphicx}	% Including figure files
\usepackage{amsmath}	% Advanced maths commands
\usepackage{amssymb}	% Extra maths symbols
\usepackage{ifxetex}
\usepackage{morefloats}

\newcommand\cRoneP{R$_1^{\prime}$}
\newcommand\cRtwoP{R$_2^{\prime}$}
\newcommand\cRonePL{R$_1^{\prime}$L}
\newcommand\cRtwoPL{R$_2^{\prime}$L}
\newcommand\cRoneRtwoP{R$_1$R$_2^{\prime}$}
\newcommand\cRoneRtwo{R$_1$R$_2$}
\newcommand\cRonetwoP{R$_{12}^{\prime}$}
\newcommand\cRonePRtwoP{R$_1^{\prime}$R$_2^{\prime}$}
\newcommand\cRone{R$_1$}
\newcommand\cRtwo{R$_2$}
\newcommand\cRoneL{R$_1$L}
\newcommand\cRtwoL{R$_2$L}
\newcommand\cRP{R$^{\prime}$}
\newcommand\RoneP{(R$_1^{\prime}$)}
\newcommand\RtwoP{(R$_2^{\prime}$)}
\newcommand\RonePL{(R$_1^{\prime}$L)}
\newcommand\RtwoPL{(R$_2^{\prime}$L)}
\newcommand\RoneRtwoP{(R$_1$R$_2^{\prime}$)}
\newcommand\RoneRtwo{(R$_1$R$_2$)}
\newcommand\RonetwoP{(R$_{12}^{\prime}$)}
\newcommand\RonePRtwoP{(R$_1^{\prime}$R$_2^{\prime}$)}
\newcommand\Rone{(R$_1$)}
\newcommand\Rtwo{(R$_2$)}
\newcommand\RoneL{(R$_1$L)}
\newcommand\RtwoL{(R$_2$L)}
\newcommand\RP{(R$^{\prime}$)}
\newcommand\Lff{L$_{45}$}
\newcommand\urs{($\underline{\rm{r}}$s)}
\newcommand\rus{(r$\underline{\rm{s}}$)}
\newcommand\uurs{($\underline{\rm{r}}$s)}
\newcommand\ruus{(r$\underline{\rm{s}}$)}
\newcommand\uAB{($\underline{\rm{A}}$B)}
\newcommand\AuB{(A$\underline{\rm{B}}$)}
\newcommand\rdark{r$_{dark}$}
\newcommand\rpl{(r$^{\prime}$l)}
\newcommand\nrpl{(nr$^{\prime}$l)}
\newcommand\rpul{(r$^{\prime}$$\underline{\rm{l}}$)}
\newcommand\cRPL{(R$^{\prime}$L)}
\newcommand\ccRPL{R$^{\prime}$L}
\newcommand\rul{(r$\underline{\rm{l}}$)}
\newcommand\ruul{r$\underline{\rm{l}}$}
\newcommand\cRUL{(R$\underline{\rm{L}}$)}
\newcommand\cuRL{($\underline{\rm{R}}$L)}
\newcommand\wurl{($\underline{\rm{r}}$l)}
\newcommand\uurl{$\underline{\rm{r}}$l}

\title[Galactic Rings Revisited. I]{Galactic Rings Revisited. I. CVRHS Classifications of 3962 Ringed Galaxies from the Galaxy Zoo 2 Database} 

\author[Ronald J. Buta]{
Ronald J. Buta\thanks{E-mail: rbuta@ua.edu}
\\
Department of Physics \& Astronomy,University of Alabama, Box
870324, Tuscaloosa, AL 35487
}
\date{Accepted XXX. Received YYY; in original form ZZZ}

\pubyear{2016}

\begin{document}
\label{firstpage}
\pagerange{\pageref{firstpage}--\pageref{lastpage}}
\maketitle

\begin{abstract}
Rings are important and characteristic features of disc-shaped
galaxies. This paper is the first in a series which re-visits galactic
rings with the goals of further understanding the nature of the
features and for examining their role in the secular evolution of
galaxy structure. The series begins with a new sample of 3962 galaxies
drawn from the Galaxy Zoo 2 citizen science database, selected because
zoo volunteers recognized a ring-shaped pattern in the morphology as
seen in Sloan Digital Sky Survey colour images. The galaxies are
classified within the framework of the Comprehensive de Vaucouleurs
revised Hubble-Sandage (CVRHS) system. It is found that zoo volunteers
cued on the same kinds of ring-like features that were recognized in
the 1995 Catalogue of Southern Ringed Galaxies (CSRG). This paper
presents the full catalogue of morphological classifications,
comparisons with other sources of classifications, and some histograms
designed mainly to highlight the content of the catalogue. The
advantages of the sample are its large size and the generally good
quality of the images; the main disadvantage is the low physical
resolution which limits the detectability of linearly small rings such
as nuclear rings. The catalogue includes mainly inner and outer disc
rings and lenses. Cataclysmic (``encounter-driven") rings (such as ring
and polar ring galaxies) are recognized in less than 1\% of the
sample.

\end{abstract}


%%%%%%%%%%%%%%%%%%%%%%%%%%%%%%%%%%%%%%%%%%%%%%%%%%

%%%%%%%%%%%%%%%%% BODY OF PAPER %%%%%%%%%%%%%%%%%%

\section{Introduction}

One approach to understanding galaxy evolution is systematic
examination of the fine details of galaxy morphology. Hidden in these
details are not only clues to the basic factors that shape galaxies,
but also how different galaxy types might be related in an evolutionary
sense. This is the domain of the emerging paradigm of galaxy secular
evolution, which is believed to be the predominant mechanism of change
in the present epoch (Kormendy \& Kennicutt 2004).

Rings are features that have the potential to tell us a great deal
about galaxy evolution and dynamics (Kormendy 1979, 2012; Buta \&
Combes 1996). Rings are mainly a phenomenon of disc galaxies, which by
virtue of their high angular momentum content are prone to the
development of well-defined internal perturbations such as bars, ovals,
and global spirals. These patterns are thought to rotate at a uniform
rate, the pattern speed $\Omega_p$, while the stars and gas clouds in a
galaxy rotate differentially with an angular velocity $\Omega(r)$ and a
radial oscillation frequency $\kappa(r)$, where $r$ is the
galactocentric radius. This will set up orbital resonances at positions
where (in the first order or ``epicyclic approximation") $\kappa = m(\Omega - \Omega_p)$,
$m$ being an integer. A galaxy whose shape is dominated by
resonances can look very distinctive compared to ones which are not (e. g.,
NGC 3081, Buta et al. 2004). Resonances can lead to slow changes in the structure 
of a galaxy, making them an important factor in the secular evolutionary 
process (Kormendy 2012).

Rings have attracted attention because they are fairly common features
of disc galaxies, and because the morphological and star formation
characteristics of distinct types of rings seem to link them to
specific resonances, mainly the inner and outer Lindblad resonances
(ILR and OLR, respectively, for which $m$=$\pm$2) and the inner and
outer 4:1 resonances (I4R and O4R, respectively, for which
$m$=$\pm$4). Resonances are not the only way of explaining the
properties of galactic rings. The recently proposed ``manifold theory"
(Athanassoula et al. 2009a,b; 2010) has also had considerable success
in explaining ring morphologies. Deciding which of these views is 
most favored will depend on observations beyond the scope of this
paper. The emphasis here will be on the resonance interpretations of
the rings. 

Normal galactic rings occur in at least three basic types: nuclear,
inner, and outer rings. Nuclear rings are the small rings often found
at the centers of early-to-intermediate type barred galaxies. In the
most extensive study to date, the Atlas of Nuclear Rings (AINUR),
Comer\'on et al. (2010) found that nuclear rings occur in 20\%$\pm$2\%
of galaxies in the type range S0$^-$ to Sd, have typical sizes ranging
from 0.2-2kpc, and may be physically smaller in galaxies with longer
bars. Nuclear rings can also be the sites of intense starbursts (e. g.,
Benedict et al. 2002).  Inner rings are intermediate-sized (major axis
diameter $\approx$2-12kpc) features which envelop a bar if present,
while outer rings are large, diffuse structures about twice the size of
a bar or inner ring. In many cases we can see that a ring is actually
made of tightly-wrapped spiral arms, leading to the concept of
``pseudorings."

Considerable information on the intrinsic shapes, relative bar-ring
orientations, and relative sizes of inner and outer rings, pseudorings,
and lenses is provided by the Catalogue of Southern Ringed galaxies
(CSRG; Buta 1995=B95), the Atlas of Resonance Rings as known in the
S$^4$G (ARRAKIS; Comer\'on et al. 2014), and the Near-Infrared S0
Survey (NIRS0S; Laurikainen et al. 2011). The CSRG is based on a visual
search for ringed galaxies using film copies of the photographic
ESO/SRC IIIa-J ($\approx$$B$-band) Southern Sky Survey, while the
ARRAKIS is based on digital images taken at the mid-infrared wavelength
of 3.6$\mu$m for the Spitzer Survey of Stellar Structure in Galaxies
(S$^4$G, Sheth et al. 2010).  The NIRS0S is a compilation of 2.2$\mu$m
groundbased digital images of 206 nearby early-type galaxies, of which
193 are in the type range S0$^-$ to Sa.

Both the ARRAKIS and the NIRS0S catalogues are statistically
well-defined samples. In the ARRAKIS study, Comer\'on et al. (2014)
found that inner rings and pseudorings occur in 35\% $\pm$ 1\% of the
sample galaxies, with a highest abundance of 60\% for stages S0$^+$ to
Sb, and that outer rings occur in 16\% $\pm$ 1\% of the galaxies,
rising to 40\% when restricted to stages S0$^+$ to Sab. For stages
later than Sb, outer rings are much rarer than inner rings in the
S$^4$G sample, dropping below the 10\% level. In the NIRS0S study,
Laurikainen et al. (2011) found that 61\% of the barred and 38\% of the
nonbarred galaxies in the sample have lenses, and that multiple lenses
are common among S0 to S0/a galaxies.

Although the CSRG, S$^4$G, and NIRS0S provide a great deal of
information on galactic rings and related features, all have
limitations.  The CSRG only includes galaxies south of declination
$-$20$^o$, and as a result very few galaxies in the catalogue have
publicly available digital imaging. Also, the S$^4$G and the NIRS0S
have sharp limits to inclusion, the former emphasizing extreme
late-type, very nearby galaxies and the latter emphasizing mainly the
brightest S0 galaxies. The Sloan Digital Sky Survey (SDSS. Gunn et al.
1998; York et al. 2000) provides an opportunity to compile a new
catalogue of rings that would complement these previous catalogues. The
survey is built around the North Polar Cap, and at the time of Data
Release 7 (DR 7; Abazajian et al. 2009) covered the sky from about 7$^h$ to
18$^h$ right ascension, and 0$^o$ to +60$^o$ declination.

In this series of papers, the structure of ringed galaxies is revisited
using a mostly new sample of galactic rings drawn from the SDSS-based
Galaxy Zoo 2 citizen science morphological database (Willett et al.
2013).  The present paper provides a classification of 3962 GZ2
galaxies in the Comprehensive de Vaucouleurs revised Hubble-Sandage
(CVRHS) system, a Hubble-based classification system which uses the
precepts of the de Vaucouleurs (1959) three-dimensional classification
volume, expanded to include other features such as inner, outer, and
nuclear lenses, nuclear rings and bars, bar ansae, barlenses, X
patterns, thick discs, extraplanar discs, and other features.  The
system is described in the de Vaucouleurs Atlas of Galaxies (Buta et al.
2007=deVA) and is extended further by Buta et al. (2015).  The main
goal of this classification is to provide a northern hemisphere sample
of ringed galaxies comparable in scope and size to the CSRG.  This
catalogue will be called the ``GZ2 Catalogue of Northern Ringed
Galaxies," or GZ2-CNRG.

Section 2 describes the Galaxy Zoo 2 project and how it was used to
compile the GZ2-CNRG.  Section 3 discusses the sample selection and
classification procedure. An analysis of the morphological information
is presented in section 4, while section 5 presents small tables and
illustrations of examples of different morphologies from the catalog. A
summary is given in section 6.

\section{Galaxy Zoo 2}

The Galaxy Zoo Project (Lintott et al. 2008) has been and continues to
be a remarkably effective amateur-professional collaboration designed
to provide basic classifications for the hundreds of thousands of
relatively well-resolved galaxies imaged in the SDSS (DR 7). The colour
images examined by zoo volunteers are constructed from separate images
obtained in green ($g$-band, 477 nm), red ($r$-band, 623 nm), and
near-infrared ($i$-band, 762.5 nm) filters. In these images, isophotes
of only a few percent of the night sky brightness (corresponding to
surface brightnesses of $\approx$26.0, 25.5, and 25.0 mag arcsec$^{-2}$
in $g$, $r$, and $i$, respectively) are readily detectable in an image
display. The typical full width at half maximum of the point spread
function is 3.1 $\pm$ 0.5 pix, 3.0 $\pm$ 0.4 pix, and 2.9 $\pm$ 0.4 pix
for $g$, $r$, and $i$, respectively, where 1 pixel =
0\rlap{.}$^{\prime\prime}$396.

The philosophy of Galaxy
Zoo is that human visual classification is superior to computer
classification with respect to reliable identification of major and
minor classes, and has the added advantage that it can allow
recognition of special cases of interest (e.g., Hanny's Voorwerp,
Lintott et al. 2009; Schawinski et al. 2010) that a computer program
likely would overlook. Computer classification also often ignores the
finer details of galaxy morphology because these are difficult to
characterize compared to the more basic galactic components. The main
advantage of computer classification is its ability to classify in a
reasonable amount of time extremely large samples that are well beyond
the capability of the small number of morphological ``experts" in the
world. To overcome this problem for visual classification, Galaxy Zoo
used a web-based interface to enlist the help of amateur galaxy
morphologists around the world.  Several hundred thousand of these
``citizen scientists" responded to the project, which led to many
important findings that have been reviewed by Fortson et al. (2010) and
Masters et al. (2010).

The first open zoo classification, called Galaxy Zoo 1 (or GZ1), used a
small number of buttons to collect information on very broad classes of
morphology, such as whether a galaxy is a spiral or an elliptical. This
was followed by Galaxy Zoo 2 (or GZ2), which used more buttons to
collect more specific and more finely-detailed information, such as
whether a bulge, bar, or ring was present. The collected data from
84,000 participants were combined into a mean catalogue (Willett et al.
2013) of visual morphological information for nearly 300,000 galaxies,
a sample more than 20 times larger than the largest comprehensive
visual survey to date (Nair \& Abraham 2010).

While a sample this large is beyond the reach of a single professional
galaxy morphologist, smaller subsets are not, and GZ2 provides the
means of drawing samples of specific characteristics worthy of
follow-up examination (e.g., Masters et al. 2011). One such
characteristic is the presence of a ring. A button asked the
participant if there was ``anything odd" about a galaxy, and if so,
``is the odd feature a ring?" Rings attracted attention early in the
project through the Galaxy Zoo Forum, where classifiers would post
their favorite cases. A ring galaxy thread was started, and although
``ring galaxy" has a specific meaning in galaxy morphology (a likely
product of a galaxy collision; Appleton \& Struck-Marcell 1996), the
thread took a more generic view. Examining this thread, which reached
many hundreds of webpages, it was clear that the bulk of the features
being recognized were ordinary inner and outer rings and pseudorings.
The application of the CVRHS system to the GZ2 sample has allowed much
of the ambiguity in ring types to be cleared up.

The great value of GZ2 is that it has allowed a catalogue like the
CSRG, which involved searching for ringed galaxies over a wide swath of
sky, to be made in far less time than the nearly 10 years it took to
compile the CSRG. Being highly focussed on ringed galaxies, the
GZ2-CNRG, like the CSRG, is not the kind of catalogue that can be used
for general statistical studies, such as determining the abundances of
different ring types, or for any kind of study that requires complete
samples. The value of the new catalogue is that it is useful for
identifying good examples of rings, extreme examples of ring and
related morphologies, and for selecting small subsets of ringed
galaxies for more detailed follow-up analysis. The fact that the
catalogue is also based on digital images which are all publicly
available virtually guarantees that extensive follow-up will be
possible. For example, in the second paper in the series, Buta
(2017=B17) carries out an analysis of $ugri$ SDSS images for a sample
of 33 cases from the catalogue, focussing mainly on the special
resonant subclasses of outer rings and pseudorings described by Buta \&
Crocker (1991 = BC91). The results of that study are summarized in
section 5.

\section{Sample Selection and Classification}

The sample was selected from the GZ2 database on the basis of the
likelihood of having a ring given the citizen participants' selection
of the appropriate button. Willett et al. (2013) describe how the
average classifications were determined in GZ2, and at the request of
the author, Zoo members K. Masters and A. Smith sent web links to SDSS
colour images of 3,977 GZ2 galaxies.\footnote{All of the images used can
be accessed individually at the URL
http://s3.amazonaws.com/zoo2/filen.jpg, where filen is the number in
column 3 of Table~\ref{tab:catalog}.} Although ideally visual
classification should be based on blue light images, which in the SDSS
are best approximated by the $g$-band, it was impractical to download
these images for such a large sample. Instead, the SDSS colour images
were used for all classifications with the understanding that
systematic effects could occur when departing from the standard
waveband of galaxy classification. Buta et al. (2015) showed that
systematic differences between CVRHS classifications made in blue light
and those made in the mid-infrared in the same system are relatively
small. SDSS colour images include both $g$-band and $i$-band
contributions.

The classification procedure involved carefully examining each colour image,
and recording the stage, family, inner variety, and outer variety using
the notations summarized in Table 1 of Buta et al. (2015).
Although SDSS colour images are not in the units (mag arcsec$^{-2}$) used
by Buta et al. (2015) for classifying galaxies in S$^4$G images, a combination of a
linear transformation at faint light levels and an asinh transformation
at higher light levels (Lupton et al. 2004) still makes effective
classification images. The asinh stretch of the intensity scales makes
both the high and low surface brightness regions relatively accessible,
just as an image in mag arcsec$^{-2}$ would provide (see also the deVA).

CVRHS classifications involve judging: (a) the stage E, E$^+$, S0$^-$,
S0$^o$, S0$^+$, S0/a, Sa, Sab, Sb, Sbc, Sc, Scd, Sd, Sdm, Sm, or Im,
ranging from ellipticals to S0s, spirals, and irregulars;
(b) the family SA, S$\underline{\rm A}$B, SAB, SA$\underline{\rm B}$, or SB, indicating
the presence or absence of a bar;
(c) the inner variety (r), ($\underline{\rm r}$s), (rs), (r$\underline{\rm s}$),
(s), ($\underline{\rm r}$l), (rl), (r$\underline{\rm l}$) or (l), indicating the presence or
absence of an inner ring, pseudoring, or lens; and
(d) the outer variety, including mainly (R$^{\prime}$), (R), (RL), (R$^{\prime}$L), (L), (R$_1$),
(R$_1^{\prime}$), (R$_1$R$_2^{\prime}$), or (R$_2^{\prime}$) types, indicating the presence
of an outer ring, pseudoring, or lens. Buta et al. (2015)
describe how these parts are determined and put together. Of the four parts, the stage
is considered the most fundamental because, for spirals at least, it correlates well with physical
parameters such as colour, HI content, and mean surface brightness (e.g., Buta et al.
1994). For S0 galaxies, CVRHS stage is based on development of structure (de Vaucouleurs 1959). For spirals,
the stage is based on the degree of central concentration, the resolution in the arms, and
the degree of openness of the spiral pattern (Sandage 1961).

To check the internal consistency of CVRHS classifications, Buta et al.
(2015) carried out the mid-infrared classification of nearly 2400
S$^4$G galaxies in mostly two independent phases. Phase 1 was the
initial classification of the galaxies, while Phase 2 was an
independent re-classification made nearly a year after the first
examination, without reference to the first examination. The final
catalogue of S$^4$G classifications was based on an unweighted average
of the two phases. For the GZ2 sample, this procedure was less
practical because of the sample size. As a consequence, only one phase
was carried out for the GZ2-CNRG.  It is likely that the internal
consistency of the GZ2-CNRG classifications is similar to that of the
S$^4$G classifications. Defining internal consistency of morphological
classifications in terms of intervals, where 1 stage interval is a
difference like, e.g., Sab to Sb; 1 family interval is a difference
like SAB to SA$\underline{\rm B}$; 1 inner variety interval corresponds
to a difference like ($\underline{\rm r}$s) to (r); and 1 outer variety
interval corresponds to a difference like (R) to ($\underline{\rm
R}$L), Buta et al. (2015) showed that for CVRHS classifications,
$\sigma (T)$ = 0.7 stage intervals, $\sigma (F)$ = 0.7 family
intervals, $\sigma (IV)$ = 0.8 inner variety intervals, and $\sigma
(OV)$ = 1.3 outer variety intervals.

Radial velocities for the GZ2 galaxies were drawn from the NASA/IPAC
Extragalactic Database (NED). The median radial velocity and redshift
are 18600 km s$^{-1}$ and 0.062, respectively, for 3658 sample
galaxies. The distribution of radial velocities is listed in
Table~\ref{tab:radvels}. This very high median parameter accounts
for the poor resolution of many of the images. Nevertheless, most
of the galaxies were still classifiable and the GZ2 selection still
reliably yielded rings even in the high redshift half of the sample.

\begin{table}
\centering
\caption{Distribution of Radial Velocities for 3658 GZ2 Ringed Galaxies}
\label{tab:radvels}
\begin{tabular}{rr}
\hline
Range           &        Number \\
(km s$^{-1}$)   &               \\
\hline
    0- 5000     &          124 \\
 5000-10000     &          642 \\
10000-15000     &          650 \\
15000-20000     &          604 \\
20000-25000     &          644 \\
25000-30000     &          414 \\
30000-35000     &          279 \\
35000-40000     &          151 \\
40000-45000     &           96 \\
45000-50000     &           29 \\
50000-55000     &           17 \\
55000-60000     &            6 \\
60000-75000     &            2 \\
\hline
\end{tabular}
\end{table}

Table~\ref{tab:catalog} lists the CVRHS classifications for 3962
galaxies from the original sample of 3977, excluding 15 cases which
appeared to be duplicates. For each galaxy, the name from NED or an
SDSS positional name is given in col. (1); col. (2) lists the number in
the Principal Galaxy Catalogue (Paturel et al. 2003), which goes far
beyond the original PGC (Paturel et al. 1989).  Col. 3 lists the GZ2
file number, which ranges from 1 to about 245,000. This number is not
so much an id as it is a useful internal reference. Col. 4 lists the
redshift of each galaxy, if known, and is useful for judging how
reliable an interpretation might be. Col. 5 lists the stage
classification on the RC3 numerical scale; Col. 6 gives the full CVRHS
type, while Col. 7 provides some brief notes on each object, often to
highlight an exceptional case of some particular morphological aspect.
In these notes, the term ``$L_{45}$ dark-spacer" refers to a galaxy
showing the dark gaps between inner and outer rings that are the
principal focus of paper 2 (B17).

The notes are also used to identify cases where the resolution seemed
poor, which thus made the classification difficult and likely less
accurate than for better resolved cases. Features where resolution is
important are the arm-doubling in two opposing quadrants used to
identify R$_2^{\prime}$ outer pseudorings and the dimples used
to identify R$_1^{\prime}$ outer pseudorings (see Figure 1 of B17); the
distinction of lenses from rings; the strength or even presence of a
bar; the detection of bar ansae; and the detection of nuclear rings and
secondary bars.

\begin{figure} \includegraphics[width=\columnwidth]{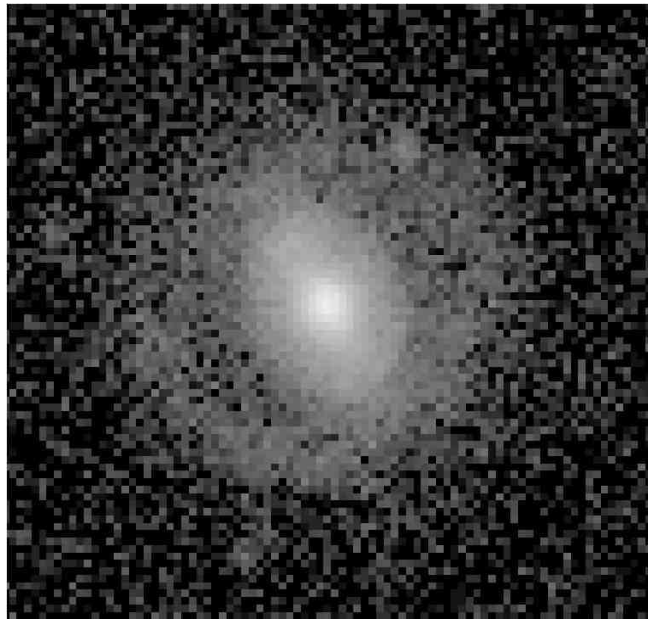}
\caption{Galaxy PGC 1224876 (15338.jpg) is a typical ``poorly resolved"
case in Table~\ref{tab:catalog}. At its redshift of
0.082, its structure is still discernable and classifiable. The image
is $g$-band and is in units of mag arcsec$^{-2}$.  } \label{fig:15338}
\end{figure}

An example of a typical ``poorly resolved" case is shown in
Figure~\ref{fig:15338}. This is a $g$-band image converted to
units of mag arcsec$^{-2}$. The full width at half maximum (FWHM) of the
point spread function (psf) on this image is 3.1 pix or
1\rlap{.}$^{\prime\prime}$23. Based on the galaxy's redshift of 0.082
and a Hubble constant of 73 km s$^{-1}$ Mpc$^{-1}$,
1$^{\prime\prime}$ = 1.6kpc, implying 1 FWHM = 2.0kpc. The galaxy is
$\approx$55 pix across, corresponding to 18 resolution elements. In
spite of the low resolution, the type of the galaxy, 
(R$_1^{\prime}$)S$\underline{\rm A}$B(l)a, is still discernible.

For comparison, the S$^4$G is a sample restricted to galaxies nearer
than 40 Mpc, and the survey's images have a FWHM of the psf of
1\rlap{.}$^{\prime\prime}$8 (Buta et al. 2010). At the limit of this
survey, the resolution would be 0.35 kpc per resolution element, or a
factor of nearly 6 better than the resolution of a distant GZ2-CNRG
object like PGC 1224876.

Table~\ref{tab:resol} allows us to evaluate the meaning of ``poor
resolution" using nuclear rings, which are among the chief
morphological features that can be lost to resolution. The table is
focussed on four GZ2-CNRG cases which are illustrated in section 5:
NGC 3767 (Figure~\ref{fig:inner-examps}), and NGC 5132, NGC 5945, and
IC 816 (Figure~\ref{fig:cataclysmics}). The analysis is based on $g$-
and $i$-band images downloaded using the SDSS DAS Query form. In each
case, a nuclear ring is seen in a $g-i$ colour index map, but the
feature is not necessarily evident in the SDSS colour image or
classified the same way in a $g$-band image. The four galaxies have
nearly the same redshift ($<z>$ = 0.022), which is well below the
GZ2-CNRG median redshift of 0.062.

To evaluate the detectability of the nuclear features in $g$-band
images, each was visually mapped on a computer monitor using IRAF
routine TVMARK. These mappings were then fitted with an ellipse to get
the semi-major axis radius $a$(nf) of the nuclear feature (nf) in
pixels. For each image also, the mean FWHM
of the psf in pixels was also determined from several
foreground stars using IRAF routine IMEXAM. Using distances from NED,
the semi-major axis radii in kpc were determined and listed in row 11 of
Table~\ref{tab:resol}. These are found to range from 1.3 to 3.2 kpc,
well above the average nuclear ring radius of 0.75 kpc determined by
Buta \& Crocker (1993) and the average size found in the AINUR
(Comer\'on et al. 2010). The AINUR differs from the GZ2-CNRG in that
the former catalogue is a well-defined statistical all-sky sample that includes
all galaxies having total $B$-band magnitudes $B_T$ $\leq$ 12.5.
Additional requirements for inclusion in the AINUR were not too high an
inclination, and the availability of a high resolution image of the
central region.

Table~\ref{tab:resol} shows that $a$(nf)/psf (psf=$<$FWHM$>$)
ranges from 2.2 to 3.6. Of the four cases, the nuclear features are
seen as nuclear rings in NGC 5132 and 5945 in both the $g$-band and the
SDSS colour images. These galaxies have $a$(nf)/psf at the higher end,
$\approx$3.7. In contrast, IC 816 has $a$(nf)/psf = 2.5; in this
galaxy, the apparent nuclear ring seen in a $g-i$ colour index map appears
as a nuclear lens in both the $g$-band image and the SDSS colour image.
The case of NGC 3767, which has $a$(nf)/psf = 2.2, may indicate a
limit. Like IC 816, the $g-i$ nuclear ring appears as a nuclear lens in
the $g$-band image, but the feature is less recognizable in the SDSS
colour image and did not become part of the classification in
Table~\ref{tab:catalog}.

Also compiled in Table~\ref{tab:resol} are the same parameters for the
inner features (if) and outer features (of) of the same four galaxies.
Relative to the point spread function, $a$(if)/psf ranges from 15.8 in
NGC 3767 to 38.3 for NGC 5945, while $a$(of)/psf ranges from 35.2 for
NGC 3767 to 69.8 for NGC 5945. For IC 816, which has a very
well-defined inner ring, we can deduce that in order for its inner ring
to have $a$(if)/psf=2.5 in an SDSS $g$-band image, it would have to be
viewed from a redshift of 0.18. By the same token, in order for the
outer R$_1$ ring of NGC 5945 to have $a$(of)/psf = 2.5, the redshift would
have to be about 0.5.  Thus, while inner rings will likely become too
small in angular size to detect near the high redshift ($z$ $\approx$
0.2) end of the GZ2-CNRG sample, outer rings should still be detectable
as rings.

The detectability of outer rings will also depend on surface
brightness. The analysis of B17 provided bar minor axis luminosity
profiles in the $g$-band for 50 galaxies having R$_1$, R$_1^{\prime}$,
R$_2^{\prime}$, and R$_1$R$_2^{\prime}$ outer features. These profiles
allow us to estimate the typical surface brightnesses of these
features perpendicular to the bar/oval axis. Using the data tables
which gave Figure A3 of B17, the peak or a representative surface
brightness of the outer ``bump" in each profile was extracted. These
were then divided into ``R$_1$" and ``R$_2$" groups based on the
classifications in Table~\ref{tab:catalog}. If the classification is
R$_1$R$_2^{\prime}$, then an object was placed into the bin of the
stronger feature. In two cases (IC 2628 and UGC 4596), the R$_1$ and
R$_2^{\prime}$ components are equally prominent, and each galaxy was
placed into both bins. This gave for 17 R$_1$ cases: $<\mu>$(R$_1$) =
24.8 $\pm$ 0.2 mag arcsec$^{-2}$ (standard deviation 0.8 mag
arcsec$^{-2}$); and for 14 R$_2$ cases: $<\mu>$(R$_2$) = 23.7 $\pm$ 0.1
mag arcsec$^{-2}$ (standard deviation 0.5 mag arcsec$^{-2}$). Thus, for
the subsample of the GZ2-CNRG selected by B17, R$_2$ outer rings and
pseudorings are on average higher in surface brightness than R$_1$
rings by about 1 mag arcsec$^{-2}$.

From this analysis, we can deduce that in a sample of galaxies like the
GZ2-CNRG, with a median redshift of 0.06, nuclear rings and lenses will
be very underabundant, and that the nuclear features we do see will be
rare cases on the high end of the linear and relative sizes of such
features. The small-size end of inner rings (as found in the ARRAKIS
study) will also be increasingly lost in the higher redshift half of
the sample. Outer rings, however, are discernible across the full
redshift range.  

\begin{table*}
\centering
\setcounter{table}{1}
\caption{CVRHS Classifications for the Galaxy Zoo 2 Ring Sample. Col. 1: NED
Name or SDSS name; col. 2: Principal Galaxy Catalogue number; col. 3: GZ2
image file name; col. 4: redshift; col. 5: numerical stage index; col. 6:
classification; col. 7: notes. The full table is available in Appendix A and online.}
 \label{tab:catalog}
 \begin{tabular}{lrrcrll}
 \hline
 2MASX/SDSS & PGC & file & $z$ & $T$ &                   CVRHS Type & Notes \\
 Other name &  &  &  &  &  &  \\
 1 & 2 & 3 & 4 & 5 & 6 & 7 \\
 \hline
{\bf $\rightarrow$ RA: 7$^h$ $\leftarrow$} & & & & \\                                                                                                                                                             
 J07242070+4209004  & 2193308 & 108110 & 0.057     &    2.0 & SB(r,bl)ab:                                                                            & poorly resolved; (r)  \\                                   
\phantom{PGC 00}"    & "\phantom{00} & "\phantom{00} & "\phantom{0} & "\phantom{0} & \phantom{00000}" & dark-spacer; strong bar  \\                                                                               
 J07250545+4125376  & 2180937 & 107884 & 0.109     &    1.0 & (R$_2^{\prime}$L)SB(s)a                                                                & very poorly resolved  \\                                   
 J07271769+4347053  & 2229046 & 109317 & 0.056     &    0.0 & (R$^{\prime}$)SA(l)0/a                                                                 & poorly resolved, but a good  \\                            
\phantom{PGC 00}"    & "\phantom{00} & "\phantom{00} & "\phantom{0} & "\phantom{0} & \phantom{00000}" & case  \\                                                                                                  
 J07285313+4118282  & 2179018 & 107367 & 0.058     &    0.0 & (R$_1$)SAB$_a$(rs)0/a                                                                  & face-on with very oval (rs);  \\                           
\phantom{PGC 00}"    & "\phantom{00} & "\phantom{00} & "\phantom{0} & "\phantom{0} & \phantom{00000}" & large \Lff\ dark-spacer  \\                                                                               
 J07301470+3551069  & 3721760 &  11644 & .....     &    1.0 & (R$^{\prime}$)SB(r)a                                                                   & excellent bar and (r)  \\                                  
 J07312520+3646309  & 2086754 &   9571 & 0.059     &    1.0 & (R$_2^{\prime}$)S$\underline{\rm A}$B(l)a                                              & \Lff\ dark-spacer  \\                                      
 J07312646+3557323  & 3721973 &   9171 & .....     &    1.0 & (R$_2^{\prime}$)SAB$_a$(l)a                                                            & better resolved  \\                                        
 J07313762+4329525  & 2222182 & 108904 & 0.126     &    0.0 & (R$_2^{\prime}$)SAB(l)0/a:                                                             & poorly resolved  \\                                        
 J07322401+4012443  & 2162405 & 106708 & 0.116     &    2.0 & (R$_1^{\prime}$)SAB(r)ab:                                                              & very poorly resolved; \RoneP\  \\                          
\phantom{PGC 00}"    & "\phantom{00} & "\phantom{00} & "\phantom{0} & "\phantom{0} & \phantom{00000}" & dimpled  \\                                                                                               
 J07345763+4144261  & 2186051 & 105775 & 0.081     &    3.0 & SA(rs)b:                                                                               & strong blue ansae  \\                                      
 J07354716+3810311  & 2119877 &   9585 & 0.073     &    1.0 & (R$_2^{\prime}$)S$\underline{\rm A}$B(l)a:                                             & poorly resolved  \\                                        
 J07363185+3830585  & 3441140 &   9587 & 0.073     &    2.0 & (R$^{\prime}$)SAB(r)ab:                                                                & star superimposed  \\                                      
 J07364249+2918181  & 1862128 &  27499 & 0.089     & $-$1.0 & (RL)S$\underline{\rm A}$B(l)0$^+$                                                      & not a dark-spacer  \\                                      
 J07374632+4657062  & 3723004 & 109152 & 0.127     &    1.0 & SA(r)a:                                                                                & very poorly resolved; small  \\                            
\phantom{PGC 00}"    & "\phantom{00} & "\phantom{00} & "\phantom{0} & "\phantom{0} & \phantom{00000}" & close companion  \\                                                                                       
 J07375367+2951031  & 1880523 &  27511 & 0.097     &    2.0 & (R$_1^{\prime}$)SAB$_a$(r$^{\prime}$l)ab                                               & \rpl\ is circular; \RoneP\ is  \\                          
\phantom{PGC 00}"    & "\phantom{00} & "\phantom{00} & "\phantom{0} & "\phantom{0} & \phantom{00000}" & very elongated  \\                                                                                        
 \hline
 \end{tabular}
 \end{table*}

\begin{table*}
\centering
\caption{Effects of resolution on morphological features in
SDSS images of four galaxies. Row 1: the redshift from NED;
row 2: the mean full width at half maximum (FWHM) of the point spread function
(psf) on downloaded $g$-band images in pixels, where 1 pix
= 0\rlap{.}$^{\prime\prime}$396; rows 3-5: the semi-major axis radius of
nuclear features (nf), inner features (if), and outer features (of)
detected in the four galaxies, in pixels, based on visual mappings of
the features on a monitor and ellipse fits to the mapped points;
rows 6-8: the semi-major axis radii in rows 3-5 relative to the mean
FWHM of the psf on the images; rows 9-11: how the nuclear feature is
classified on the $g$-band image, in a $g-i$ colour index map, and as
seen in an SDSS colour image, respectively; row 9: inner feature
classification; row 10: outer feature classification; rows 11-13:
linear diameters of features based on distances from NED. The feature radii
are based on deprojected images for NGC 5132, NGC 5945, and IC 816;
no deprojection was needed for NGC 3767, which is already nearly
face-on. 
}
\label{tab:resol}
\begin{tabular}{lcccc}
\hline
Parameter & NGC 3767 & NGC 5132 & NGC 5945 & IC 816 \\
\hline
$z$           &  0.021   & 0.024      &  0.018     &  0.024   \\
$<FWHM>$ psf $g$-band (pix) & 3.65 & 4.40      &  2.89      &   3.53   \\  
$a$(nf)(pix)  &    8.1   &   16.9     &   10.1     &   8.9    \\
$a$(if)(pix)  &   57.8   &   98.2     &  110.8     &  68.7    \\
$a$(of)(pix)  &  101.7   &  157.6     &  201.8     & 174.1:   \\
$a$(nf)/psf   &    2.2   &    3.8     &    3.5     &   2.5    \\
$a$(if)/psf   &   15.8   &   22.3     &   38.3     &  19.4    \\
$a$(of)/psf   &   35.2   &   35.8     &   69.8     &  49.3    \\
nf morphology $g$-band & nl &   nr       &    nr      &   nl     \\
nf morphology $g-i$    & nr &   nr       &    nr      &   nr     \\
nf morphology SDSS colour & .... &    nr    & nr  & nl   \\
if morphology &   l      &  $\underline{\rm r}$s & rs & $\underline{\rm r}$s \\
of morphology &  R$_1$   &  R$_1^{\prime}$  & R$_1$  & R$^{\prime}$? \\
$a$(nf) (kpc) &   1.3    &  3.2  &  1.5   &  1.7 \\
$a$(if) (kpc) &   9.6    & 18.9  & 16.4   & 12.8 \\
$a$(of) (kpc) &  16.8    & 30.3  & 29.9   & 32.4: \\
\hline  
\end{tabular}
\end{table*}

\begin{figure}
\includegraphics[width=\columnwidth]{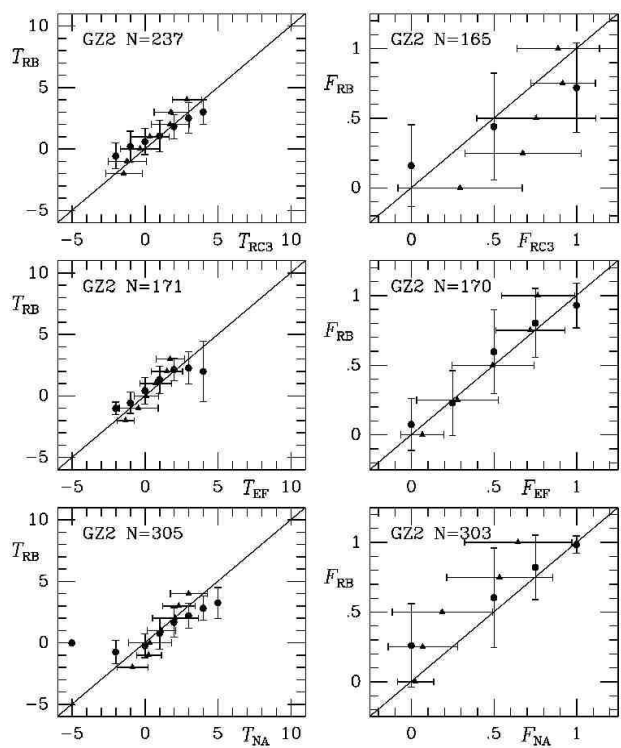}
\caption{Comparison of Table 2 stages and families with other sources:
RC3 (de Vaucouleurs et al. 1991); EFIGI (Baillard et al. 2011); and Nair \& Abraham
(2010)}
\label{fig:comps}
\end{figure}

Figure~\ref{fig:comps} compares the numerical stage indices $T$ (listed
specifically in Table~\ref{tab:catalog}) and family indices $F$ (not
listed specifically in Table~\ref{tab:catalog}) with those from other
sources: RC3 (de Vaucouleurs et al.  1991), the EFIGI Catalogue
(Baillard et al. 2011), and Nair \& Abraham (2010=NA2010). The
catalogues were matched using PGC numbers from cross-indices based on
NED. The different catalogues have samples defined in different ways,
which accounts for the unequal sample sizes. Also, many RC3 galaxies
which have a stage classification do not necessarily have a family
classification.

The stage ranges from $T$ = $-$5 (type E) to $T$=10 (type Im) on the
RC3 scale. The family code is that used by Baillard et al. (2011) but
is mapped here to mean $F$ = 0.00, 0.25, 0.50, 0.75, and 1.00 for SA,
S$\underline{\rm A}$B, SAB, SA$\underline{\rm B}$, and SB,
respectively. This breakdown was only possible for the EFIGI sample;
RC3 only gives SA, SAB, and SB classifications, i.e., no underlines,
and NA2010 do not have an S$\underline{\rm A}$B category. The
classifications listed in RC3 are based almost entirely on photographic
image material, while the EFIGI and NA2010 catalogues are based on SDSS
images (colour images for EFIGI, and $g$-band images for NA2010).

Table~\ref{tab:fourway} collects the results of a four-way comparison
between the Table~\ref{tab:catalog}, RC3, EFIGI, and NA2010
comparisons. For each comparison (six total), the rms dispersion
$\sigma_{ij} = \sqrt{{1\over N}\Sigma(T_j-T_i)^2}$ is listed and linear
least squares was used to derive the individual $\sigma_i$.  On
average, the external agreement among sources is $\sigma_i$$\approx$1.0
stage intervals, which is typical of visual classification (Naim et
al. 1995).

\section{Catalogue Contents}

\subsection{Cue for GZ2}

The first question to ask is what features did the GZ2 volunteers cue
on when they selected the ``Is the peculiar feature a ring?" button on
the GZ2 interactive webpages? That is, how effective was the GZ2
approach for selecting galaxies with rings? As described in the previous
section, the selected galaxies have virtually the same kinds of ring
phenomena as those that are contained in the CSRG, i.e, mainly inner
and outer rings, and their pseudoring and lens counterparts. This does
not mean that every galaxy in the SDSS footprint that actually has one
or more of the standard ring types actually made it into the GZ2-CNRG;
the ``Is there anything odd?" button had other choices, and not all
cases may have been recognized. Also, galaxies imaged in the SDSS after
DR7 will not necessarily be included.

We have seen from Table~\ref{tab:resol} that nuclear rings are not
expected to be present in significant numbers in
Table~\ref{tab:catalog} because the features are small and SDSS colour
images do not reveal them very well. Only for relatively nearby
galaxies or distant galaxies with much larger than average rings are
such features detectable in the GZ2 sample in SDSS colour images.  Inner
rings and pseudorings are much larger than nuclear rings, being about
the size of a bar if present. Except for weak inner pseudorings, these
are well-represented in the catalogue. Outer rings and pseudorings are
also well-represented, including especially the BC91 resonant
subclasses R$_1$, R$_1^{\prime}$, R$_2^{\prime}$, and
R$_1$R$_2^{\prime}$. In a sample where the median redshift is nearly
0.06, the ``cue" for most GZ2 participants likely was an outer ring or
pseudoring, because these would be detectable at greater distances due
to their larger sizes.  This was confirmed by the analysis in the
previous section. Willett et al. (2013) also comment that ``In the
absence of specific instructions on different types of ring ..., GZ2
classifiers are much more likely to identify outer rings."

\begin{table}
\centering
\caption{External Agreement Between Classifications.
Each column gives the rms dispersion between the two
sources ($i$ and $j$) indicated. The number in parentheses next to each
value is the number of galaxies in each comparison. The individual
$\sigma_i$ are derived from a linear least squares analysis. Sources:
EFIGI=Baillard et al. (2011); NA2010=Nair \& Abraham (2010); RC3=de
Vaucouleurs et al. (1991)}
\label{tab:fourway}
\begin{tabular}{cccccc}
\hline
Source & & RB & EFIGI & {NA2010} & {RC3} \\
& $i\rightarrow$ & 1 & 2 & 3 & 4 \\
& $j$ &  &  & & \\
& $\downarrow$ & & & & \\
\hline
       &      &             &             &             &             \\
EFIGI  &    2 &   1.22(171) &  .......... &  .......... & ..........  \\
NA2010 &    3 &   1.46(305) &   1.20( 60) &  ..........  & .......... \\
RC3    &    4 &   1.43(237) &   1.36(173) &   1.67( 65)  & .......... \\
       &      &             &             &             &  \\
$\sigma_i(T)$  & & 0.94 & 0.66 & 1.11 & 1.18  \\
\hline
\end{tabular}
\end{table}

\begin{table*}
\caption{Inventory of outer features. Col.1: classification; col. 2:
number of galaxies}
\label{tab:orinvents}
\begin{tabular}{lrlrlrlrlrlr}
\hline
Type & $n$ & Type & $n$ & Type & $n$ & Type & $n$ & Type & $n$ & Type & $n$ \\
 1 & 2 & 1 & 2 & 1 & 2 & 1 & 2 & 1 & 2 & 1 & 2 \\
\hline
(R$^{\prime}$)                                     & 734 & 
(R$^{\prime}$,L,R$_1^{\prime}$)                    &   1 & 
sum R                                              & 401 & 
(L:)                                               &   2 & 
(R$_1$L:)                                          &   1 & 
(R$_1^{\prime}$R$_2^{\prime}$L)                    &   1 \\ 
(R$^{\prime}$:)                                    &  26 & 
sum R$^{\prime}$                                   & 793 & 
($\underline{\rm R}$L)                             &   4 & 
(L?)                                               &   1 & 
sum R$_1$L                                         &  45 & 
(R$_1$2P)                                          &  10 \\ 
(R$^{\prime}$?)                                    &   6 & 
(R)                                                & 372 & 
(RL)                                               & 225 & 
(L,R)                                              &   8 & 
(R$_1^{\prime}$)                                   & 468 & 
sumR$_1$R$_2^{\prime}$                             & 218 \\ 
(R$^{\prime}$,L)                                   &   2 & 
(R:)                                               &  11 & 
(RL:)                                              &   6 & 
(L,L)                                              &   1 & 
(R$_1^{\prime}$:)                                  &  10 & 
(R$_2$)                                            &   3 \\ 
(R$^{\prime}$,R)                                   &   9 & 
(R?)                                               &   4 & 
(RL,R)                                             &   3 & 
(L,R$_1$)                                          &   1 & 
sum R$_1^{\prime}$                                 & 478 & 
(R$_2^{\prime}$)                                   & 505 \\ 
(R$^{\prime}$,R$^{\prime}$)                        &   5 & 
(RR)                                               &   2 & 
(RL,R$^{\prime}$)                                  &   1 & 
(L,R$^{\prime}$)                                   &   2 & 
(R$_1^{\prime}$L)                                  &   8 & 
(R$_2^{\prime}$:)                                  &   2 \\ 
(R$^{\prime}$,RL)                                  &   3 & 
(R,R$^{\prime}$)                                   &   3 & 
sum RL                                             & 239 & 
(L,R$_1^{\prime}$)                                 &   1 & 
sum R$_1^{\prime}$L                                &   8 & 
sum R$_2^{\prime}$                                 & 510 \\ 
(R$^{\prime}$,R$^{\prime}$L)                       &   1 & 
(R,R$^{\prime}$L)                                  &   1 & 
(R$^{\prime}$L)                                    & 155 & 
(L,R$^{\prime}$,R$^{\prime}$)                      &   1 & 
(R$_1^{\prime}$,R$_1$R$_2^{\prime}$)               &   1 & 
(R$_2^{\prime}$L)                                  &   4 \\ 
(R$^{\prime}$,R$_1$L)                              &   1 & 
(R,L)                                              &   3 & 
(R$^{\prime}$L,R)                                  &   1 & 
sum L                                              & 127 & 
(R$_1$R$_2^{\prime}$)                              & 188 & 
sum R$_2^{\prime}$L                                &   4 \\ 
(R$^{\prime}$,R$^{\prime}$,L)                      &   2 & 
(R:,L)                                             &   1 & 
(R$^{\prime}$L,R$_1$)                              &   1 & 
(R$_1$)                                            & 222 & 
(R$_1$R$_2^{\prime}$:)                             &   1 & 
.......... &   0 \\ 
(R$^{\prime}$,R,R)                                 &   1 & 
(R,R$_1^{\prime}$)                                 &   1 & 
sum R$^{\prime}$L                                  & 157 & 
(R$_1$:)                                           &   5 & 
(R$_1$R$_2^{\prime}$L)                             &   3 & 
.......... &   0 \\ 
(R$^{\prime}$,R$^{\prime}$,R)                      &   1 & 
(R:,R$_1^{\prime}$)                                &   1 & 
(R$\underline{\rm L}$)                             &   9 & 
sum R$_1$                                          & 227 & 
(R$_1$R$_2$L)                                      &   2 & 
.......... &   0 \\ 
(R$^{\prime}$,R$^{\prime}$,R$^{\prime}$)           &   1 & 
(R,R$_2^{\prime}$)                                 &   2 & 
(L)                                                & 101 & 
(R$_1$L)                                           &  44 & 
(R$_1^{\prime}$R$_2^{\prime}$)                     &  12 & 
.......... &   0 \\ 
\hline
\end{tabular}
\end{table*}

\begin{table*}
\caption{Inventory of inner features. Col.1: classification; col. 2:
number of galaxies}
\label{tab:irinvents}
\begin{tabular}{lrlrlrlrlrlr}
\hline
Type & $n$ & Type & $n$ & Type & $n$ & Type & $n$ & Type & $n$ & Type & $n$\\
 1 & 2 & 1 & 2 & 1 & 2 & 1 & 2 & 1 & 2 & 1 & 2 \\
\hline
(s)                       &   391 & 
(rs,nd)                   &     1 & 
($\underline{\rm r}$s,$\underline{\rm r}$s,l)               &     1 & 
sum $\underline{\rm r}$s                   &   508 & 
(rl)                      &   228 & 
sum r$^{\prime}$l                   &   211 \\ 
(s:)                      &    38 & 
(rs,nr)                   &     1 & 
($\underline{\rm r}$s,bl)                  &   101 & 
(r)                       &   605 & 
(rl:)                     &     1 & 
(r$\underline{\rm l}$)                     &     4 \\ 
(s,bl)                    &     3 & 
(rs,r)                    &     4 & 
($\underline{\rm r}$s,bl,nl)               &     1 & 
(r:)                      &     5 & 
(rl,bl)                   &    28 & 
(r$\underline{\rm l}$,bl)                  &     1 \\ 
(s,bl,nl)                 &     1 & 
(rs,r,bl)                 &     1 & 
($\underline{\rm r}$s,bl,nr)               &     1 & 
(r,bl)                    &   168 & 
(rl,l)                    &     1 & 
sum r$\underline{\rm l}$                   &     5 \\ 
(s,nb)                    &     1 & 
(rs,r,l)                  &     1 & 
($\underline{\rm r}$s,l)                   &     2 & 
(r,l)                     &     6 & 
(rl,nl)                   &     2 & 
(l)                       &   973 \\ 
(s:,bl)                   &     1 & 
(rs,rs)                   &     4 & 
($\underline{\rm r}$s,nb)                  &     2 & 
(r,nb)                    &     1 & 
(rl,tb)                   &     1 & 
(l:)                      &    17 \\ 
(sl)                      &     1 & 
(rs,rs,bl)                &     1 & 
($\underline{\rm r}$s,nd)                  &     1 & 
(r,nl)                    &     3 & 
(p,rl)                    &     4 & 
(l:,bl)                   &     1 \\ 
sum s                     &   436 & 
(rs,rs,l)                 &     1 & 
($\underline{\rm r}$s,nd?)                 &     1 & 
(r,nr?)                   &     1 & 
sum rl                    &   265 & 
(l,$\underline{\rm r}$s)                   &     2 \\ 
(r$\underline{\rm s}$)                     &     4 & 
(rs,s)                    &     1 & 
($\underline{\rm r}$s,nl)                  &     3 & 
(r,nrl)                   &     3 & 
(r$^{\prime}$l)                    &   189 & 
(l,bl)                    &     9 \\ 
(r$\underline{\rm s}$,bl)                  &     1 & 
(p,rs)                    &     4 & 
($\underline{\rm r}$s,nr)                  &     2 & 
(r,r)                     &     1 & 
(r$^{\prime}$l:)                   &     3 & 
(l,nrl)                   &     2 \\ 
sum r$\underline{\rm s}$     &     5 & 
(p,rs,bl)                 &     2 & 
($\underline{\rm r}$s,nrl)                 &     1 & 
(r,tb)                    &     1 & 
(r$^{\prime}$l,bl)                 &     6 & 
(l,r,$\underline{\rm r}$s)                 &     1 \\ 
(rs)                      &   308 & 
(p,rs,rs)                 &     1 & 
($\underline{\rm r}$s,r)                   &     4 & 
(p,r)                     &    12 & 
(r$^{\prime}$l,nl)                 &     1 & 
(l,r$_d$)                   &     1 \\ 
(rs:)                     &     4 & 
(p,rs,rs,bl)              &     1 & 
($\underline{\rm r}$s,r,bl)                &     1 & 
(p,r,bl)                  &    15 & 
(r$^{\prime}$l,ns)                 &     1 & 
(l,r$\underline{\rm s}$)                   &     1 \\ 
(rs,$\underline{\rm r}$s)                  &     3 & 
sum rs                    &   365 & 
($\underline{\rm r}$s,rl)                  &     1 & 
(p,r,nl)                  &     1 & 
(r$^{\prime}$l,r)                  &     3 & 
(l,rs,bl)                 &     1 \\ 
(rs,bl)                   &    21 & 
($\underline{\rm r}$s)                     &   348 & 
($\underline{\rm r}$s,tb)                  &     1 & 
sum r                     &   822 & 
($\underline{\rm r}$$^{\prime}$l)                   &     2 & 
(l,tb)                    &     1 \\ 
(rs,r$^{\prime}$l)                 &     1 & 
($\underline{\rm r}$s,$\underline{\rm r}$s)                 &     1 & 
(p,$\underline{\rm r}$s)                   &    13 & 
($\underline{\rm r}$l)                     &     7 & 
(p,r$^{\prime}$l)                  &     5 & 
(ls)                      &     3 \\ 
(rs,l)                    &     5 & 
($\underline{\rm r}$s,$\underline{\rm r}$s,bl)              &     1 & 
(p,$\underline{\rm r}$s,bl)                &    22 & 
sum $\underline{\rm r}$l                   &     7 & 
(p,r$^{\prime}$l,bl)               &     1 & 
sum l                     &  1012 \\ 
\hline
\end{tabular}
\end{table*}

\begin{table*}
\caption{Inventory of Morphologies. Col.1: classification; Col. 2: number of galaxies; Col. 3: percentage of the total $N$ listed;
Col. 4: mean RC3 stage index for this classification; Col. 5: standard deviation of $<T>$}
\label{tab:invents}
\begin{tabular}{lrrrrlrrrr}
\hline
Type & $n$ & \%$N$ & $<T>$ & $\sigma_T$ & Type & $n$ & \%$N$ & $<T>$ & $\sigma_T$ \\
 1 & 2 & 3 & 4 & 5 & 1 & 2 & 3 & 4 & 5 \\
\hline
E               &    0 &  0.0 &   -5 &      & \uurl            &    7 &  0.2 & -0.7 &  0.5 \\
E$^+$           &    0 &  0.0 &   -4 &      & rl              &  265 &  7.3 &  0.4 &  1.5 \\
S0$^-$          &    3 &  0.1 &   -3 &      & rpl             &  211 &  5.8 &  1.0 &  1.4 \\
S0$^o$          &   49 &  1.2 &   -2 &      & \ruul            &    5 &  0.1 &  1.0 &  1.6 \\
S0$^+$          &  709 & 18.0 &   -1 &      & l               & 1012 & 27.8 &  0.8 &  1.5 \\
S0/a            &  747 & 18.9 &    0 &      & N               & 3636 &      &      &      \\
Sa              &  776 & 19.7 &    1 &      &                 &      &      &      &      \\
Sab             &  688 & 17.4 &    2 &      & \cRP             &  793 & 24.7 &  2.2 &  1.4 \\
Sb              &  753 & 19.1 &    3 &      & R               &  401 & 12.5 & -0.2 &  1.2 \\
Sbc             &  120 &  3.0 &    4 &      & RL              &  239 &  7.4 & -0.5 &  1.0 \\
Sc              &   70 &  1.8 &    5 &      & \ccRPL           &  157 &  4.9 &  0.8 &  1.0 \\
Scd             &   19 &  0.5 &    6 &      & L               &  127 &  4.0 & -0.3 &  1.4 \\
Sd              &    9 &  0.2 &    7 &      & \cRone           &  227 &  7.1 & -0.1 &  1.1 \\
Sdm             &    3 &  0.1 &    8 &      & \cRoneL          &   45 &  1.4 &  0.0 &  1.1 \\
Sm              &    2 &  0.1 &    9 &      & \cRoneP          &  478 & 14.9 &  1.6 &  1.0 \\
Im              &    0 &  0.0 &   10 &      & \cRonePL         &    8 &  0.2 &  0.5 &  1.1 \\
N               & 3948 &      &      &      & \cRoneRtwoP      &  221 &  6.9 &  1.4 &  1.0 \\
                &      &      &      &      & \cRtwoP          &  510 & 15.9 &  2.2 &  1.3 \\
SA              & 1119 & 28.4 &  0.9 &  1.7 & \cRtwoPL         &    4 &  0.1 &  0.2 &  0.5 \\
S$\underline{\rm A}$B           &  526 & 13.3 &  0.9 &  1.6 & N               & 3210 &      &      &      \\
SAB             & 1362 & 34.6 &  1.4 &  1.6 &                 &      &      &      &      \\
SA$\underline{\rm B}$           &  354 &  9.0 &  1.4 &  1.7 & ansae           &  722 & 18.2 &  1.0 &  1.5 \\
SB              &  580 & 14.7 &  1.3 &  1.7 & bl              &  389 &  9.8 &  0.8 &  1.4 \\
N               & 3941 &      &      &      & p               &   81 &  2.0 &  2.1 &  0.8 \\
                &      &      &      &      & nr              &    5 &  0.1 &  1.6 &  0.5 \\
s               &  436 & 12.0 &  2.3 &  1.6 & nrl             &    6 &  0.2 &  0.3 &  1.0 \\
r$\underline{\rm s}$            &    5 &  0.1 &  2.6 &  2.1 & nl              &   12 &  0.3 &  1.1 &  1.8 \\
rs              &  365 & 10.0 &  2.2 &  1.5 & RG              &    5 &  0.1 & .... &  ... \\
$\underline{\rm r}$s            &  508 & 14.0 &  1.6 &  1.3 & PRG             &   10 &  0.3 & .... &  ... \\
r               &  822 & 22.6 &  0.8 &  1.6 & N               & 3962 &      &      &      \\
\hline
\end{tabular}
\end{table*}

In fact, any sample of galaxies selected on the basis of ring
morphologies will have a number of built-in biases: (1) a bias towards
face-on galaxies, because rings are most easily detectable in low
inclination systems; (2) a bias towards including the intrinsically
largest rings of any type (a Malmquist-like bias); (3) a bias towards
selecting the highest surface brightness features; (4) a bias towards
selecting the highest contrast features, and (5) biases due to
integrated brightness limits and distance. 

Point 1 is especially true for an optical imaging sample like the
GZ2-CNRG. Inner rings tend to have a younger stellar population than do
bars, and are likely located in the thin part of the disc of any
galaxy. At high inclination, both foreshortening and extinction could
make it difficult to detect an inner ring. Point 2 affects all types of
rings, the extremes depending on the actual dispersion of the linear
sizes of the different ring types. Point 3 can become an especially
important issue with increasing redshift, owing to the significant
dispersion in surface brightness of galactic rings and because of the
$(1+z)^{-4}$ effect (Hubble \& Tolman 1935) on surface brightness.
Point 4 in a way affects how inner and outer features are classified,
either as well-defined rings, ring-lenses, or even as lenses. The
highest contrast rings are local maxima in surface intensity, not
merely ``bumps" in a luminosity profile. In principle, resolution could
affect the contrast of a ring and make it less evident, even in high
surface brightness regions.  Point 5 results from the natural
limitations imposed by the telescope, the detector, and the procedure
used for the survey. This could lead to systematic effects in
classification with increasing redshift, as discussed by Willett et al.
(2013).

\subsection{Inventory and Histograms}

Tables~\ref{tab:orinvents} and ~\ref{tab:irinvents} provide
inventories of outer and inner varieties, respectively. These parts
of CVRHS classification are more complex than family and stage
owing to the number of possible features placed between the bracketts.
Both tables highlight multiple inner and outer features, the latter
including at least one case with up to three distinct outer
pseudorings. Subtle doubled inner rings and pseudorings are also seen.
Several examples are illustrated and discussed in section 5. Inner
varieties have the added complexity of including nuclear features as
well as dust rings (r$_d$), triaxial bulges (tb), and plumes (p) or
secondary spiral arcs. 

\begin{figure}
\includegraphics[width=\columnwidth]{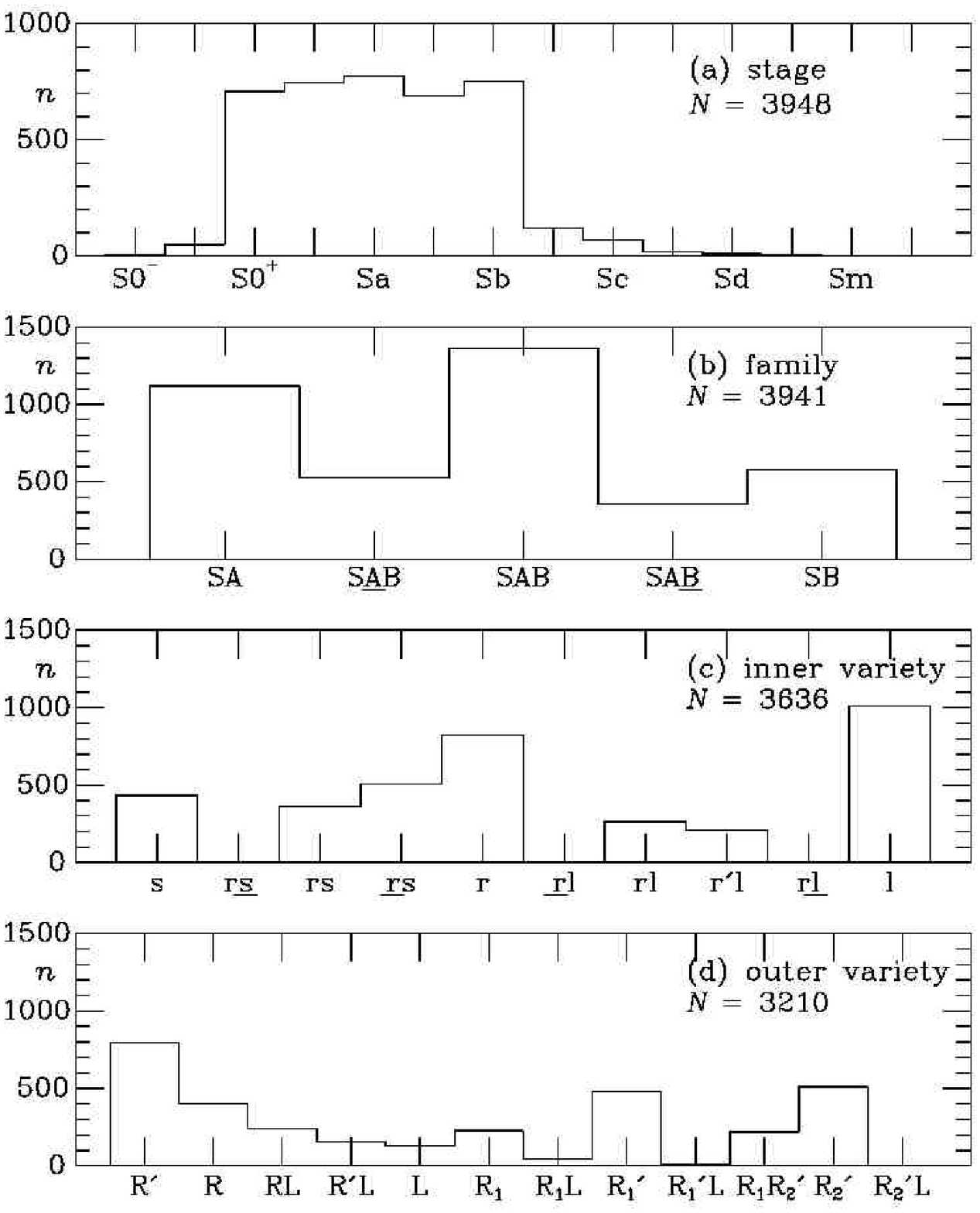}
\caption{Histograms of morphological features in the GZ2 sample, based
on Table~\ref{tab:invents}.}
\label{fig:histos}
\end{figure}

For the remainder of this section, only the sums in
Tables~\ref{tab:orinvents} and ~\ref{tab:irinvents} will be used. These are
transferred into Table~\ref{tab:invents} with additional inventories of
stage, family, and other morphological features the survey has revealed
for GZ2-CNRG galaxies; most are shown as the histograms in
Figure~\ref{fig:histos}. For family, inner variety, and outer variety,
Table~\ref{tab:invents} also gives the mean and standard deviation of
the de Vaucouleurs stage index $T$ for each listed subset.  The
distribution of stages shows a pronounced restriction to the CVRHS
stage range of S0$^+$ to Sb, accounting for nearly 93\% of the galaxies
where a stage classification could be made. The likely reason for this
is that zoo volunteers cued mainly on outer rings and not inner rings
when they selected the "Is the odd feature a ring?" button. As noted in
section 1, Comer\'on et al. (2014) found that the relative frequency of
outer rings and pseudorings drops significantly for stages later than
Sb, while inner rings remain relatively abundant to stage Sd at least.
The effect is compounded by the fact that inner rings are also largest
in both physical and relative size in the range S0$^+$ to Sb (de
Vaucouleurs \& Buta 1980; Buta \& de Vaucouleurs 1982; Comer\'on et al.
2014). These effects conspire to make late-type galaxies
under-represented in the catalogue.

The distribution of bar classifications in Figure~\ref{fig:histos} shows
a definite emphasis on weaker apparent bar strengths. Only 15\% of
3941 galaxies are classified in Table~\ref{tab:catalog} as type SB,
while 28\% are classified as SA. Weak bars are most common, at 57\%.
Many of these weaker bar features are in the form of ansae along
the apparent major axis of an oval, and it is likely that many of
those classified as SA also have an oval that is simply more difficult
to recognize.

\begin{figure}
\includegraphics[width=\columnwidth]{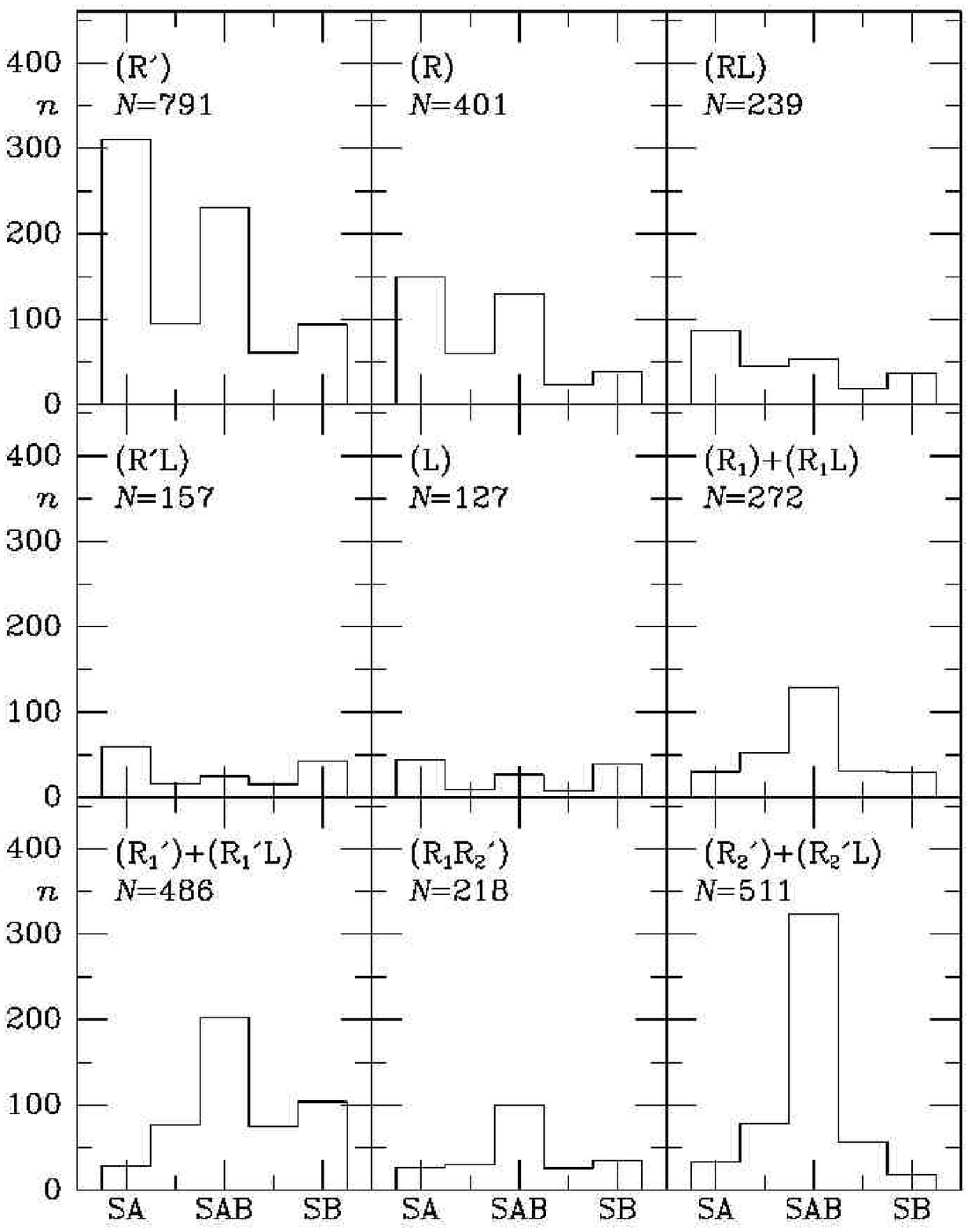}
\caption{Histograms of the distributions by family for different 
outer variety morphological features in the GZ2 sample. Each frame
gives the number of objects $N$ in the histogram. These may differ from
Table~\ref{tab:invents} because not all of the objects have a family
classification.}
\label{fig:combined-outer}
\end{figure}

\begin{figure}
\includegraphics[width=\columnwidth]{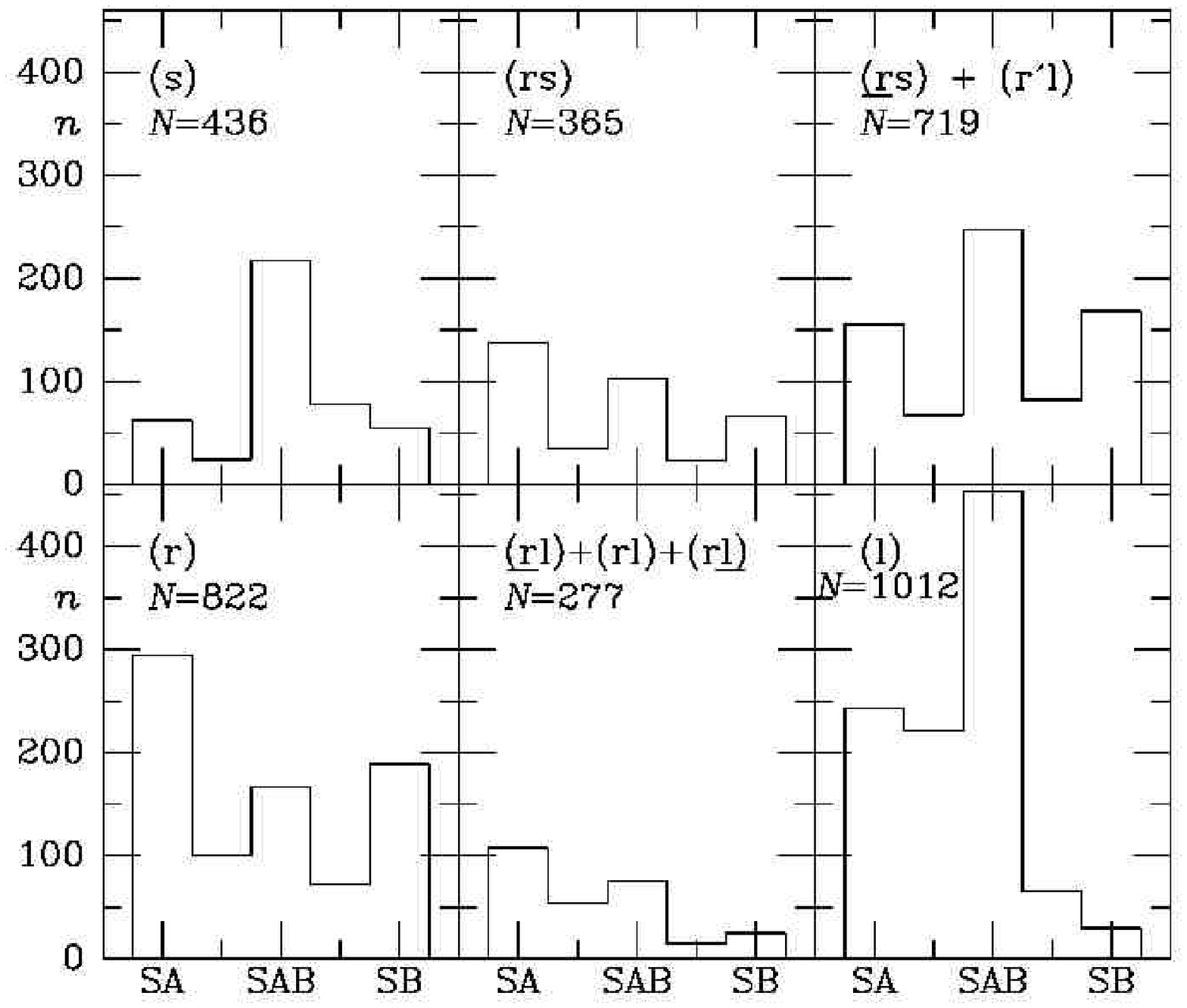}
\caption{Histograms of the distributions by family for different inner
variety morphological features in the GZ2 sample Each frame gives the
number of objects $N$ in the histogram. These may differ from
Table~\ref{tab:invents} because not all of the objects have a family
classification.}
\label{fig:combined-inner}
\end{figure}

Lenses are formally defined as morphological features of galaxies which
have a shallow brightness gradient interior to a sharp edge (Kormendy
1979). Although lenses were not the features that the sample was cued
on, they are still well-represented because many ringed galaxies also
have lenses. There is a lens analogue to each type of ring, and all
three are included in the CVRHS classifications in
Table~\ref{tab:catalog}.  An interesting result in
Table~\ref{tab:invents} is the predominance of inner lenses (l) among
the inner features recognized.  It is likely that this is partly a
resolution effect; a low contrast inner ring viewed at poor resolution
might be misclassified as an inner lens.  Well-defined inner rings and
pseudorings [types (rs), ($\underline{\rm r}$s), and (r)] make up
nearly 47\% of the 3636 galaxies having a variety classification.

Among outer features, the most common pattern recognized is an outer
pseudoring, (R$^{\prime}$), accounting for 25\% of 3210 features
classified in the categories listed in Table~\ref{tab:invents}.  These
are ring-like patterns formed from outer spiral arms that could not
be assigned to one of the outer resonant subclasses.  By comparison,
closed outer rings are recognized in 13\% of the cases.  More subtle
versions of these features, called outer ring-lenses (RL) and outer
pseudoring-lenses (R$^{\prime}$L), are at 8\% and 5\%, respectively.
Outer lenses (L), which are outer features with a more uniform
brightness distribution (Kormendy 1979), are found in 4\% of the
cases.

\begin{figure}
\includegraphics[width=\columnwidth]{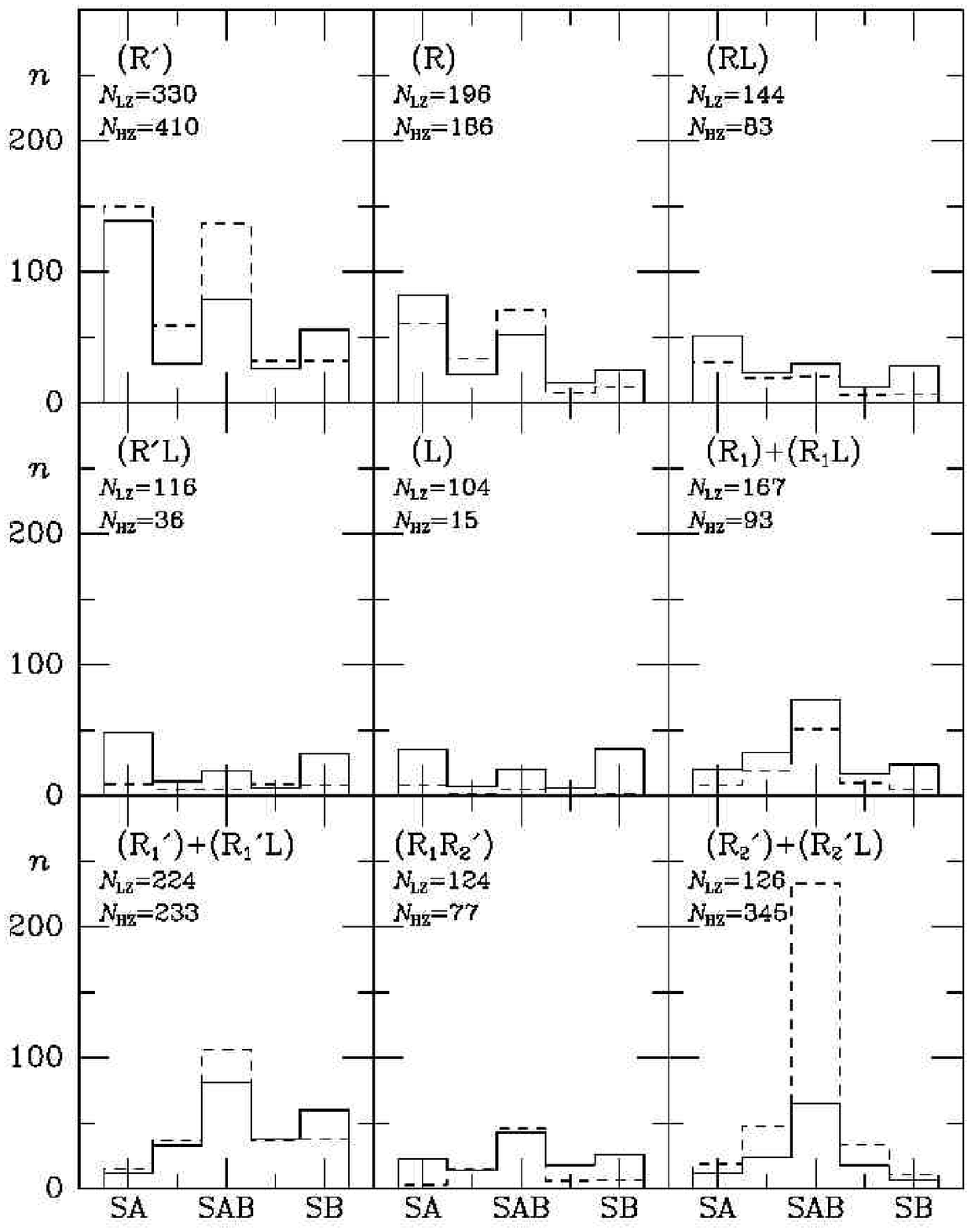}
\caption{Same as Figure~\ref{fig:combined-outer}, but divided into near
(solid histograms) and far (dashed histograms) redshift subsamples at
the median radial velocity of 18600 km s$^{-1}$. Those galaxies
lacking a redshift measurement are excluded from the histograms}
\label{fig:combined-outer-vel}
\end{figure}

\begin{figure}
\includegraphics[width=\columnwidth]{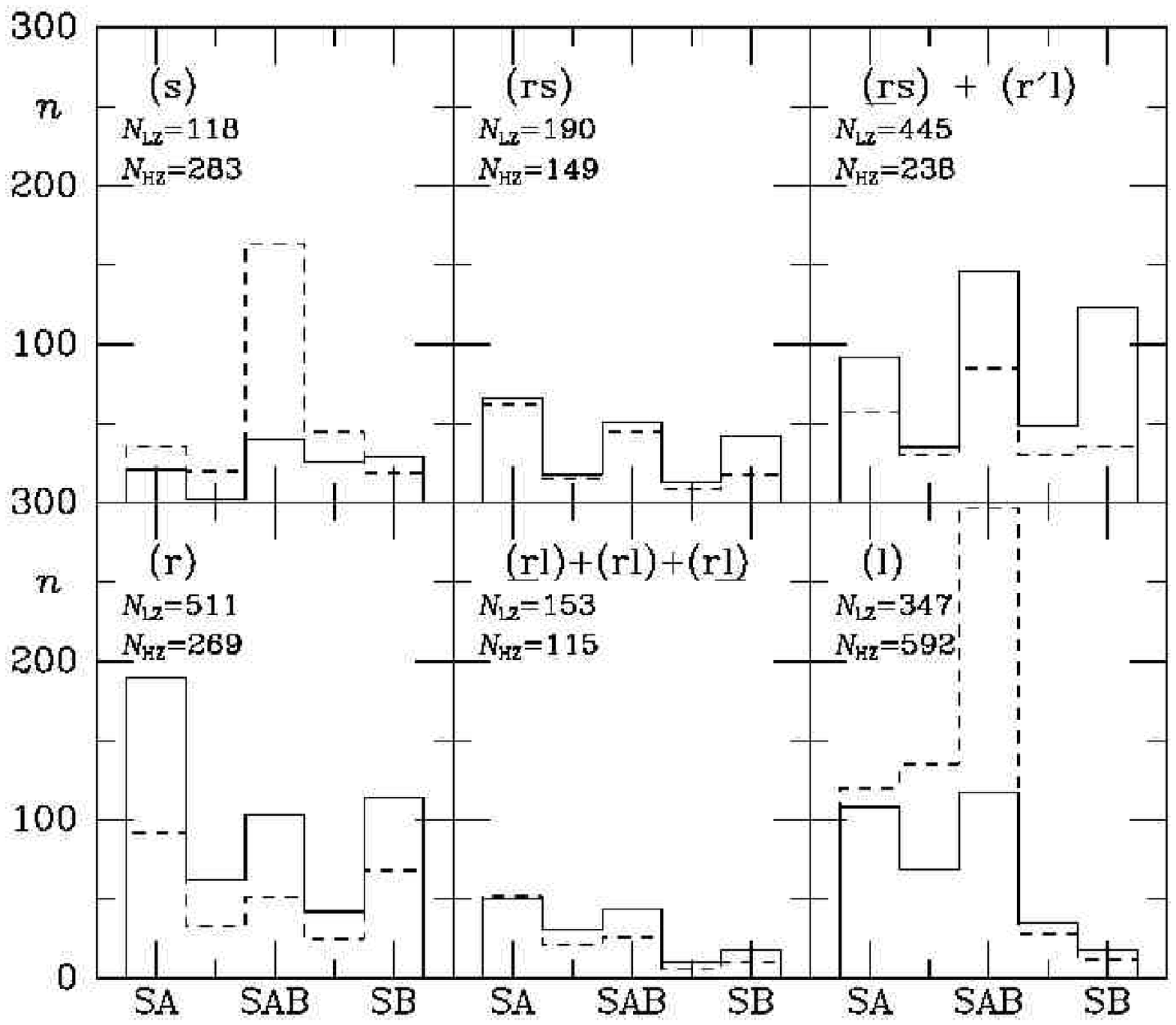}
\caption{Same as Figure~\ref{fig:combined-inner}, but divided into near
(solid histograms) and far (dashed histograms) redshift subsamples at
the median radial velocity of 18600 km s$^{-1}$. Those galaxies
lacking a redshift measurement are excluded from the histograms}
\label{fig:combined-inner-vel}
\end{figure}

The outer resonant subclass morphologies R$_1$, R$_1^{\prime}$,
R$_2^{\prime}$, R$_1$R$_2^{\prime}$, and R$_1$L are recorded in
46.5\% of the 3210 galaxies for which an outer variety could be
recognized. In the CSRG, R$_1$ and R$_1^{\prime}$ features are more
abundent than R$_2^{\prime}$ in the ratio n$_1$:n$_2$ = 0.64:0.36 =
1.78, similar to that found in the ARRAKIS study. From Table 4 we can derive
$n$(R$_1$+R$_1$L+R$_1^{\prime}$+R$_1^{\prime}$L+0.5xR$_1$R$_2^{\prime}$)
= $n_1$ = 869 and $n$(R$_2^{\prime}$ + R$_2^{\prime}$L + R$_2$ +
0.5xR$_1$R$_2^{\prime}$) = $n_2$ = 625, giving a ratio $n_1/n_2$ of
1.39. This assigns half of the R$_1$R$_2^{\prime}$ cases to R$_1$ and
half to R$_2^{\prime}$. It was shown in section 3 that R$_2^{\prime}$
rings have a higher average surface brightness than R$_1$ and
R$_1^{\prime}$ rings. Also, R$_2^{\prime}$ pseudorings occur in later
types on average ($<T>$=2.2) than R$_1^{\prime}$ pseudorings
($<T>$=1.6). These effects together could in part account for the
lower value of $n_1/n_2$ in the GZ2-CNRG.

The most distinctive cases showing outer resonant
morphologies are galaxies classified as (R$_2^{\prime}$)SAB(s) and
(R$_2^{\prime}$)SAB(l). An Sb example is NGC 210, which is not in the
GZ2-CNRG sample but is described in the Hubble Atlas (Sandage 1961) and
the deVA.

Some of these issues can be examined further in the histograms in
Figures~\ref{fig:combined-outer}-~\ref{fig:combined-inner}, which show the
distributions of family classifications for different ring or lens
feature classifications. With no restrictions on radial velocity, these
show some interesting family preferences for different features. For
example, no category of inner or outer features shown has its highest
$n$ in the SB bin. Instead, the (R), (RL), (R$^{\prime}$), and
(R$^{\prime}$L) categories have the highest number of galaxies in the
SA bins while outer resonant subclass features have their highest $n$ values in
the SAB bin. In fact, the (R$_2^{\prime}$) category has a very high $n$
in the SAB bin, with very few classified as SA or SB types. Only the
(R$_1^{\prime}$) category shows a comparable number of SB and SAB
classifications.

\subsection{Redshift Dependences}

To evaluate what these trends mean, it is instructive to break up the
sample into low redshift and high redshift halves. We take the dividing
line to be 18600 km s$^{-1}$, the median radial velocity of
the sample. Figures~\ref{fig:combined-outer-vel} and
~\ref{fig:combined-inner-vel} show the results, with solid lines for the
near subsample and dashed lines for the far subsample. Each graph also
shows the number of low redshift ($N_{LZ}$) and high redshift
($N_{HZ}$) galaxies in each subset. As for the full samples, the near
samples of outer resonant subclass features prefer SAB classifications over SB
classifications. Most noteworthy, however, is how drastically different
the near and far histograms are for the (R$_2^{\prime}$) category. The high
redshift sample has more than 2.5 times as many galaxies as does the
low redshift sample in this case. A similar effect is not seen in the
other outer resonant subclass categories, but a less extreme version is seen in
the (R$^{\prime}$) category.

There are several possible reasons for the excess of (R$_2^{\prime}$)
cases in the high redshift half of the sample: (1) There is a
classification bias causing many distant outer rings to be placed
incorrectly into the R$_2^{\prime}$ category; (2) R$_2^{\prime}$
pseudorings are larger in a relative sense, have higher surface
brightness, occur in later types, and have more star formation than
R$_1$ or R$_1^{\prime}$ rings, making them easier to recognize at
larger distances; (3) the high redshift subset samples a larger volume
of space than the near subset, and thus includes more examples of these
features.

All of these points could contribute to the observed excess of
R$_2^{\prime}$ features. For example, one category that would be
difficult to recognize at higher redshifts with SDSS images would be
(R$_1$R$_2^{\prime}$), the combined outer ring/pseudoring type thought
to be related to the two main families of periodic orbits near the OLR.
If the R$_1$ component is weak, such galaxies will be classified as
(R$_2^{\prime}$) at higher redshifts. 

\begin{table*}
\centering
\caption{GZ2-CNRG galaxies having different kinds of outer features. 
Col. 1: name of galaxy in
Table~\ref{tab:catalog}; col. 2: GZ2 image file name; col. 3:
classification from Table~\ref{tab:catalog}. }
\label{tab:list1}
\begin{tabular}{lrllrl}
\hline
Name & file.jpg & Type & Name & file.jpg & Type \\
 1 & 2 & 3 & 1 & 2 & 3 \\
\hline
                     &        &          &                      &        & \\ 
\noalign{\centerline{(a) Outer Rings (R)}}
                     &        &          &                      &        & \\ 
CGCG 007-032         &   6354 & (R)SA(r)0$^+$                                                                    & MESSIER 096          & 217755 & (R)SAB($\underline{\rm r}$s)ab                                                   \\
CGCG 038-119         &  18572 & (R)SAB$_a$(l)0$^+$                                                               & NGC 2859             &  84158 & (R)SAB$_a$(r$\underline{\rm l}$,bl)0$^+$                                         \\
CGCG 095-086         & 182256 & (R)SA(r)b                                                                        & NGC 3419             &  81341 & (R)SA(l)0$^+$                                                                    \\
CGCG 097-024         & 181013 & (R)SA$\underline{\rm B}$$_a$(l)0/a                                               & NGC 3945             &  35457 & (R)SAB$_a$(rl,bl)0$^+$                                                           \\
CGCG 107-049         & 134546 & (R)SA(rl)0$^+$                                                                   & NGC 4274             & 137726 & (R)SA$\underline{\rm B}$($\underline{\rm r}$s)ab                                 \\
CGCG 124-039         & 154666 & (R)S$\underline{\rm A}$B(rs,$\underline{\rm r}$s)a                               & NGC 4513             &   8697 & (R)E3 or (R)SA(l)0$^+$                                                           \\
CGCG 194-013         &  54624 & (R)SA(r)0$^+$                                                                    & UGC 04414            & 118577 & (R)SB(rs,bl)0/a                                                                  \\
CGCG 240-005         &  54789 & (R)SB(r,bl)0/a                                                                   & UGC 04498            &  45123 & (R)S$\underline{\rm A}$B$_a$(rl)0$^+$                                            \\
CGCG 267-055         &  48754 & (R)SA(r$^{\prime}$l)0/a                                                          & UGC 05025            & 188515 & (R)SA(rl)0$^+$                                                                   \\
CGCG 276-039         & 182927 & (R)SAB$_a$(r)0$^+$                                                               & UGC 05489            & 164782 & (R)SA(r)a                                                                        \\
IC 3003              & 130594 & (R)SAB(r,bl)0$^+$                                                                & UGC 05859            & 213092 & (R)SAB$_a$(r,bl)0$^+$                                                            \\
                     &        &          &                      &        & \\ 
\noalign{\centerline{(b) Outer Ring-Lenses (RL)}}
                     &        &          &                      &        & \\ 
CGCG 033-057         &  76542 & (RL)SB$_a$(l,bl)0$^+$                                                            & NGC 4227             & 132953 & (RL)SA(r)0$^+$                                                                   \\
CGCG 062-038         & 187438 & (RL)SB$_a$0$^+$                                                                  & NGC 4229             & 130971 & (RL)SB$_a$0$^+$                                                                  \\
CGCG 184-046         & 114568 & (RL)SB$_a$0$^+$                                                                  & NGC 4292             & 197910 & (RL)SAB(rs,bl)0/a                                                                \\
CGCG 210-023         &  73551 & (RL)SAB$_a$($\underline{\rm r}$s,bl)0/a                                          & NGC 4454             &   4806 & (RL)SAB(r,bl)0/a                                                                 \\
CGCG 241-027         & 240408 & (RL)SB(r$^{\prime}$l)0$^o$                                                       & NGC 4728             & 154884 & (RL)SAB$_a$(l)0$^+$                                                              \\
IC 2557              & 206992 & (RL)SAB(r)0$^+$                                                                  & NGC 4859             & 154587 & (RL)SA(r)0$^+$                                                                   \\
IC 3238              & 111978 & (RL)SA(l)0$^o$                                                                   & NGC 5052             & 136177 & (RL)SB$_a$(r$^{\prime}$l)0/a                                                     \\
NGC 2679             &  78662 & (RL)SAB(rl,bl)0$^+$                                                              & NGC 5261             &  31341 & (RL)SA(l)0$^+$                                                                   \\
NGC 2950             & 191488 & (RL)SB$_a$(l)0$^+$                                                               & UGC 04587            & 193313 & (RL)SA(r)0$^+$                                                                   \\
NGC 3300             &  81770 & (RL)SB$_a$(rs)0$^+$                                                              & UGC 09980            &  37546 & (RL)SAB$_a$($\underline{\rm r}$s)0$^+$                                           \\
NGC 3540             & 117778 & (RL)SB($\underline{\rm r}$s,bl)0$^+$                                             & UGC 10361            & 182880 & (RL)SB$_a$0$^+$                                                                  \\
                     &        &          &                      &        & \\ 
\noalign{\centerline{(c) Outer Lenses (L)}}
                     &        &          &                      &        & \\ 
CGCG 038-068         &  57633 & (L)SAB$_a$(r,bl)0$^+$                                                            & NGC 4643             &  14926 & (L)SB($\underline{\rm r}$s,bl)0/a                                                \\
CGCG 047-120         &  20648 & (L)SB(r,bl)0/a                                                                   & NGC 4745             & 154887 & (L)SB($\underline{\rm r}$s)0$^+$ pec                                             \\
CGCG 159-114         & 158541 & (L)SB($\underline{\rm r}$s,bl)0/a                                                & NGC 4919             & 158420 & (L)SA(rl)0$^+$                                                                   \\
CGCG 190-017         & 128957 & (L)SB(rs,bl)0$^+$                                                                & NGC 5321             & 123036 & (L)SB(r)0$^o$                                                                    \\
CGCG 224-100         &  14287 & (L)SAB(r,bl)0$^+$                                                                & NGC 5636             &  30251 & (L)SAB(rs)a                                                                      \\
NGC 4245             & 137723 & (L)SB($\underline{\rm r}$s,bl)0/a                                                & NGC 5770             &  20906 & (L)SB(r,bl)0$^+$                                                                 \\
NGC 4264             & 198847 & (L)SAB($\underline{\rm r}$s)0$^+$                                                & NGC 6243             &  97141 & (L)SA(r)0/a                                                                      \\
NGC 4300             & 185315 & (L)SA$_x$(r)0/a                                                                  & NGC 6310             &   8024 & (L)SA(r)0$^+$ sp                                                                 \\
NGC 4309             & 221805 & (L)SAB(r)0/a                                                                     & UGC 05380            &  29337 & (L)SB(r,bl)0/a                                                                   \\
NGC 4429             &  76175 & (L)SAB$_{ax}$0/a                                                                 & UGC 08225            & 215258 & (L)SA($\underline{\rm r}$s)a                                                     \\
                     &        &          &                      &        & \\ 
\noalign{\centerline{(d) Outer Pseudorings (R$^{\prime}$)}}
                     &        &          &                      &        & \\ 
CGCG 007-039         &  22057 & (R$^{\prime}$)SA(r$^{\prime}$l)b                                                 & NGC 3900             & 154463 & (R$^{\prime}$)SA(r)a / E                                                         \\
CGCG 033-007         &  25483 & (R$^{\prime}$)SA(r,l)ab                                                          & NGC 4290             &  86952 & (R$^{\prime}$)SB$_a$(rs)b                                                        \\
CGCG 034-058         &  74035 & (R$^{\prime}$)SA($\underline{\rm r}$s)b                                          & NGC 5779             & 200218 & (R$^{\prime}$)SA$\underline{\rm B}$(rl,bl)0/a                                    \\
CGCG 093-080         & 188347 & (R$^{\prime}$)SA(rs)b pec                                                        & NGC 6491             &  43420 & (R$^{\prime}$)SA(l)ab                                                            \\
CGCG 121-094         & 149592 & (R$^{\prime}$)SA($\underline{\rm r}$l)0/a                                        & UGC 04000            & 106009 & (R$^{\prime}$)SAB$_a$($\underline{\rm r}$s,bl)b pec                              \\
CGCG 154-032         & 133545 & (R$^{\prime}$)SA(r)b                                                             & UGC 04145            & 149170 & (R$^{\prime}$)SAB(r$^{\prime}$l)ab                                               \\
CGCG 266-056         &  63657 & (R$^{\prime}$)S$\underline{\rm A}$B(l)ab                                         & UGC 04864            & 185788 & (R$^{\prime}$)SA(rs,rs,l)b                                                       \\
MCG +07-23-032       & 224506 & (R$^{\prime}$)SAB$_a$(rs)bc                                                      & UGC 05080            &  52664 & (R$^{\prime}$)SA(l)0/a                                                           \\
NGC 2740             & 193203 & (R$^{\prime}$)SAB($\underline{\rm r}$s,bl)b                                      & UGC 06049            &  26413 & (R$^{\prime}$)SA(l)b                                                             \\
                     &        &          &                      &        & \\ 
\noalign{\centerline{(e) Multiple Outer Features }}
                     &        &          &                      &        & \\ 
CGCG 068-049         & 222706 & (RR)SA(r)0$^+$                                                                   & NGC 3594             &  43818 & (RL,R)SA$\underline{\rm B}$(r)0/a                                                \\
CGCG 074-143         &  91487 & (L,R)SA(rl)0$^+$                                                                 & NGC 4022             & 159222 & (L,R)SB$_a$(l)0$^o$                                                              \\
CGCG 108-061         & 141138 & (R,R$^{\prime}$)SA(r)ab                                                          & NGC 4391             &  29868 & (L,R)SA(l)0$^o$                                                                  \\
CGCG 197-002         &  72327 & (L,R)S$\underline{\rm A}$B(rl)0$^+$                                              & NGC 4457             &  16642 & (RL,R)S$\underline{\rm A}$B(l)0/a                                                \\
IC 0472              & 109367 & (RR)SA$\underline{\rm B}$$_a$(r$\underline{\rm s}$,bl)ab                         & NGC 5210             & 222265 & (R,R$^{\prime}$L)SA($\underline{\rm r}$s)a                                       \\
MCG +09-20-064       &  48658 & (R,L)SAB(r)0/a                                                                   & NGC 5977             & 178798 & (RL,R)SB(rs,bl)0/a                                                               \\
NGC 3013             & 115500 & (L,R)SAB($\underline{\rm r}$s,bl)0/a                                             & NGC 6028             & 141823 & (R$^{\prime}$,R$^{\prime}$,R$^{\prime}$)SAB$_a$(l)a                              \\
NGC 3151             &  84046 & (R)SB$_a$(r$^{\prime}$l)0$^+$                                           & SBS 0927+493         &  49936 & (L,R)SB$_a$0/a                                                                   \\
\hline
\end{tabular}
\end{table*}

\begin{table*}
\centering
\caption{GZ2-CNRG galaxies having outer resonant subclass features}
\label{tab:list2}
\begin{tabular}{lrllrl}
\hline
Name & file.jpg & Type & Name & file.jpg & Type \\
 1 & 2 & 3 & 1 & 2 & 3 \\
\hline
                     &        &          &                      &        & \\ 
\noalign{\centerline{(a) R$_1$ Outer Rings}}
                     &        &          &                      &        & \\ 
CGCG 004-035         & 243732 & (R$_1$)SAB$_a$(l)0/a                                                             & IC 0676              &  75285 & (R$_1$)SB(r$^{\prime}$l)0/a                                                      \\
CGCG 045-099         &  30192 & (R$_1$)SAB$_a$(rs)0/a                                                            & IC 2248              & 205327 & (R$_1$)SA(r)0$^+$                                                                \\
CGCG 065-002         &  82105 & (R$_1$)SB$_a$($\underline{\rm r}$s,bl)ab                                         & IC 3808              & 113413 & (R$_1$)SA($\underline{\rm r}$s)0/a                                               \\
CGCG 092-056         & 187092 & (R$_1$)SAB$_a$(l)0/a                                                             & MCG +06-32-024       & 101620 & (R$_1$)SAB(r)0/a                                                                 \\
CGCG 122-055         & 150005 & (R$_1$)SA$\underline{\rm B}$$_a$(l)0/a                                           & NGC 3767             & 181326 & (R$_1$)SA$\underline{\rm B}$(l)0$^+$                                             \\
CGCG 153-031         & 123705 & (R$_1$)SA(r)0$^+$                                                                & NGC 4221             &  30031 & (R$_1$)SAB(r,bl)0$^+$                                                            \\
CGCG 158-063         & 153033 & (R$_1$)S$\underline{\rm A}$B(r,bl)0$^+$                                          & NGC 5945             &  66991 & (R$_1$)SB(rs,nr)ab                                                               \\
CGCG 195-006         & 234553 & (R$_1$)SB$_a$(r$\underline{\rm s}$)a                                             & UGC 08661            & 171318 & (R$_1$)SAB($\underline{\rm r}$s)0/a                                              \\
CGCG 248-051         &  86613 & (R$_1$)S$\underline{\rm A}$B$_a$(r)0$^+$                                         & UGC 10320            & 121583 & (R$_1$)S$\underline{\rm A}$B(rs)a                                                \\
CGCG 250-035         & 201777 & (R$_1$)SAB($\underline{\rm r}$s,bl)0$^+$                                     & UGC 10374            & 184471 & (R$_1$)SAB$_a$(r,bl)0/a                                                          \\
                     &        &          &                      & & \\
\noalign{\centerline{(b) R$_1$L Outer Ring-Lenses}}
                     &        &          &                      & & \\
ARK 189              &  60240 & (R$_1$L)SA$\underline{\rm B}$($\underline{\rm r}$s)ab                            & IC 3199              &  63145 & (R$_1$L)SAB$_a$(rl,bl)0/a                                                        \\
CGCG 039-184         & 197254 & (R$_1$L)SAB(r)0$^+$                                                              & KUG 1000+468         &  51435 & (R$_1$L)SAB($\underline{\rm r}$s)0/a                                             \\
CGCG 067-004         & 215672 & (R$_1$L)SB$_a$(s,bl,nl)0/a                                                       & KUG 1119+250         & 170809 & (R$_1$L)SAB(r,bl)0$^+$                                                           \\
CGCG 088-041         & 120622 & (R$_1$L)SAB($\underline{\rm r}$s,bl)a                                            & NGC 3150             & 207351 & (R$_1$L)SAB($\underline{\rm r}$s)0$^+$                                           \\
CGCG 098-074         & 236780 & (R$_1$L)SB(r$^{\prime}$l,bl)0/a                                                  & NGC 3380             & 150955 & (R$_1$L)SAB(l,rs,bl)ab                                                           \\
CGCG 239-014         &  50813 & (R$_1$L)S$\underline{\rm A}$B(r)0/a                                              & NGC 3398             &  49023 & (R$_1$L)SAB($\underline{\rm r}$s)ab                                              \\
CGCG 262-055         & 108568 & (R$_1$L)S$\underline{\rm A}$B$_a$(rl)0/a                                         & NGC 5370             & 200519 & (R$_1$L)SB$_a$($\underline{\rm r}$s,bl)0/a                                       \\
GIN 449              &  67875 & (R$_1$L)S$\underline{\rm A}$B(rl)0/a                                             & UGC 08521            &  15008 & (R$_1^{\prime}$L)SB(p,rs,bl)ab                                                   \\
                     &        &          &                      &        & \\ 
\noalign{\centerline{(c) R$_1^{\prime}$ Outer Pseudorings}}
                     &        &          &                      &        & \\ 
CGCG 045-068         & 198356 & (R$_1^{\prime}$)SAB(rs)0/a                                                       & NGC 4113             & 132570 & (R$_1^{\prime}$)SA$\underline{\rm B}$($\underline{\rm r}$s,nr)ab                 \\
CGCG 065-033         &  80346 & (R$_1^{\prime}$)S$\underline{\rm A}$B(r)b                                        & NGC 4614             & 159683 & (R$_1^{\prime}$)SA$\underline{\rm B}$$_a$(r$^{\prime}$l,ns)a                     \\
CGCG 156-027         & 131471 & (R$_1^{\prime}$)SB(rs,bl)a                                                       & NGC 5377             &  55032 & (R$_1^{\prime}$)SAB$_{xa}$(r$^{\prime}$l)a                                       \\
CGCG 156-057         & 151499 & (R$_1^{\prime}$)SB($\underline{\rm r}$s,r,bl)a                                   & NGC 5554             &  41993 & (R$_1^{\prime}$)SB(r,bl)a                                                        \\
CGCG 225-014         &  33332 & (R$_1^{\prime}$)SAB$_a$(rs)0/a                                                   & UGC 04335            & 105878 & (R$_1^{\prime}$)SB$_a$(p,r,bl)a                                                  \\
CGCG 262-067         & 109406 & (R$_1^{\prime}$)SB($\underline{\rm r}$s,bl)ab                                    & UGC 04602            & 225823 & (R$_1^{\prime}$)SAB(rs)ab                                                        \\
IC 0588              &  19783 & (R$_1^{\prime}$)SAB(r)0/a                                                        & UGC 06176            & 167753 & (R$_1^{\prime}$)SA$\underline{\rm B}$$_a$(s)0/a                                  \\
IC 1047              & 168843 & (R$_1^{\prime}$)SA$\underline{\rm B}$(p,$\underline{\rm r}$s,bl)a                & UGC 06645            & 159621 & (R$_1^{\prime}$)SB(p,rs,rs)ab                                                    \\
IC 2473              & 114686 & (R$_1^{\prime}$)SB(r,bl)ab                                                       & UGC 07065            &   4324 & (R$_1^{\prime}$)SB(r,bl)b                                                        \\
IC 3012              &  76157 & (R$_1^{\prime}$)S$\underline{\rm A}$B($\underline{\rm r}$s)ab                    & UGC 08794            & 167980 & (R$_1^{\prime}$)SAB(rl)ab                                                        \\
NGC 3536             & 150119 & (R$_1^{\prime}$)SAB$_a$(p,r,bl)ab                                                & UGC 10255            & 234984 & (R$_1^{\prime}$)SB(p,$\underline{\rm r}$s,bl)ab                                  \\
NGC 3855             & 134455 & (R$_1^{\prime}$)SB($\underline{\rm r}$s)ab                                       & UGC 10819            &  10640 & (R$_1^{\prime}$)S$\underline{\rm A}$B$_a$(rl)a                                   \\
                     &        &          &                      &        & \\ 
\noalign{\centerline{(d) R$_1$R$_2^{\prime}$ Combined Outer Rings/Pseudorings}}
                     &        &          &                      &        & \\ 
CGCG 126-058         & 162376 & (R$_1$R$_2^{\prime}$)SA(r)a                                                      & MCG +08-15-053       &   6959 & (R$_1$R$_2^{\prime}$)SA$\underline{\rm B}$(r,bl)0/a                              \\
CGCG 156-030         & 149710 & (R$_1$R$_2^{\prime}$)S$\underline{\rm A}$B(r$^{\prime}$l)ab                      & MCG +12-16-025       &  42947 & (R$_1$R$_2^{\prime}$)SB$_a$($\underline{\rm r}$s,bl)ab                           \\
CGCG 185-014         & 114576 & (R$_1$R$_2^{\prime}$)SAB(rs)ab                                                   & NGC 2766             & 115891 & (R$_1$R$_2^{\prime}$)S$\underline{\rm A}$B(r)a                                   \\
CGCG 191-066         & 101609 & (R$_1$R$_2^{\prime}$)S$\underline{\rm A}$B(rl)a                                  & NGC 5701             &  41051 & (R$_1$R$_2^{\prime}$)SB(r$^{\prime}$l,bl)a                                       \\
IC 1141              & 172197 & (R$_1$R$_2^{\prime}$)S$\underline{\rm A}$B$_a$(r)ab                              & UGC 04406            & 109750 & (R$_1$R$_2^{\prime}$)S$\underline{\rm A}$B($\underline{\rm r}$l)0/a             \\
IC 2568              & 116004 & (R$_1$R$_2^{\prime}$)SB$_{ax}$(s)a                                               & UGC 04596            & 150808 & (R$_1$R$_2^{\prime}$)SA(rr)b                                                     \\
IC 2628              & 209826 & (R$_1$R$_2^{\prime}$)SAB(l)a                                                     & UGC 04613            &  22171 & (R$_1$R$_2^{\prime}$)SB(r,bl)b pec                                               \\
IC 2968              & 237222 & (R$_1$R$_2^{\prime}$)S$\underline{\rm A}$B(rs)a                                  & UGC 08343            & 167932 & (R$_1$R$_2^{\prime}$)SA(l)ab                                                     \\
IC 4447              & 127728 & (R$_1$R$_2^{\prime}$)SA(r)a                                                      & UGC 08730            & 139559 & (R$_1$R$_2^{\prime}$)SB$_a$(rl)ab                                                \\
MCG +05-34-082       & 133126 & (R$_1$R$_2^{\prime}$)SAB$_a$(l)b                                                 & UGC 10168            & 191175 & (R$_1$R$_2^{\prime}$)SAB$_a$(r$^{\prime}$l)a                                     \\
                     &        &          &                      &        & \\ 
\noalign{\centerline{(e) R$_2^{\prime}$ Outer Pseudorings}}
                     &        &          &                      &        & \\ 
CGCG 074-080         &  88789 & (R$_2^{\prime}$)SAB$_a$(l)a                                                      & MCG +07-22-023       & 213087 & (R$_2^{\prime}$)SA$\underline{\rm B}$$_a$(l)ab                                   \\
CGCG 078-060         & 178170 & (R$_2^{\prime}$)SAB$_a$(rl)b                                                     & MCG +09-29-043       &  43467 & (R$_2^{\prime}$)SAB(s)ab                                                         \\
CGCG 094-065         & 186365 & (R$_2^{\prime}$)SAB$_a$(l)ab                                                     & NGC 2962             &  59486 & (R$_2^{\prime}$)SAB(r$^{\prime}$l)a                                              \\
CGCG 138-039         & 100962 & (R$_2^{\prime}$)SAB(r)b                                                          & NGC 4935             &  97771 & (R$_2^{\prime}$)SAB$_a$(rs)b                                                     \\
CGCG 161-048         & 132312 & (R$_2^{\prime}$)SA$\underline{\rm B}$(s)0/a                                      & NGC 5211             & 242145 & (R$_2^{\prime}$)SA(rs)b                                                          \\
CGCG 167-025         &  99446 & (R$_2^{\prime}$)SAB$_a$(r$^{\prime}$l)b                                          & NGC 5613             & 226229 & (R$_2^{\prime}$)SAB$_a$(r,nrl)a                                                  \\
CGCG 207-027         &  24315 & (R$_2^{\prime}$)SA$\underline{\rm B}$$_a$(l)a                                    & NGC 5674             &  41028 & (R$_2^{\prime}$)SA$\underline{\rm B}$$_a$(s)c                                    \\
IC 2348              & 128011 & (R$_2^{\prime}$)SA(r)b                                                           & UGC 04771            &  35305 & (R$_2^{\prime}$)SAB$_a$(l)b                                                      \\
IC 4669              & 202698 & (R$_2^{\prime}$)SAB$_a$(r$^{\prime}$l,nl)b                                       & UGC 05881            & 156962 & (R$_2^{\prime}$)SAB$_a$(r$^{\prime}$l)ab                                         \\
MCG +07-18-040       &  49605 & (R$_2^{\prime}$)SB$_a$(l)b                                                       & UGC 09418            & 136387 & (R$_2^{\prime}$)SAB$_a$(r$^{\prime}$l)b                                          \\
\hline
\end{tabular}
\end{table*}

\begin{table*}
\centering
\caption{GZ2-CNRG galaxies having special morphological features.}
\label{tab:list3}
\begin{tabular}{lrllrl}
\hline
Name & file.jpg & Type & Name & file.jpg & Type \\
 1 & 2 & 3 & 1 & 2 & 3 \\
\hline
                     &        &          &                      &        & \\ 
\noalign{\centerline{(a) Extremely oval SB inner rings}}
                     &        &          &                      &        & \\ 
CGCG 008-010         &  29334 & (R$_1^{\prime}$R$_2^{\prime}$)S$\underline{\rm A}$B(r)a                          & MCG +06-32-024       & 101620 & (R$_1$)SAB(r)0/a                                                                 \\
CGCG 008-073         &  29011 & (R$_1$)SA$\underline{\rm B}$(rl)0/a                                              & MCG +06-35-004       &  96143 & (R$_1^{\prime}$)SA$\underline{\rm B}$(r)a                                        \\
CGCG 012-021         & 244114 & (R$^{\prime}$)SB($\underline{\rm r}$s)ab                                         & MCG +08-29-001       &  72185 & SA$\underline{\rm B}$(r,bl)0$^+$                                                 \\
CGCG 039-184         & 197254 & (R$_1$L)SAB(r)0$^+$                                                              & NGC 2766             & 115891 & (R$_1$R$_2^{\prime}$)S$\underline{\rm A}$B(r)a                                   \\
CGCG 098-074         & 236780 & (R$_1$L)SB(r$^{\prime}$l,bl)0/a                                                  & NGC 3380             & 150955 & (R$_1$L)SAB(l,rs,bl)ab                                                           \\
CGCG 108-166         & 138236 & (R$_1$)SA$\underline{\rm B}$(r,bl)0$^+$                                          & NGC 3398             &  49023 & (R$_1$L)SAB($\underline{\rm r}$s)ab                                              \\
CGCG 156-057         & 151499 & (R$_1^{\prime}$)SB($\underline{\rm r}$s,r,bl)a                                   & NGC 4113             & 132570 & (R$_1^{\prime}$)SA$\underline{\rm B}$($\underline{\rm r}$s,nr)ab                 \\
CGCG 242-069         & 238439 & (R$_1$)SAB(r)0/a                                                                 & PGC 54897            & 172898 & (R$_1^{\prime}$)SA$\underline{\rm B}$$_a$(r$^{\prime}$l)ab                       \\
IC 0588              &  19783 & (R$_1^{\prime}$)SAB(r)0/a                                                        & PGC 58013            & 227267 & (R$_1$L)S$\underline{\rm A}$B($\underline{\rm r}$s)ab                            \\
IC 0816              & 230298 & (R$^{\prime}$?)SB($\underline{\rm r}$s,bl,nl)0/a                                 & PGC 1857116          & 150974 & (R$_1^{\prime}$)SAB$_a$($\underline{\rm r}$s)a                                   \\
IC 1092              & 217549 & (R)SB($\underline{\rm r}$s,bl)0/a                                                & UGC 04335            & 105878 & (R$_1^{\prime}$)SB$_a$(p,r,bl)a                                                  \\
                     &        &          &                      &        & \\ 
\noalign{\centerline{(b) Nearly circular SB inner rings}}
                     &        &          &                      &        & \\ 
CGCG 033-053         &  25813 & (R$^{\prime}$,R$^{\prime}$L)SB($\underline{\rm r}$s,bl)a                         & MCG +06-27-035       & 130411 & SAB(r,bl)0$^+$                                                                   \\
CGCG 047-120         &  20648 & (L)SB(r,bl)0/a                                                                   & MCG +07-34-045       &  72028 & SAB($\underline{\rm r}$l,bl)0$^+$                                                \\
CGCG 102-062         & 175604 & SB(r,bl)0$^+$                                                                    & NGC 3527             & 150113 & (R$_1^{\prime}$)SB($\underline{\rm r}$s,bl)a pec                                 \\
CGCG 119-082         & 120556 & SB(r,bl)0$^+$                                                                    & NGC 4608             &  62741 & (RL)SB(rl,bl)0/a                                                                 \\
CGCG 158-063         & 153033 & (R$_1$)S$\underline{\rm A}$B(r,bl)0$^+$                                          & NGC 4643             &  14926 & (L)SB($\underline{\rm r}$s,bl)0/a                                                \\
CGCG 180-022         &  54681 & (R$^{\prime}$L)SB(r,bl)a                                                         & NGC 5335             &  19625 & SB($\underline{\rm r}$s,bl)ab                                                    \\
CGCG 184-023         & 114055 & (R$_1^{\prime}$)S$\underline{\rm A}$B(rl)ab                                      & NGC 5370             & 200519 & (R$_1$L)SB$_a$($\underline{\rm r}$s,bl)0/a                                       \\
CGCG 185-056         & 121992 & (R$_1^{\prime}$)SB(p,r,bl)ab                                                     & NGC 5686             &  93551 & SB$_a$(r,bl)0$^o$                                                                \\
CGCG 196-067         &  71778 & (R$_1$R$_2^{\prime}$)SB(r,bl)a                                                   & NGC 5770             &  20906 & (L)SB(r,bl)0$^+$                                                                 \\
CGCG 213-027         & 226717 & SAB(r,bl)0$^+$                                                                   & UGC 05380            &  29337 & (L)SB(r,bl)0/a                                                                   \\
                     &        &          &                      &        & \\ 
\noalign{\centerline{(c) Strong SB or SAB inner ring galaxies with very faint outer discs}}
                     &        &          &                      &        & \\ 
CGCG 119-082         & 120556 & SB(r,bl)0$^+$                                                                    & PGC 38118            &   1742 &  SB(r)0$^+$                                                                      \\
IC 0816              & 230298 & (R$^{\prime}$?)SB($\underline{\rm r}$s,bl,nl)0/a                                 & PGC 3801031          &  11296 &  (R$_1$)S$\underline{\rm A}$B(r)0$^+$                                            \\
J143842.06+174555.5  & 162727 & (R)SB(r)0$^+$                                                                    & PGC 3867832          &  14360 &  SB(r,bl)0$^+$                                                                   \\
MCG +06-27-035       & 130411 & SAB(r,bl)0$^+$                                                                   & PGC 2103294          &  10514 &  (R$_1^{\prime}$)SB($\underline{\rm r}$s)ab                                      \\
MCG +07-34-045       &  72028 & SAB($\underline{\rm r}$l,bl)0$^+$                                                & VIII Zw 391          &  15091 & (RL:)SB(r,bl)0$^+$                                                               \\
MCG +08-29-001       &  72185 & SA$\underline{\rm B}$(r,bl)0$^+$                                                 & VV 673 NED03         & 118022 & SB(r)0$^+$                                                                       \\
                     &        &          &                      &        & \\ 
\noalign{\centerline{(d) Lenses and ring-lenses in nonbarred galaxies}}
                     &        &          &                      &        & \\ 
CGCG 014-041         &   3622 & (RL)SA(l)0$^+$                                                                   & NGC 4919             & 158420 & (L)SA(rl)0$^+$                                                                   \\
CGCG 018-026         & 242798 & (R)S$\underline{\rm A}$B$_a$(l)0$^+$                                             & NGC 5057             & 131603 & (RL)SA(l)0$^+$                                                                   \\
CGCG 218-012         & 225284 & (RL)SA(l)0$^o$                                                                   & NGC 5225             & 219722 & SA($\underline{\rm r}$s)ab                                                       \\
IC 2516              & 205167 & (RL)SA(l)0$^+$                                                                   & NGC 5501             &  14749 & SA(l)0$^+$                                                                       \\
NGC 3419             &  81341 & (R)SA(l)0$^+$                                                                    & NGC 5602             &  65859 & (RL)SA(rl)0$^+$                                                                  \\
                     &        &          &                      &        & \\ 
\noalign{\centerline{(e) Lenses and pseudoring-lenses in barred galaxies}}
                     &        &          &                      &        & \\ 
CGCG 033-057         &  76542 & (RL)SB$_a$(l,bl)0$^+$                                                            & NGC 3767             & 181326 & (R$_1$)SA$\underline{\rm B}$(l)0$^+$                                             \\
CGCG 207-027         &  24315 & (R$_2^{\prime}$)SA$\underline{\rm B}$$_a$(l)a                                    & NGC 5701             &  41051 & (R$_1$R$_2^{\prime}$)SB(r$^{\prime}$l,bl)a                                       \\
CGCG 208-013         &  48446 & (R$_1$R$_2^{\prime}$)SAB$_a$(l)a                                                 & PGC 53315            & 245210 & (R)S$\underline{\rm A}$B$_a$(l)0/a                                               \\
MCG +07-18-040       &  49605 & (R$_2^{\prime}$)SB$_a$(l)b                                                       & PGC 1224876          &  15338 & (R$_1^{\prime}$)S$\underline{\rm A}$B(l)a                                        \\
MCG +08-16-006       &   6599 & (RL)SB$_a$(l)ab                                                                  & UGC 04771            &  35305 & (R$_2^{\prime}$)SAB$_a$(l)b                                                      \\
                     &        &          &                      &        & \\ 
\noalign{\centerline{(f) Bars with strong yellow ansae}}
                     &        &          &                      &        & \\ 
CGCG 126-071         & 170208 & (R$^{\prime}$)SA$\underline{\rm B}$$_a$(r$^{\prime}$l)0/a                        & MCG +10-13-043       &  10109 & (R)SAB$_a$(l)0/a                                                                 \\
CGCG 151-055 NED02   & 206564 & (R$_1^{\prime}$)SAB$_a$(r$^{\prime}$l)b                                          & NGC 3151             &  84046 & (R)SB$_a$(r$^{\prime}$l)0$^+$                                                    \\
CGCG 184-046         & 114568 & (RL)SB$_a$0$^+$                                                                  & NGC 3300             &  81770 & (RL)SB$_a$(rs)0$^+$                                                              \\
CGCG 186-082         & 127537 & (R$_1$)SB$_a$(l)0/a                                                              & NGC 4996             &   1814 & (R)SB$_a$(s,bl)0$^+$                                                             \\
CGCG 195-006         & 234553 & (R$_1$)SB$_a$(r$\underline{\rm s}$)a                                             & NGC 5052             & 136177 & (RL)SB$_a$(r$^{\prime}$l)0/a                                                     \\
CGCG 225-014         &  33332 & (R$_1^{\prime}$)SAB$_a$(rs)0/a                                                   & UGC 04771            &  35305 & (R$_2^{\prime}$)SAB$_a$(l)b                                                      \\
IC 2568              & 116004 & (R$_1$R$_2^{\prime}$)SB$_{ax}$(s)a                                               & UGC 05859            & 213092 & (R)SAB$_a$(r,bl)0$^+$                                                            \\
IC 3075              & 165525 & (R)SA$\underline{\rm B}$$_a$(l)0$^+$                                             & UGC 05881            & 156962 & (R$_2^{\prime}$)SAB$_a$(r$^{\prime}$l)ab                                         \\
IC 3199              &  63145 & (R$_1$L)SAB$_a$(rl,bl)0/a                                                        & UGC 08730            & 139559 & (R$_1$R$_2^{\prime}$)SB$_a$(rl)ab                                                \\
IC 3207              & 170000 & (R)SA$\underline{\rm B}$$_a$(l)0/a                                               & UGC 10361            & 182880 & (RL)SB$_a$0$^+$                                                                  \\
\hline
\end{tabular}
\end{table*}

\begin{table*}
\centering
\caption{Features of special interest.}
\label{tab:list4}
\begin{tabular}{lrllrl}
\hline
Name & file.jpg & Type & Name & file.jpg & Type \\
 1 & 2 & 3 & 1 & 2 & 3 \\
\hline
                     &        &          &                      &        & \\ 
\noalign{\centerline{(a) Galaxies with blue bar/inner ring ansae}}
                     &        &          &                      &        & \\ 
CGCG 067-004         & 215672 & (R$_1$L)SB$_a$(s,bl,nl)0/a                                                       & PGC 2175916          & 209036 & (R$_2^{\prime}$)S$\underline{\rm A}$B(r)b                                        \\
CGCG 092-056         & 187092 & (R$_1$)SAB$_a$(l)0/a                                                             & PGC 2215577          & 192720 & (R$_1$)SAB$_a$($\underline{\rm r}$s)ab                                           \\
CGCG 151-013         & 226516 & (R)SB$_a$0$^+$                                                                   & PGC 2257366          & 214424 & (R$_2^{\prime}$)SA$\underline{\rm B}$(s)a                                        \\
PGC 54897            & 172898 & (R$_1^{\prime}$)SA$\underline{\rm B}$$_a$(r$^{\prime}$l)ab                       & PGC 2280736          & 232708 & (R$_1^{\prime}$)SA$\underline{\rm B}$$_a$(l,tb)0/a                               \\
PGC 1322879          & 222601 & (R$^{\prime}$)SA$\underline{\rm B}$$_a$(s)ab                                     & PGC 2315502          & 208594 & (R$_1^{\prime}$)SA$\underline{\rm B}$$_a$(r$^{\prime}$l)ab                       \\
PGC 1373953          &  75695 & (R:)SA$\underline{\rm B}$$_a$(rs)ab                                              & PGC 3124127          &  31096 & (R$_1^{\prime}$)SAB$_a$(s)ab                                                     \\
PGC 1581060          & 144884 & (R$_1^{\prime}$)SB$_a$(s)a                                                       & PGC 3549313          & 219961 & (R$_1^{\prime}$)SAB$_a$(rs)b pec                                                 \\
PGC 1671158          & 164903 & (R$_1^{\prime}$)SAB$_a$(l)ab                                                     & PGC 3565814          & 232738 & (R$^{\prime}$:)SAB($\underline{\rm r}$s)a                                        \\
PGC 1749837          & 154388 & (R$_1$)SAB(rl)0/a                                                                & PGC 3813421          & 140867 & (R$_1^{\prime}$)SAB$_a$(l)a                                                      \\
PGC 1809074          & 149677 & (R$_1$)SB$_a$(l)0$^+$                                                            & UGC 08237            & 200463 & (R$_1^{\prime}$)SAB$_a$(r)ab                                                     \\
                     &        &          &                      &        & \\ 
\noalign{\centerline{(b) SA galaxies with blue star-forming rings}}
                     &        &          &                      &        & \\ 
CGCG 121-040         & 150833 & (R$^{\prime}$,L)SA(r)a                                                           & PGC 1368674          & 176996 & SA(rl)0$^+$                                                                      \\
CGCG 163-012         & 128432 & SA(r)0$^+$                                                                       & PGC 1689688          & 145428 & SA(rs)ab: pec                                                                    \\
CGCG 223-043         &  66834 & SA($\underline{\rm r}$s)0/a pec                                                  & PGC 1707956          & 159580 & SA(rl)0$^+$ pec                                                                  \\
CGCG 225-102         & 147840 & (R$^{\prime}$)SA(l)ab                                                            & PGC 1744837          & 159509 & SA(r)0/a pec                                                                     \\
CGCG 266-009         &  44226 & (R$^{\prime}$)SA($\underline{\rm r}$s)ab [c]                                     & PGC 1890317          &  78912 & (R$^{\prime}$)SA(r)a                                                             \\
IC 0748              &  60123 & SA(r)0/a                                                                         & PGC 2080212          & 119849 & (R)SA0$^+$                                                                       \\
IC 0753              & 242594 & S$\underline{\rm A}$B(r)0$^+$                                                    & PGC 2114071          & 228378 & (R)E0 or SA(r)0$^+$                                                              \\
IC 1007              &  31442 & SA(r)0$^+$                                                                       & PGC 2136644          &  70238 & (R$^{\prime}$)SA(r)a pec                                                         \\
IC 1033              & 219774 & (L)SA(r)0/a                                                                      & PGC 2224134          & 240743 & SA($\underline{\rm r}$s,nl)ab                                                    \\
KUG 1310+262B        & 157711 & SA(r)0$^+$                                                                       & PGC 2403740          & 147759 & (R$^{\prime}$)SA(l)ab                                                            \\
KUG 1320+325         & 130869 & (R$^{\prime}$L)SA(rs)bc                                                          & PGC 3087375          & 107183 & SA(rl)0$^+$                                                                      \\
NGC 2775             &  74171 & SA(r$_d$,r,l)0/a                                                                 & PGC 3569185          & 232138 & SA(r)0$^+$                                                                       \\
NGC 3011             & 119108 & (R$^{\prime}$$\underline{\rm L}$)SA(r)0/a                                        & PGC 3764033          & 181605 & (R)SA0$^+$                                                                       \\
NGC 3884             & 174683 & (R$^{\prime}$L)SA($\underline{\rm r}$s)a                                         & PGC 3874173          & 232242 & (R$^{\prime}$L)SA($\underline{\rm r}$s)0/a                                       \\
NGC 4138             & 225465 & SA(r)0/a                                                                            & PGC 48308            & 116890 & SA(r)0/a [bc]                                                                    \\
NGC 4580             &  31250 & (R$^{\prime}$)SA(rs,$\underline{\rm r}$s)ab                                      & UGC 04572            &  50522 & (L,L)SA(rl)0$^+$                                                                 \\
NGC 5178             & 210654 & (R$^{\prime}$)SAB($\underline{\rm r}$s)a                                         & UGC 05025            & 188515 & (R)SA(rl)0$^+$                                                                   \\
NGC 5499             & 226199 & SA(rr)0/a                                                                        & UGC 08431            & 125723 & (R$^{\prime}$)SA(l)ab                                                            \\
NGC 6155             &  33127 & (RL)SA(rs)c                                                                      & UGC 10070            &  71286 & SA(r)ab                                                                          \\
                     &        &          &                      &        & \\ 
\noalign{\centerline{(c) Cataclysmic (''encounter-driven") rings}}
                     &        &          &                      &        & \\ 
CGCG 068-056         &  75525 & SA0$^o$ + PR/IR                                                                  & PGC 1357732          & 231463 & PR? or (R$^{\prime}$,R)SAb:                                                      \\
CGCG 180-023         &  54690 & RG                                                                               & PGC 1691234          & 165557 & RG pec                                                                           \\
CGCG 252-017         & 147545 & (R$^{\prime}$)SAb / E0                                                           & PGC 2121614          & 115615 & RG                                                                               \\
J090012.44+385234.2  &  50364 & (R)E0:                                                                           & PGC 3498384          & 233364 & S0 sp + PR?                                                                      \\
J125253.45+531427.1  & 205140 & (R)S$\underline{\rm A}$B0$^+$: or (R)E1                                          & PGC 3746765          & 162868 & RG                                                                               \\
MCG +06-33-026       & 213020 & S0$^-$ sp + PR                                                                   & PGC 4547778          & 165449 & (R)E0 or (R)SA0$^+$                                                              \\
MCG +10-19-014       &  87233 & SAB($\underline{\rm r}$s)ab or PRG?                                              & UGC 06334            & 151964 & (R)Sab or SA0$^+$ + PR                                                           \\
MRK 1477             & 224664 & (R$_2^{\prime}$)SAB(s)c + PR                                                     & UGC 06652            & 122172 & S0$^o$ sp + IR                                                                   \\
PGC 1156194          & 190164 & (R)SABb pec (IR?)                                                                & UGC 07683 NOTES01    &   8698 & RG (RE)                                                                          \\
PGC 1278613          & 198250 & (R)E2: or (R)SA0$^o$:                                                            & UGC 08139            & 116847 & (R$^{\prime}$)SAB($\underline{\rm r}$s)b pec / IRG?                              \\
                     &        &          &                      &        & \\ 
\noalign{\centerline{(d) Nuclear rings, ring-lenses, and lenses}}
                     &        &          &                      &        & \\ 
CGCG 047-129         &  30270 & (R$_1^{\prime}$)SAB$_a$(l,nrl)0/a                                                & NGC 5132             &  98173 & (R$_1^{\prime}$)SB($\underline{\rm r}$s,bl,nr)a                                  \\
CGCG 061-049 NED01   &  80691 & (R$_1^{\prime}$)SAB(r,nr?)a                                                      & NGC 5613             & 226229 & (R$_2^{\prime}$)SAB$_a$(r,nrl)a                                                  \\
CGCG 067-004         & 215672 & (R$_1$L)SB$_a$(s,bl,nl)0/a                                                       & NGC 5945             &  66991 & (R$_1$)SB(rs,nr)ab                                                               \\
CGCG 105-052         & 162766 & (R$_1$)SB(l,nrl)0$^+$                                                            & PGC 1807472          &  66658 & SB(r,nl)0/a                                                                      \\
CGCG 163-037         & 136353 & SA(r,nl)bc                                                                       & PGC 1809390          & 120076 & (R$^{\prime}$?)SAB(r,nrl)0/a:                                                    \\
CGCG 225-075         & 146800 & SA($\underline{\rm r}$s,nr$^{\prime}$)a pec                                      & PGC 2130373          &  69346 & SAB($\underline{\rm r}$s,nl)b                                                    \\
CGCG 266-056         &  63657 & (R$^{\prime}$)S$\underline{\rm A}$B(l)ab                                         & PGC 2224134          & 240743 & SA($\underline{\rm r}$s,nl)ab                                                    \\
IC 0816              & 230298 & (R$^{\prime}$?)SB($\underline{\rm r}$s,bl,nl)0/a                                 & PGC 2282221          & 105062 & (L)SA(rl,nl)0$^+$                                                                \\
IC 4669              & 202698 & (R$_2^{\prime}$)SAB$_a$(r$^{\prime}$l,nl)b                                       & PGC 3549988          & 220826 & (R$_1^{\prime}$)SA(p,r,nl)b                                                      \\
MCG +06-22-002       & 206930 & (R$^{\prime}$:)SAB($\underline{\rm r}$s,nl)a                                     & PGC 94018            & 136639 & (L)S$\underline{\rm A}$B(rl,nl)0$^+$                                             \\
NGC 4043             & 197603 & (R$^{\prime}$L)SA($\underline{\rm r}$s,nrl)0/a                                   & UGC 05885            & 131764 & (R$_1^{\prime}$)SAB$_a$($\underline{\rm r}$s,nr)ab                               \\
NGC 4245             & 137723 & (L)SB($\underline{\rm r}$s,bl)0/a                                                & UGC 08431            & 125723 & (R$^{\prime}$)SA(l)ab                                                            \\
NGC 4290             &  86952 & (R$^{\prime}$)SB$_a$(rs)b                                                        & UGC 08521            &  15008 & (R$_1^{\prime}$L)SB(p,rs,bl)ab                                                   \\

\hline
\end{tabular}
\end{table*}

The histograms for inner features in Figure~\ref{fig:combined-inner} show
again a preference for weak bars, although the SB bins are high for
($\underline{\rm r}$s) and (r) galaxies. Most interesting is the strong
preference for weak bars among galaxies classified as having inner
lenses (l). When the sample is split into high and low redshift subsets
(Figure~\ref{fig:combined-inner-vel}), we see that the strong SAB peak for
inner lenses is caused by an excess of (l) classifications in the high
redshift half. The same trend is found for (s)-variety classifications.
The high redshift excesses in (l), (s), and (R$_2^{\prime}$)
classifications are all related because (R$_2^{\prime}$)SAB(s) and
(R$_2^{\prime}$)SAB(l) are among the most common classifications in
Table~\ref{tab:catalog}, as already noted above.

Other small subcategories listed in Table~\ref{tab:invents} include
``ring galaxies" (RG; Appleton \& Struck-Marcell 1996) and ``polar ring
galaxies" (PRG; Schweizer et al. 1983). The term ``cataclysmic" was
used by Buta et al. (2015) to account for the likely violent
(collisional or disruptive) origin for such rings, as opposed to the
bulk of the rings in Table~\ref{tab:catalog}, which are likely passive
features resulting from internal perturbations and long-term secular
evolution. Although only 5 likely or possible ring galaxies and 10
likely or possible polar ring/inclined ring galaxies are included in
Table~\ref{tab:catalog}, the resolution of ambiguous cases could
increase these numbers.

A type of feature not yet recognized in many catalogues is the barlens,
classified with the type symbol (bl). Barlenses were recognized by
Laurikainen et al. (2011, 2013) as broad components seen in face-on
bars that could in some cases be mistaken to be part of a more
spherical classical bulge. Photometrically, these round (and sometimes
oval), inner components are best interpreted as lenses that are smaller
than the bar (and by default smaller than a typical inner lens) but
generally larger than a nuclear lens. Athanassoula et al. (2015; see
also Athanassoula 2016 and Salo \& Laurikainen 2016) concluded, based
on numerical simulations, that barlenses are nothing more than the
face-on view of the thick inner part of a typical bar which, if seen
edge-on, would show a boxy/peanut/X shape. Herrera-Endoqui et al. (2016)
measured the SDSS $ugriz$ colours of barlenses and the outer parts of
bars for 47 S$^4$G galaxies, and concluded that the colours are
basically the same, suggesting that barlenses are in fact part of the
bar. An actual, nearly edge-on view of a thick barlens flanked by the
thin outer ends of a bar is seen in a 3.6$\mu$m S$^4$G image of NGC
4216 (Buta 2012; Buta et al.  2015).  Barlenses are recorded in 389
(9.8\%) of the GZ2 sample galaxies and are most easily recognized when
the bar is strong. Barlenses are also recognized in the S$^4$G mid-IR
morphological catalogue of Buta et al. (2015) and in the NIRS0S
catalogue (Laurikainen et al. 2011).

A special category in Table~\ref{tab:invents} is the ``plume"
classification, symbolized by (p), as in a classification such
as (R$_1^{\prime}$)SB(p,$\underline{\rm r}$s,nr)ab. This
refers to short secondary spiral arcs located just off 
the leading ends of the bar; the protoype is NGC 1433 
where the features are exceptionally prominent (Buta 1986).
Plumes are recognized in 81 cases
(2\%) of the catalogue and are included mainly to show that NGC 1433,
while an exceptional case, is not unique. Using the ``gap method" to
locate the CR, B17 examined the locations
of plumes in several galaxies relative to resonances, and concluded
that the features lie in the region between the CR and the O4R (see also
Rautiainen et al. 2004).

\section{Examples of Morphologies}

Tables~\ref{tab:list1}-\ref{tab:list4} provide lists of representative
examples of the various morphologies contained in the catalog. Some
are illustrated in this section using SDSS colour images. These were
downloaded from the NED website, and in some cases have better
resolution than the GZ2 images used for the CVRHS classifications.

\begin{figure*}
\includegraphics[width=\textwidth]{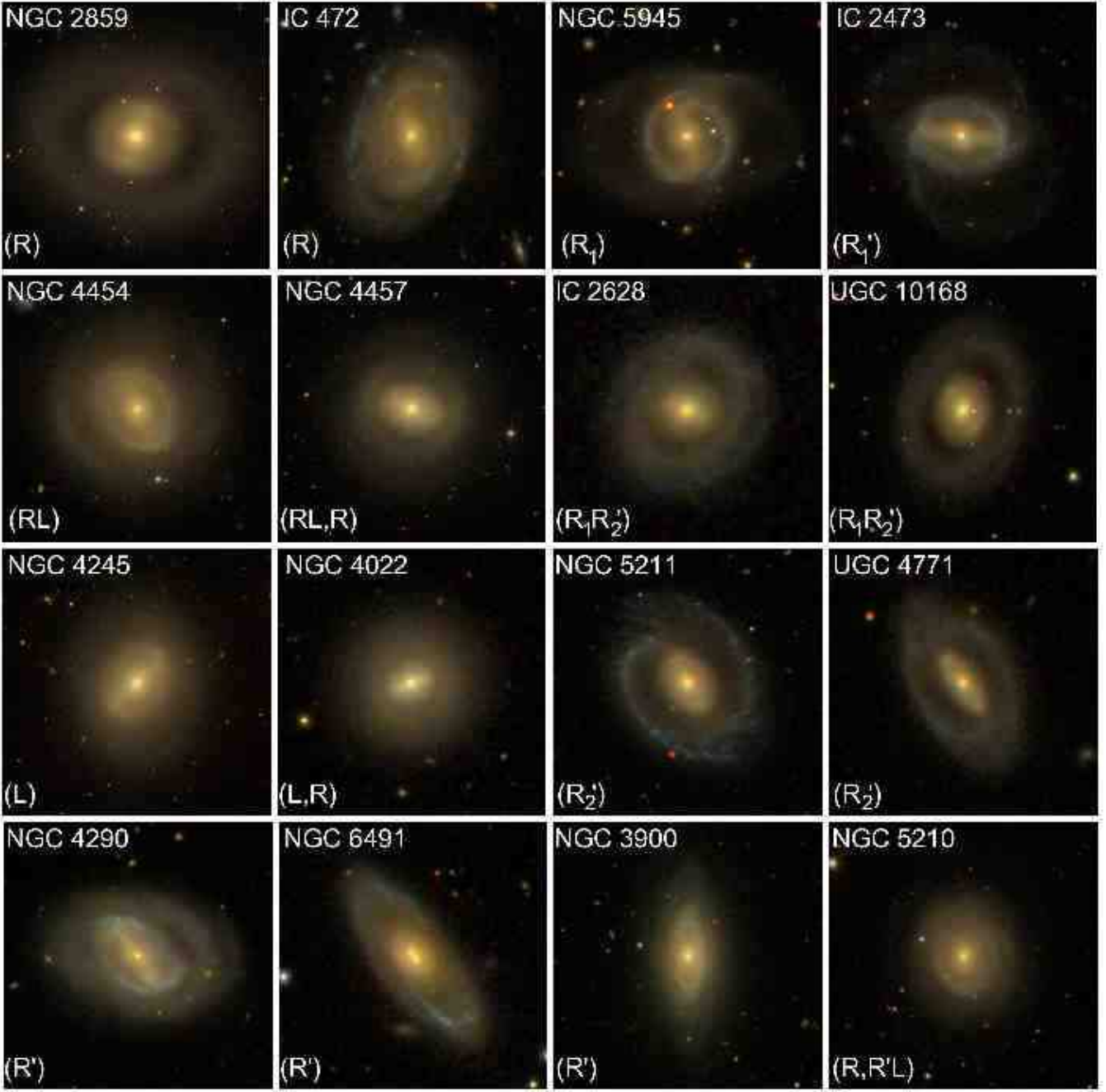}
\caption{Examples of outer varieties from the GZ2-CNRG, based on SDSS
colour images. The images are oriented with north at the top and east to
the left, and the fields shown have the following lengths on a side:
6\rlap{.}$^{\prime}$8 (NGC 4457);
5\rlap{.}$^{\prime}$1 (NGC 2859); 
4\rlap{.}$^{\prime}$7 (NGC 3900, 4245); 
3\rlap{.}$^{\prime}$4 (NGC 4454, 5210, 5945); 
3\rlap{.}$^{\prime}$0 (NGC 4290, NGC 5211); 
2\rlap{.}$^{\prime}$5 (UGC 10168); 
2\rlap{.}$^{\prime}$1 (IC 472, IC 2473); 
1\rlap{.}$^{\prime}$7 (NGC 4022, NGC 6491, UGC 4771); 
1\rlap{.}$^{\prime}$0 (IC 2628).
}
\label{fig:outer-examps}
\end{figure*}

Tables~\ref{tab:list1} and ~\ref{tab:list2}
focus on outer variety features,
including regular outer rings (R), outer pseudorings (R$^{\prime}$),
outer ring-lenses (RL), outer lenses (L), and the outer resonant
subclasses (R$_1$), (R$_1^{\prime}$), (R$_2^{\prime}$), and
(R$_1$R$_2^{\prime}$). One to three examples from each subsection
of these tables are shown in Figure~\ref{fig:outer-examps}. The figure
highlights only specific features; the full classifications are in
Tables~\ref{tab:list1} and ~\ref{tab:list2}.

All of the illustrated examples are fairly normal; the six outer
resonant ones are examined in more detail in B17. Also shown in the
figure are three examples of unusual double outer features:  (RL,R),
(L,R), and (R,R$^{\prime}$L). Outer varieties specified in this manner,
with a comma between the recognized features, means that the features
in parentheses are ordered with decreasing size from left to right. In
the case of (R$_1$R$_2^{\prime}$), no comma is used and the order is
reversed because the features were historically considered part of a
single pattern. The two outer features in NGC 4457 were also recognized
by Buta et al. (2015) on a deep 3.6$\mu$m image of the galaxy. NGC 4457
and 4022 highlight the possibility of a distinct class of objects where
a clear outer ring lies nested within a larger outer lens or ring-lens
of about the same shape. NGC 5210 has a double outer feature that is
different from these: beyond a distinct R$^{\prime}$L, which is a type
of feature consisting of an outer lens with a subtle embedded spiral,
there is a very faint, nearly detached outer ring. Other possible
combinations are listed in Table~\ref{tab:orinvents}.

A special case in Figure~\ref{fig:outer-examps} is IC 472, an unusual
Sab galaxy having a knotty outer ring which envelops an inner spiral.
This spiral includes a very open inner pseudoring of type
(r$\underline{\rm s}$), and is one of very few classified as such in
the GZ2-CNRG. Such open inner pseudorings are rare among early-type
spiral galaxies, but are fairly common among extreme late-type
(Scd-Sdm) barred spirals (Buta et al. 2015). A deprojected $g$-band
image of this galaxy reveals that the outer ring could have a
significant oval shape with a major axis misaligned with the bar by as
much as 20$^o$. The galaxy has a much fainter ring beyond the main (R)
feature whose nature is unclear.

The six outer resonant cases in Figure~\ref{fig:outer-examps} are
well-defined examples.  Almost 1500 cases like these are included in
the whole catalogue. Outer resonant subclass rings and pseudorings were
the primary focus of the B17 study, where a method was outlined to
derive the location of the corotation resonance (CR).  As noted by B17,
the CR, which occurs where $\Omega_p = \Omega$, is probably the most
important resonance governing the structure of these galaxies. By
placing the radius of the CR in the dark gaps between inner and outer
features, and assuming flat rotation curves, the structure of these
galaxies could be interpreted in a consistent way in terms of a few
specific disc resonances. The outer ring and pseudoring features in 50
galaxies (29 from the GZ2-CNRG and 21 from the CSRG) were visually
mapped and fitted with ellipses whose dimensions were compared with the
predicted locations of resonances.  The results favored a link between
R$_1$ and R$_1^{\prime}$ features and the O4R, and between
R$_2^{\prime}$ features and the OLR. The combined type,
R$_1$R$_2^{\prime}$, is therefore implied to be a double resonance
feature. The study also favored a link between inner rings,
pseudorings, lenses, and the ends of bars with the I4R.

It was noted in section 1 that resonances are only one way of viewing
the nature of these features. B17 also examines the implications of the
alternative ``manifold" theory (Athanassoula et al. 2009a,b; 2010),
which interprets outer resonant subclass rings and pseudorings in terms
of chaotic orbits near the $L_1$ and $L_2$ unstable Lagrangian points
in the bar field. Rather than scattering along unstable orbits around
these points, stars or particles are focussed by tubes called
``manifolds" to follow certain shapes. These shapes strongly resemble
the observed outer resonant subclasses of rings and spirals. Although
B17 examines the sample in terms of manifold theory, lacking more
detailed information it is difficult to decide which theory -
resonances or manifolds - is most likely to explain the properties of
observed rings and pseudorings.

\begin{figure*}
\includegraphics[width=\textwidth]{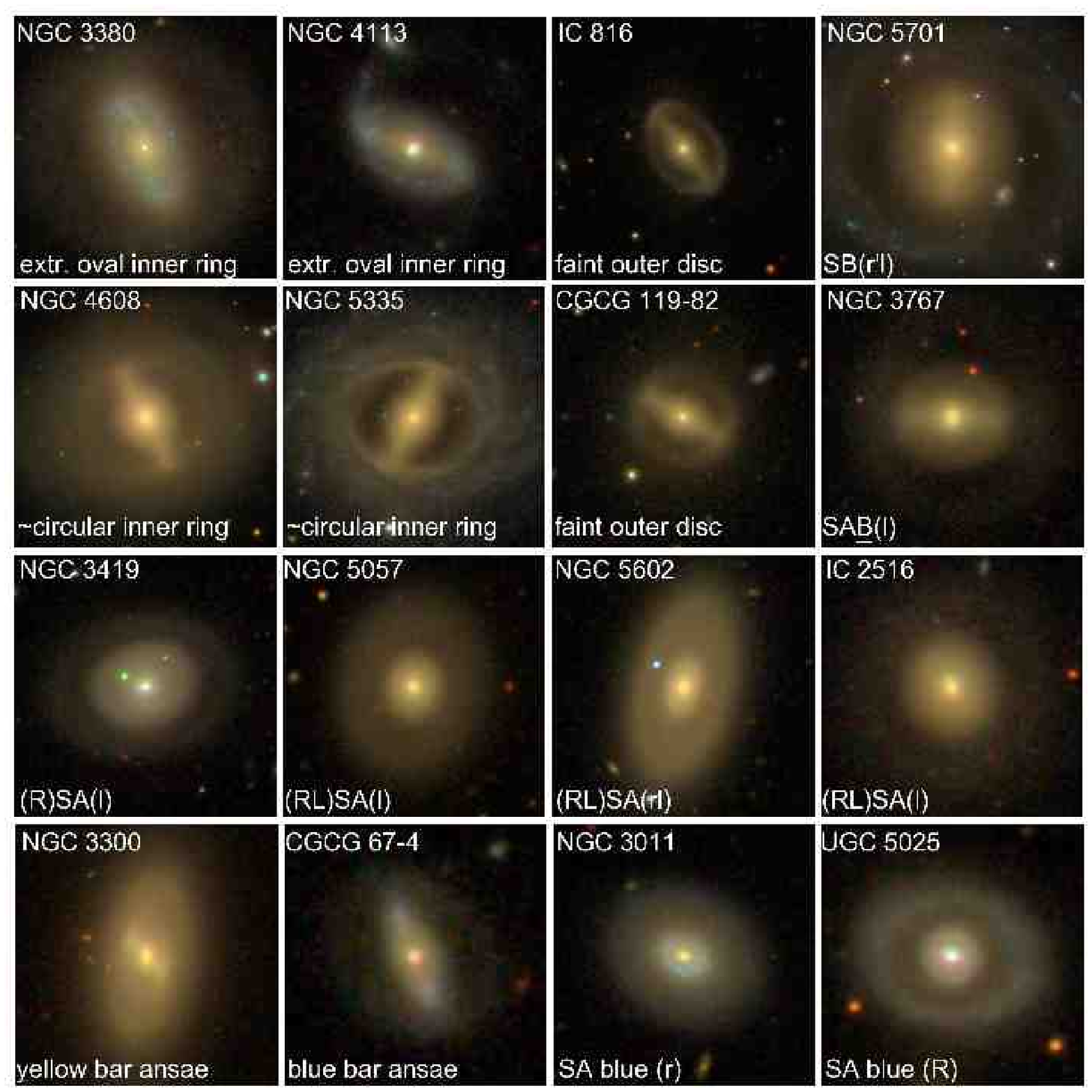}
\caption{Examples of inner varieties and other morphological features
from the GZ2-CNRG, based on SDSS colour images. 
The images are oriented with north at the top and east to
the left, and the fields shown have the following lengths on a side:
3\rlap{.}$^{\prime}$4 (NGC 4608, NGC 5701);
2\rlap{.}$^{\prime}$5 (IC 816);
1\rlap{.}$^{\prime}$7 (NGC 3300, NGC 3380, NGC 3419, NGC 3767, NGC 5057, NGC 5335, NGC 5602, CGCG 119-82);
1\rlap{.}$^{\prime}$3 (NGC 3011, NGC 4113);
1\rlap{.}$^{\prime}$2 (IC 2516);
0\rlap{.}$^{\prime}$85 (CGCG 67-4, UGC 5025).
}
\label{fig:inner-examps}
\end{figure*}

Table~\ref{tab:list3}a lists interesting cases where a nearly face-on
SB or SAB galaxy has a significantly elongated inner ring (intrinsic
axis ratio $q_o$ $<$ 0.7). Some of these ``extremely oval" inner rings
(Buta 2014) are noticeably cuspy.  In contrast, Table~\ref{tab:list3}b
lists a comparable number of nearly circular SB and SAB inner rings.
Examples of both highly-elongated and nearly circular inner rings are
shown in the four upper left panels of Figure~\ref{fig:inner-examps}.
The wide range of intrinsic axis ratios of inner features is important
because the factors determining intrinsic inner ring shape are not
well-understood. Numerical simulations have suggested that one factor
could be the strength of a bar (B17 and references therein). Also, the
way star forming regions are distributed around inner rings appears to
be sensitive to intrinsic ring shape (Crocker et al. 1996; Grouchy et
al. 2010). More than 1700 inner rings and pseudorings are included in
the catalogue.

While the inner rings and pseudorings in SB and SAB galaxies are
generally fairly bright, the outer rings seen in the same galaxies can
be bright or very faint to the point of near invisibility.
Table~\ref{tab:list3}c lists 12 examples where the outer ring or disc
structure is barely detected at the depth of SDSS images.  Two examples
(IC 816 and CGCG 119-82) are shown in Figure~\ref{fig:inner-examps}.
Both are remarkable also for the apparent strength of their bar and
inner ring.

The catalogue includes important examples of lenses in both nonbarred
(Table~\ref{tab:list3}d) and barred (Table~\ref{tab:list3}e) galaxies.
Inner lenses in SB galaxies have been attributed to ongoing bar
evolution (Kormendy 1979; Laurikainen et al. 2011, 2013), which makes
the lenses seen in nonbarred galaxies especially interesting from a
secular evolutionary point of view (Kormendy 1984).
Table~\ref{tab:list3}d shows that SA galaxies with an inner lens also
often have an outer feature. Four of the best examples in the catalogue
are shown in the third row of Figure~\ref{fig:inner-examps}. All likely
triggered the "Is the odd feature a ring?" button in GZ2 because of
these outer features, not the inner lenses. The SDSS images of NGC
5057, NGC 5602, and IC 2516 show that both the inner lenses and the
outer ring-lenses in these galaxies are made of a very old stellar
population. NGC 3419 is different from these in having a well-defined
outer ring as opposed to a more diffuse outer ring-lens, and a
whiter-coloured inner lens.

\begin{figure*}
\includegraphics[width=\textwidth]{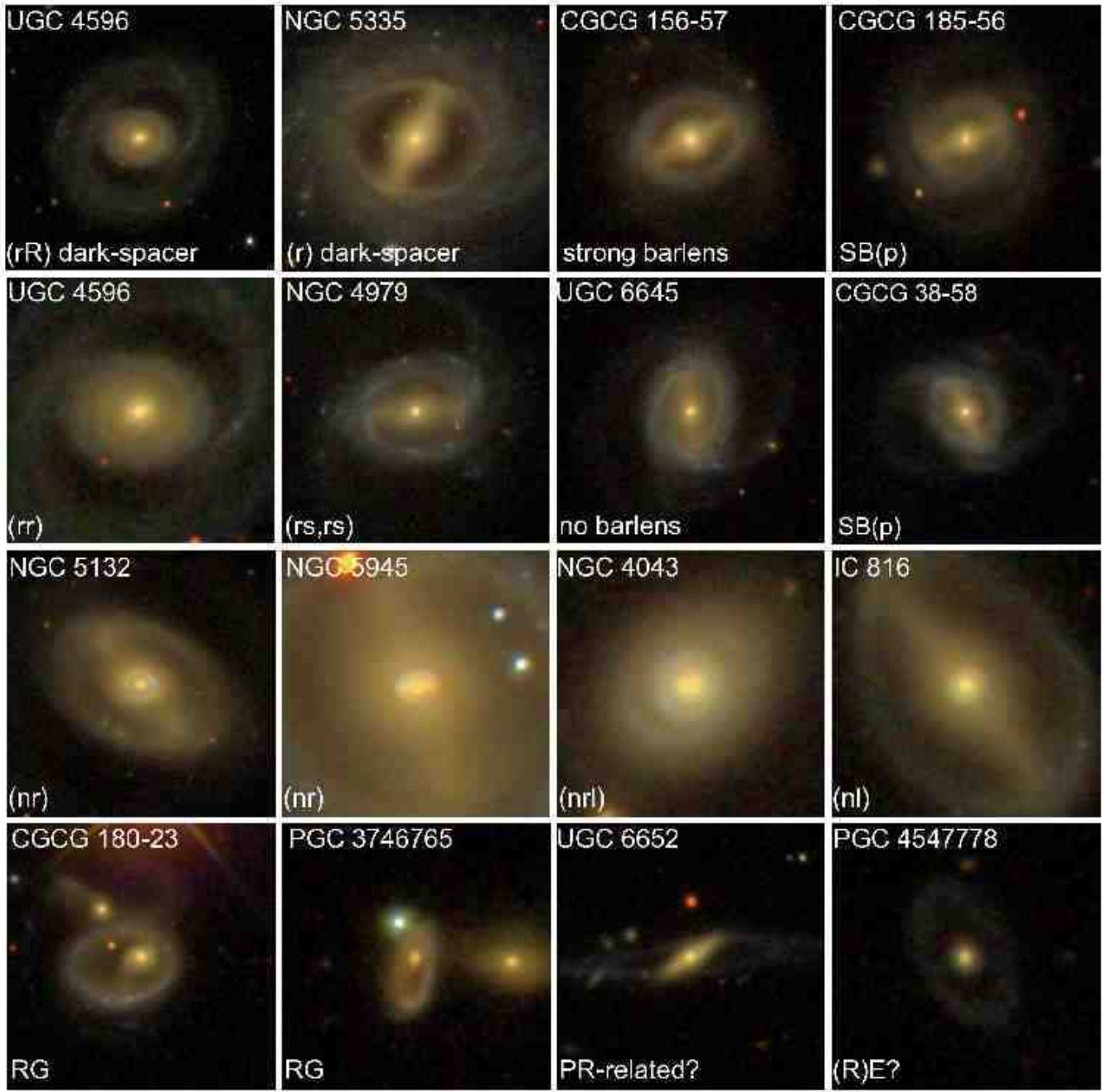}
\caption{Examples of other morphologies from the GZ2-CNRG, based on SDSS
colour images. The images are oriented with north at the top and east to
the left, and the fields shown have the following lengths on a side:
1\rlap{.}$^{\prime}$7 (NGC 4979, NGC 5132, NGC 5335, UGC 4596, UGC 6645, UGC 6652);
1\rlap{.}$^{\prime}$3 (CGCG 38-58, CGCG 156-57, CGCG 180-23, CGCG 185-56);
0\rlap{.}$^{\prime}$85 (IC 816-nr, NGC 4043, NGC 5945-nr, PGC 3746765, PGC 4547778, UGC 4596-rr).
}
\label{fig:cataclysmics}
\end{figure*}

Two barred galaxies with inner lenses are shown in the upper right
panels of Figure~\ref{fig:inner-examps}. NGC 5701 is near enough for
the lens to show a subtle ``brackett" (rs) (de Vaucouleurs 1959; see
also the deVA), which is the basis for the classification
(r$^{\prime}$l). Comparing NGC 5701 with NGC 4608 in the same figure
highlights the dichotomy noted by Gadotti and de Souza (2003), that the
disc in the bar region of NGC 4608 appears to be absent, as if
destroyed by the bar.  In paper 2 (B17), cases like NGC 4608 (and NGC
5335 next to it in Figure~\ref{fig:inner-examps}) are called ``(r)
dark-spacers." In some cases, the dark spaces lie between inner and
outer features. These are called ``(rR) dark-spacers" in B17. Both
types are illustrated in the upper left frames of
Figure~\ref{fig:cataclysmics}. All of the illustrated outer resonant
subclass galaxies in Figure~\ref{fig:inner-examps} are of the
(rR)-type.  Figure~\ref{fig:inner-examps} also shows that bars viewed
against bright inner lenses do not appear to be as strong as those in
an (r) dark-spacer. This distinction was quantified in B17 using
relative Fourier intensity amplitudes.

Ansae (``handles") are  round, linear, or curved surface brightness
enhancements seen in the outer parts of some bars. Ansae are recognized
in 715 (18\%) of the sample galaxies using the subscript "a" after the
bar classification, as in SAB$_a$ or SB$_a$ (Buta et al. 2015).
Table~\ref{tab:list3}f lists examples with old stellar population ansae
while Table~\ref{tab:list4}a lists examples with younger stellar
population ansae. An example of each is shown in the lower left panels
of Figure~\ref{fig:inner-examps}. In general, ansae appear to be
stellar dynamical in nature, at least in early-type galaxy bars like
that in NGC 3300. The existence of blue ansae as in CGCG 67-4 suggests
a link to the way star formation is distributed around an extremely
oval inner ring.  HII regions tend to ``bunch up" around the major axis
of the ring (usually consistent with the bar axis; Buta et al. 2004),
an effect likely due to velocity crowding and slow down as gas clouds
approach corotation in the rotating reference frame (Byrd et al.
2006).

Martinez-Valpuesta et al. (2007) determined that ansae are most
prevalent among early-type barred galaxies, especially SB0 types
where the frequency is as high as 40\%. The frequency of ansae was found to
drop significantly for types later than Sb. The GZ2-CNRG cannot be used
to evaluate this issue further because, as noted in section 4.1,
the catalogue is biased towards inclusion of outer rings whose
frequency also drops significantly for types later than Sb.

One of the most important categories of ringed galaxies in the GZ2
database is nonbarred galaxies with bright star-forming rings.
Table~\ref{tab:list4}b lists 38 examples from the GZ2-CNRG, about 1/3
of which are classified as late S0s (type S0$^+$). The lower right
panels of Figure~\ref{fig:inner-examps} show two examples. NGC 3011 is
characterized by a small star-forming inner ring and an outer lens with
subtle blue spiral structure. UGC 5025 is an example where star
formation is prominent in both an inner ring-lens and a bright outer
ring. 

The lack of an apparent bar in such galaxies could argue in favor of an
environmental origin for the rings. Ilyina et al. (2014) examined the
UV bright rings in four S0 galaxies from the GALEX survey (Martin et
al. 1995; Gil de Paz et al. 2007), and found evidence of decoupled
gaseous discs and normal solar abundances in the gas. These authors
suggested that some nonbarred star-forming rings in early-type galaxies
are impact-related or contain enriched gas donated from a local
neighbor.

Two of the upper right frames in Figure~\ref{fig:cataclysmics} show a
barred galaxy with a prominent barlens (CGCG 156-57) and one which
appears to have little or no barlens (UGC 6645). Based on the NIRS0S,
Laurikainen et al. (2013) suggested that inner lenses in some nonbarred
S0 galaxies were once the barlens in a barred S0. In this scenario of
galactic evolution, the ends of a regular bar evolve into ansae which
slowly dissolve into an inner ring-lens. This process likely destroys
the bar but leaves the barlens intact. Other galaxies with prominent
barlenses in Figures~\ref{fig:outer-examps}-~\ref{fig:cataclysmics}
include NGC 2859, 3300, 3380, 3767, 4245, 4290, 4608, 4979, 5132, 5335,
5701, 5945, IC 472, 816, 2473, and CGCG 119-82, and 185-56.

Like outer features, some galaxies show doubled inner variety features.
Two are highlighted in the left-most panels of
Figure~\ref{fig:cataclysmics}.  UGC 4596 has two close inner rings near
the rim of a massive, barlike oval. In contrast, NGC 4979 has a subtle
doubled inner pseudoring. The former pair of inner rings is classified
as ``(rr)" while the latter is classified as ``(rs,rs)". Like outer
varieties, inner varieties are enclosed in parentheses and list
features from largest to smallest, separated by commas. Also showing
subtle doubled varieties in Figure~\ref{fig:cataclysmics} are CGCG
156-57 [SB($\underline{\rm r}$s,r,bl)] and UGC 6645 [SB(p,rs,rs)].
CGCG 185-56 and CGCG 38-58 in the upper right frames of
Figure~\ref{fig:cataclysmics} show subtle plumes outside a conspicuous
inner ring. Two other examples in the same figure are NGC 4979 and UGC
6645.

Table~\ref{tab:list4}c summarizes a few of the best
examples of ring and polar ring galaxies in the catalogue. Two ring
galaxies and one polar-ring related case are shown in the bottom row of
Figure~\ref{fig:cataclysmics}. For none of the listed PR examples
is the interpretation definite; all could however be related to genuine
polar ring galaxies. Most of the candidate PRGs in
Table~\ref{tab:list4}c are also included in the Sloan-based Polar Ring
Catalogue (SPRC) of Moiseev et al. (2011), a catalogue also based on
Galaxy Zoo 2. Some galaxies in the SPRC are interpreted differently in
Table~\ref{tab:catalog}. A noteworthy case is UGC 4596 (SPRC 249),
which is classified as type (R$_1$R$_2^{\prime}$)SA(rr)b in
Table~\ref{tab:catalog} and which is also presented by B17 as a
prototype of an ``(rR) dark-spacer" galaxy, owing to the darkness of
the gaps between its inner and outer ring features (see upper left
frame of Figure~\ref{fig:cataclysmics}). In the SPRC, UGC 4596 is listed
as a possible face-on polar ring or related extraplanar feature.  While
I cannot rule this interpretation out, the outer arms of UGC 4596 show
a well-defined (R$_1$R$_2^{\prime}$) morphology that favors the galaxy
being a normal resonance ring type where the bar is a massive oval.
The obvious dimpling in the R$_1$ component supports this
interpretation.

The bottom right panel of Figure~\ref{fig:cataclysmics} shows one
example of an object classified as type ``(R)E" in
Table~\ref{tab:catalog}. These are
cases where a large, blue ring envelops what appears to be an
elliptical galaxy as opposed to an S0 galaxy. A well-studied example of
such a system is Hoag's Object (Schweizer et al. 1987), although none
of the several examples of the type in Table~\ref{tab:catalog}
shows the delicate structure of that object.

There are other possible ring and polar ring related cases in
Table~\ref{tab:catalog} that could prove to be genuine. Even if these
were included, both categories of encounter-driven rings still
constitute less than 1\% of the entries, confirming their rarity among
the galactic ring population.

Finally, Table~\ref{tab:list4}d summzarizes the galaxies where a
nuclear ring (nr), nuclear ring-lens (nrl), or nuclear lens (nl) is
recognized in Table~\ref{tab:catalog}. Four examples are shown in
row 3 of Figure~\ref{fig:cataclysmics}. Only 26 examples (0.6\%
of the catalogue) show such features, but as discussed in section 3 and
Table~\ref{tab:resol}, the sample is heavily biased towards the largest
examples, and it is likely that many of these kinds of features are
missed due to poor resolution of the images in the central regions of
the galaxies and/or the limitations of SDSS colour images. Using the
well-resolved sample of nuclear rings in the AINUR, Comer\`on et al.
(2010) were able to deduce that nuclear rings occur in 20\% of nearby
galaxies. If gray-scale colour index maps had been available for every
galaxy in the GZ2-CNRG sample, the number of galaxies recognized to
have nuclear rings or nuclear star formation would likely have been
substantially higher.

Like inner rings, pseudorings, and lenses, nuclear rings, pseudorings,
and lenses are classified within the inner variety bracketts. For
example, the inner variety of NGC 5132 in the bottom left frame of
Figure~\ref{fig:cataclysmics} is SB($\underline{\rm r}$s,bl,nr),
indicating that the barlens is larger than the nuclear ring.

\section{Summary}

The main result of this paper is a catalogue of CVRHS classifications
for 3962 galaxies drawn from the Galaxy Zoo 2 citizen science
morphological database. The catalogue on one hand tells the types of
features that Zoo volunteers cued on when they selected the button ``Is
the odd feature a ring"?, and on another provides a new large sample of
ringed galaxies for further study. Not suprisingly, the bulk of the
features that led to this button selection are normal disc rings,
mostly of the inner and especially outer types as defined in the CVRHS
system. These rings are well-represented because they are common,
large, and easily recognizable to redshifts $z$ $\approx$ 0.1.

The catalogue also serves the purpose of isolating subsets of the best
examples of galactic rings for follow-up analysis, e.g., using the
science-ready $ugri$ images from the Sloan Digital Sky Survey. This
will be the subject of other papers in this series.

An inventory of the catalogue gives the following:

\noindent
- more than 90\% of the selected galaxies are in the narrow CVRHS stage
range S0$^+$ to Sb. Although rings do occur in types Sbc and later,
these tend to be intrinsically smaller, less well-defined, and
generally much less frequent than the rings in earlier types.
Rings in galaxies earlier than S0$^+$ would likely also be difficult to
recognize at high $z$.

\noindent
- The sample tends to emphasize weak bars (classified as SAB), often
in the form of ansae or broad ovals.

\noindent
- nearly half of the outer features in the catalogue are classified as
outer resonant subclass rings.  In general, the bars seen in these
categories are weak (type SAB).

\noindent
- The catalogue has many biases that limit its statistical usefulness.
Other catalogues, such as the AINUR, ARRAKIS, and the NIRS0S, provide more
reliable information on statistical quantities such as ring frequencies
or the distribution of intrinsic sizes of the different types of rings.

Although no formal study has yet been made, the outer resonant
subclasses of galactic rings, as well as regular outer rings and
pseudorings, appear to be recognizable at much higher redshifts than
Table~\ref{tab:catalog} covers. In an examination of bars in the
GZ-Candels survey, Simmons et al. (2014; their Figure 3) show at least
two barred galaxies at redshifts $z$ =1.31 and 1.97 having what appear
to be R$_1^{\prime}$ outer pseudorings, and one nonbarred galaxy at $z$
= 0.58 having an inner lens and an outer ring. Similarly, Melvin et al.
(2014) show images from Galaxy Zoo-Hubble that include two R$_1$
examples (their Figure 4g and 4i) at redshifts $z$ = 0.47 and 0.665. As
high redshift studies of bars continue, following rings like those
discussed in this series of papers may also be worthy for direct
exploration of secular evolution.

I thank the anonymous referee for many helpful comments that improved
this paper. This work was supported in part by grant RGC-2012-30 from
the Research Grants Committee, University of Alabama. The development
of Galaxy Zoo was supported in part by the Alfred P. Sloan Foundation
and by the Leverhulme Trust. I thank K. L. Masters and Arfon Smith for
sending the links to the images of the sample of GZ2 ringed galaxies. I
also thank the numerous Galaxy Zoo 2 volunteers (named at
authors.galaxyzoo.org) for making it possible for me to compile and
examine a large new sample of ringed galaxies in far less time than it
took me to compile the Catalogue of Southern Ringed Galaxies. This
research has made use of the NASA/IPAC Extragalactic Database (NED)
which is operated by the Jet Propulsion Laboratory, California
Institute of Technology, under contract with the National Aeronautics
and Space Administration. Funding for the creation and distribution of
the SDSS Archive has been provided by the Alfred P.  Sloan Foundation,
the Participating Institutions, NASA, NSF, the U. S.  Department of
Energy, the Japanese Monbukagakusho, and the Max Planck Society.  IRAF
is written and supported by the National Optical Astronomy
Observatories (NOAO) in Tucson, Arizona. NOAO is operated by the
Association of Universities for Research in Astronomy (AURA), Inc.,
under cooperative agreement with the National Science Foundation.

\section{Appendix A}

The full version of Table 2 is given here and is also available online.

 \clearpage
 \begin{table*}
 \centering
 \setcounter{table}{1}
 \caption{CVRHS Classifications for the Galaxy Zoo 2 Ring Sample. Col. 1: NED
Name or SDSS name; col. 2: Principal Galaxy Catalogue number; col. 3: GZ2
image file name; col. 4: redshift; col. 5: numerical stage index; col. 6:
classification; col. 7: notes}
 \label{tab:catalog}
\begin{tabular}{lrrcrll}
 \hline
 2MASX/SDSS & PGC & file & $z$ & $T$ &                   CVRHS Type & Notes \\
 Other name &  &  &  &  &  &  \\
 1 & 2 & 3 & 4 & 5 & 6 & 7 \\
 \hline
{\bf $\rightarrow$ RA: 7$^h$ $\leftarrow$} & & & & \\                                                                                                                                                             
 J07242070+4209004  & 2193308 & 108110 & 0.057     &    2.0 & SB(r,bl)ab:                                                                            & poorly resolved; (r)  \\                                   
\phantom{PGC 00}"    & "\phantom{00} & "\phantom{00} & "\phantom{0} & "\phantom{0} & \phantom{00000}" & dark-spacer; strong bar  \\                                                                               
 J07250545+4125376  & 2180937 & 107884 & 0.109     &    1.0 & (R$_2^{\prime}$L)SB(s)a                                                                & very poorly resolved  \\                                   
 J07271769+4347053  & 2229046 & 109317 & 0.056     &    0.0 & (R$^{\prime}$)SA(l)0/a                                                                 & poorly resolved, but a good  \\
\hline
 \end{tabular}
 \end{table*}
 \clearpage
 \begin{table*}
 \centering
 \setcounter{table}{1}
 \caption{(cont.) CVRHS Classifications for the         Galaxy Zoo 2 Ring Sample.}
 \label{tab:catalog}
 \begin{tabular}{lrrcrll}
 \hline
 2MASX/SDSS & PGC & file & $z$ & $T$ &                   CVRHS Type & Notes \\
 Other &  &  &  &  &  &  \\
 1 & 2 & 3 & 4 & 5 & 6 & 7 \\
 \hline
J074802.52+184855.5 & 1571241 & 206139 & 0.053     &    1.0 & (R$^{\prime}$L)S$\underline{\rm A}$B(l)a                                               & poorly resolved  \\                                        
 J07480718+2237280  & 1673295 &  74381 & 0.077     &    3.0 & (R$^{\prime}$)SABb:                                                                    & poorly resolved; \RP\ is  \\                               
\phantom{PGC 00}"    & "\phantom{00} & "\phantom{00} & "\phantom{0} & "\phantom{0} & \phantom{00000}" & excellent but broken on one side  \\                                                                      
 J07480882+3249337  & 2013808 &  27533 & 0.080     &    1.0 & SAB($\underline{\rm r}$s)a                                                             & not a dark-spacer; face-on;  \\                            
\phantom{PGC 00}"    & "\phantom{00} & "\phantom{00} & "\phantom{0} & "\phantom{0} & \phantom{00000}" & poorly resolved  \\
\hline
 \end{tabular}
 \end{table*}
 \clearpage
 \begin{table*}
 \centering
 \setcounter{table}{1}
 \caption{(cont.) CVRHS Classifications for the         Galaxy Zoo 2 Ring Sample.}
 \label{tab:catalog}
 \begin{tabular}{lrrcrll}
 \hline
 2MASX/SDSS & PGC & file & $z$ & $T$ &                   CVRHS Type & Notes \\
 Other &  &  &  &  &  &  \\
 1 & 2 & 3 & 4 & 5 & 6 & 7 \\
 \hline
 J07543223+4338306  & 2225603 &   9216 & 0.104     &    0.0 & (R$_1^{\prime}$)SB(l)0/a                                                               & strong bar  \\                                             
 J07544061+3420340  & 2047680 &  26541 & 0.075     &    1.0 & (R$^{\prime}$)S$\underline{\rm A}$B(l)a                                                & poorly resolved  \\                                        
J075503.83+340830.7 & 2044521 &  27320 & 0.137     &    4.0 & SA(s)bc pec                                                                            & poorly resolved  \\                                        
MRK 0382            & 22190   &  24520 & 0.034     &    4.0 & (R$_{12}^{\prime}$)SA$\underline{\rm B}$(r)bc                                          & very unusual; strong \Lff\  \\                             
\phantom{PGC 00}"    & "\phantom{00} & "\phantom{00} & "\phantom{0} & "\phantom{0} & \phantom{00000}" & dark-spacer  \\
\hline
 \end{tabular}
 \end{table*}
 \clearpage
 \begin{table*}
 \centering
 \setcounter{table}{1}
 \caption{(cont.) CVRHS Classifications for the         Galaxy Zoo 2 Ring Sample.}
 \label{tab:catalog}
 \begin{tabular}{lrrcrll}
 \hline
 2MASX/SDSS & PGC & file & $z$ & $T$ &                   CVRHS Type & Notes \\
 Other &  &  &  &  &  &  \\
 1 & 2 & 3 & 4 & 5 & 6 & 7 \\
 \hline
\phantom{PGC 00}"    & "\phantom{00} & "\phantom{00} & "\phantom{0} & "\phantom{0} & \phantom{00000}" & ansae, (l), and (L)  \\                                                                                
CGCG 287-039        & 22689   & 109621 & 0.029     &    3.0 & (R$^{\prime}$,R)SA(r)b                                                                 & !excellent with a smooth (r)  \\                           
\phantom{PGC 00}"    & "\phantom{00} & "\phantom{00} & "\phantom{0} & "\phantom{0} & \phantom{00000}" & and very knotty blue (R),  \\                                                                             
\phantom{PGC 00}"    & "\phantom{00} & "\phantom{00} & "\phantom{0} & "\phantom{0} & \phantom{00000}" & likely an \Rtwo  \\                                                                                       
CGCG 207-027        & 22694   &  24315 & 0.050     &    1.0 & (R$_2^{\prime}$)SA$\underline{\rm B}$$_a$(l)a                                          & !excellent case  \\
\hline
 \end{tabular}
 \end{table*}
 \clearpage
 \begin{table*}
 \centering
 \setcounter{table}{1}
 \caption{(cont.) CVRHS Classifications for the         Galaxy Zoo 2 Ring Sample.}
 \label{tab:catalog}
 \begin{tabular}{lrrcrll}
 \hline
 2MASX/SDSS & PGC & file & $z$ & $T$ &                   CVRHS Type & Notes \\
 Other &  &  &  &  &  &  \\
 1 & 2 & 3 & 4 & 5 & 6 & 7 \\
 \hline
\phantom{PGC 00}"    & "\phantom{00} & "\phantom{00} & "\phantom{0} & "\phantom{0} & \phantom{00000}" & resolved; bar clear; (R) distorted;  \\                                                                   
\phantom{PGC 00}"    & "\phantom{00} & "\phantom{00} & "\phantom{0} & "\phantom{0} & \phantom{00000}" & almost face-on  \\                                                                                        
SBS 0808+580B       & 23022   & 109630 & 0.027     &    1.0 & SAB$_a$($\underline{\rm r}$s)a                                                         & !excellent case; very subtle  \\                           
\phantom{PGC 00}"    & "\phantom{00} & "\phantom{00} & "\phantom{0} & "\phantom{0} & \phantom{00000}" & features  \\                                                                                              
MCG +08-15-049      & 23015   &   7174 & 0.054     &    0.0 & (RL)SA$\underline{\rm B}$$_a$(l)0/a                                                    & excellent (RL)  \\
\hline
 \end{tabular}
 \end{table*}
 \clearpage
 \begin{table*}
 \centering
 \setcounter{table}{1}
 \caption{(cont.) CVRHS Classifications for the         Galaxy Zoo 2 Ring Sample.}
 \label{tab:catalog}
 \begin{tabular}{lrrcrll}
 \hline
 2MASX/SDSS & PGC & file & $z$ & $T$ &                   CVRHS Type & Notes \\
 Other &  &  &  &  &  &  \\
 1 & 2 & 3 & 4 & 5 & 6 & 7 \\
 \hline
 J08203757+1529398  & 1486130 & 151696 & .....     &    2.0 & (R$_{12}^{\prime}$)SB(s)ab:                                                            & poorly resolved; \Rone\ part  \\                           
\phantom{PGC 00}"    & "\phantom{00} & "\phantom{00} & "\phantom{0} & "\phantom{0} & \phantom{00000}" & is bluish; \RtwoP\ part is  \\                                                                            
\phantom{PGC 00}"    & "\phantom{00} & "\phantom{00} & "\phantom{0} & "\phantom{0} & \phantom{00000}" & yellowish  \\                                                                                             
UGC 04335           & 23412   & 105878 & 0.031     &    1.0 & (R$_1^{\prime}$)SB$_a$(p,r,bl)a                                                        & !excellent (r), a perfect  \\                              
\phantom{PGC 00}"    & "\phantom{00} & "\phantom{00} & "\phantom{0} & "\phantom{0} & \phantom{00000}" & case; bar underfills (r); \RoneP\  \\
\hline
 \end{tabular}
 \end{table*}
 \clearpage
 \begin{table*}
 \centering
 \setcounter{table}{1}
 \caption{(cont.) CVRHS Classifications for the         Galaxy Zoo 2 Ring Sample.}
 \label{tab:catalog}
 \begin{tabular}{lrrcrll}
 \hline
 2MASX/SDSS & PGC & file & $z$ & $T$ &                   CVRHS Type & Notes \\
 Other &  &  &  &  &  &  \\
 1 & 2 & 3 & 4 & 5 & 6 & 7 \\
 \hline
UGC 04414           & 23700   & 118577 & 0.025     &    0.0 & (R)SB(rs,bl)0/a                                                                        & !excellent case; half of (R)  \\                           
\phantom{PGC 00}"    & "\phantom{00} & "\phantom{00} & "\phantom{0} & "\phantom{0} & \phantom{00000}" & is blue; classic brackett (rs)  \\                                                                        
\phantom{PGC 00}"    & "\phantom{00} & "\phantom{00} & "\phantom{0} & "\phantom{0} & \phantom{00000}" & around bar  \\                                                                                            
 J08272782+3649258  & 2087632 &  45560 & 0.082     &    1.0 & (R$^{\prime}$)S$\underline{\rm A}$B(l)a                                                & good case but poorly resolved  \\                          
 J08273213+1842237  & 1567697 & 148186 & 0.112     &    3.0 & (R$^{\prime}$)SABb: pec                                                                & \RP\ made of single arm  \\
\hline
 \end{tabular}
 \end{table*}
 \clearpage
 \begin{table*}
 \centering
 \setcounter{table}{1}
 \caption{(cont.) CVRHS Classifications for the         Galaxy Zoo 2 Ring Sample.}
 \label{tab:catalog}
 \begin{tabular}{lrrcrll}
 \hline
 2MASX/SDSS & PGC & file & $z$ & $T$ &                   CVRHS Type & Notes \\
 Other &  &  &  &  &  &  \\
 1 & 2 & 3 & 4 & 5 & 6 & 7 \\
 \hline
 J08330779+5009599  & 2364893 & 191851 & 0.053     &    1.0 & (R$_1^{\prime}$)SB(s)a                                                                 & good case; \Lff\ dark-spacer  \\                           
 J08331903+3124002  & 1945338 & 239825 & 0.071     & $-$1.0 & (R)SA(l)0$^+$                                                                          & excellent case; all around  \\                             
\phantom{PGC 00}"    & "\phantom{00} & "\phantom{00} & "\phantom{0} & "\phantom{0} & \phantom{00000}" & dark spacer; faint arc ansae  \\                                                                          
MCG +09-14-056      & 23990   &  10079 & 0.063     &    2.0 & (R$^{\prime}$)SA(r$^{\prime}$l)ab                                                      & .......... \\                                              
UGC 04468           & 24002   &  48447 & 0.025     &   .... & ETG (shells) pec                                                                       & very strange object with at  \\
\hline
 \end{tabular}
 \end{table*}
 \clearpage
 \begin{table*}
 \centering
 \setcounter{table}{1}
 \caption{(cont.) CVRHS Classifications for the         Galaxy Zoo 2 Ring Sample.}
 \label{tab:catalog}
 \begin{tabular}{lrrcrll}
 \hline
 2MASX/SDSS & PGC & file & $z$ & $T$ &                   CVRHS Type & Notes \\
 Other &  &  &  &  &  &  \\
 1 & 2 & 3 & 4 & 5 & 6 & 7 \\
 \hline
\phantom{PGC 00}"    & "\phantom{00} & "\phantom{00} & "\phantom{0} & "\phantom{0} & \phantom{00000}" & \RoneP\ dimpled on one side; suspect  \\                                                                  
\phantom{PGC 00}"    & "\phantom{00} & "\phantom{00} & "\phantom{0} & "\phantom{0} & \phantom{00000}" & bar - \RoneP\ misalignment;  \\                                                                           
\phantom{PGC 00}"    & "\phantom{00} & "\phantom{00} & "\phantom{0} & "\phantom{0} & \phantom{00000}" & strong (r) arc-handles  \\                                                                                
 J08413967+2647414  & 1788601 & 116235 & 0.089     &    0.0 & SA($\underline{\rm r}$s)0/a: pec                                                       & face-on; poorly resolved  \\                               
 J08414357+2930444  & 1868698 &  78649 & 0.080     & $-$1.0 & SA($\underline{\rm r}$s)0$^+$                                                          & good but poorly resolved;  \\
\hline
 \end{tabular}
 \end{table*}
 \clearpage
 \begin{table*}
 \centering
 \setcounter{table}{1}
 \caption{(cont.) CVRHS Classifications for the         Galaxy Zoo 2 Ring Sample.}
 \label{tab:catalog}
 \begin{tabular}{lrrcrll}
 \hline
 2MASX/SDSS & PGC & file & $z$ & $T$ &                   CVRHS Type & Notes \\
 Other &  &  &  &  &  &  \\
 1 & 2 & 3 & 4 & 5 & 6 & 7 \\
 \hline
\phantom{PGC 00}"    & "\phantom{00} & "\phantom{00} & "\phantom{0} & "\phantom{0} & \phantom{00000}" & obvious ansae  \\                                                                                         
 J08482689+6136288  & 2621183 & 107829 & 0.027     &    0.0 & (R$^{\prime}$:)SB(r)0/a [c]                                                            & small bulge early-type spiral  \\                          
CGCG 061-013        & 24759   & 186082 & 0.030     &    2.0 & (R$_1$R$_2^{\prime}$)SA(r)ab                                                           & !excellent \RtwoP\ and (r);  \\                            
\phantom{PGC 00}"    & "\phantom{00} & "\phantom{00} & "\phantom{0} & "\phantom{0} & \phantom{00000}" & strong \Lff\ dark-spacer;  \\                                                                             
\phantom{PGC 00}"    & "\phantom{00} & "\phantom{00} & "\phantom{0} & "\phantom{0} & \phantom{00000}" & face-on  \\
\hline
 \end{tabular}
 \end{table*}
 \clearpage
 \begin{table*}
 \centering
 \setcounter{table}{1}
 \caption{(cont.) CVRHS Classifications for the         Galaxy Zoo 2 Ring Sample.}
 \label{tab:catalog}
 \begin{tabular}{lrrcrll}
 \hline
 2MASX/SDSS & PGC & file & $z$ & $T$ &                   CVRHS Type & Notes \\
 Other &  &  &  &  &  &  \\
 1 & 2 & 3 & 4 & 5 & 6 & 7 \\
 \hline
 J08543349+0744553  & 1329827 &  79465 & 0.094     &    2.0 & SB(r)ab:                                                                               & poorly resolved, but a good  \\                            
\phantom{PGC 00}"    & "\phantom{00} & "\phantom{00} & "\phantom{0} & "\phantom{0} & \phantom{00000}" & case; (r) dark-spacer like NGC  \\                                                                        
\phantom{PGC 00}"    & "\phantom{00} & "\phantom{00} & "\phantom{0} & "\phantom{0} & \phantom{00000}" & 5335; strong bar  \\                                                                                      
CGCG 208-067        & 25013   &  45059 & 0.031     &    1.0 & (R$_1$)SAB(p,r$^{\prime}$l)a                                                           & excellent case with weak  \\                               
\phantom{PGC 00}"    & "\phantom{00} & "\phantom{00} & "\phantom{0} & "\phantom{0} & \phantom{00000}" & plumes; \Lff\ dark-spacer  \\
\hline
 \end{tabular}
 \end{table*}
 \clearpage
 \begin{table*}
 \centering
 \setcounter{table}{1}
 \caption{(cont.) CVRHS Classifications for the         Galaxy Zoo 2 Ring Sample.}
 \label{tab:catalog}
 \begin{tabular}{lrrcrll}
 \hline
 2MASX/SDSS & PGC & file & $z$ & $T$ &                   CVRHS Type & Notes \\
 Other &  &  &  &  &  &  \\
 1 & 2 & 3 & 4 & 5 & 6 & 7 \\
 \hline
 J09000228+5750067  & 3736510 &  10115 & .....     &    1.0 & (R$_1^{\prime}$)S$\underline{\rm A}$B(l)a                                              & ring faint \\                                                
J090012.44+385234.2 & ....... &  50364 & 0.075     & $-$5.0 & (R)E0:                                                                                 & \RP\ is broad zone of  \\                                  
\phantom{PGC 00}"    & "\phantom{00} & "\phantom{00} & "\phantom{0} & "\phantom{0} & \phantom{00000}" & flocculent spiral structure; bulge is  \\                                                                 
\phantom{PGC 00}"    & "\phantom{00} & "\phantom{00} & "\phantom{0} & "\phantom{0} & \phantom{00000}" & like an E0  \\                                                                                            
 J09005433+3346581  & 2038711 &  79092 & 0.051     & $-$1.0 & (RL)SB0$^+$                                                                            & very good case but poorly  \\
\hline
 \end{tabular}
 \end{table*}
 \clearpage
 \begin{table*}
 \centering
 \setcounter{table}{1}
 \caption{(cont.) CVRHS Classifications for the         Galaxy Zoo 2 Ring Sample.}
 \label{tab:catalog}
 \begin{tabular}{lrrcrll}
 \hline
 2MASX/SDSS & PGC & file & $z$ & $T$ &                   CVRHS Type & Notes \\
 Other &  &  &  &  &  &  \\
 1 & 2 & 3 & 4 & 5 & 6 & 7 \\
 \hline
 J09080001+3140055  & 1958331 &  83734 & 0.063     &    0.0 & SA($\underline{\rm r}$s)0/a                                                            & very good case  \\                                         
 J09082077+3952374  & 3454634 &  50371 & 0.085     & $-$1.0 & (RL)S$\underline{\rm A}$B0$^+$:                                                        & poorly resolved  \\                                        
 J09082576+3511335  & 3454651 &  72926 & 0.087     &    3.0 & (R$_2^{\prime}$)SAB$_a$(r$^{\prime}$l)b                                                & very good case, but poorly  \\                             
\phantom{PGC 00}"    & "\phantom{00} & "\phantom{00} & "\phantom{0} & "\phantom{0} & \phantom{00000}" & resolved  \\                                                                                              
J090828.20+183852.2 & ....... & 156087 & 0.084     &    2.0 & SA(r)ab: pec                                                                           & .......... \\
\hline
 \end{tabular}
 \end{table*}
 \clearpage
 \begin{table*}
 \centering
 \setcounter{table}{1}
 \caption{(cont.) CVRHS Classifications for the         Galaxy Zoo 2 Ring Sample.}
 \label{tab:catalog}
 \begin{tabular}{lrrcrll}
 \hline
 2MASX/SDSS & PGC & file & $z$ & $T$ &                   CVRHS Type & Notes \\
 Other &  &  &  &  &  &  \\
 1 & 2 & 3 & 4 & 5 & 6 & 7 \\
 \hline
\phantom{PGC 00}"    & "\phantom{00} & "\phantom{00} & "\phantom{0} & "\phantom{0} & \phantom{00000}" & yellow-white, mostly (L) feature; bluish  \\                                                              
\phantom{PGC 00}"    & "\phantom{00} & "\phantom{00} & "\phantom{0} & "\phantom{0} & \phantom{00000}" & star-forming \urs; much light  \\                                                                         
\phantom{PGC 00}"    & "\phantom{00} & "\phantom{00} & "\phantom{0} & "\phantom{0} & \phantom{00000}" & beyond (L)  \\                                                                                            
CGCG 121-040        & 26041   & 150833 & 0.027     &    1.0 & (R$^{\prime}$,L)SA(r)a                                                                 & (r) very bright and blue  \\                               
 J09143697+3234213  & 2003084 & 207086 & 0.049     &    1.0 & (R$^{\prime}$)SA(l)a                                                                   & poorly resolved but clear;  \\
\hline
 \end{tabular}
 \end{table*}
 \clearpage
 \begin{table*}
 \centering
 \setcounter{table}{1}
 \caption{(cont.) CVRHS Classifications for the         Galaxy Zoo 2 Ring Sample.}
 \label{tab:catalog}
 \begin{tabular}{lrrcrll}
 \hline
 2MASX/SDSS & PGC & file & $z$ & $T$ &                   CVRHS Type & Notes \\
 Other &  &  &  &  &  &  \\
 1 & 2 & 3 & 4 & 5 & 6 & 7 \\
 \hline
 J09191064+3325295  & 2031768 &  83956 & 0.023     & $-$1.0 & (RL)SA(l)0$^+$ [d] pec                                                                 & Comet Holmes look-alike;  \\                               
\phantom{PGC 00}"    & "\phantom{00} & "\phantom{00} & "\phantom{0} & "\phantom{0} & \phantom{00000}" & ``puff;" very odd; no bulge  \\                                                                           
 J09193939+3917528  & 2147227 &  52406 & 0.092     &    4.0 & (R$^{\prime}$)SA(rs)bc pec                                                             & round center plus bluish  \\                               
\phantom{PGC 00}"    & "\phantom{00} & "\phantom{00} & "\phantom{0} & "\phantom{0} & \phantom{00000}" & pseudoring; (r) versus (R) unclear  \\                                                                    
J091940.95-002151.3 & ....... &   5260 & 0.055     & $-$1.0 & (R)SAB$_a$(l)0$^+$                                                                     & !excellent (R)  \\
\hline
 \end{tabular}
 \end{table*}
 \clearpage
 \begin{table*}
 \centering
 \setcounter{table}{1}
 \caption{(cont.) CVRHS Classifications for the         Galaxy Zoo 2 Ring Sample.}
 \label{tab:catalog}
 \begin{tabular}{lrrcrll}
 \hline
 2MASX/SDSS & PGC & file & $z$ & $T$ &                   CVRHS Type & Notes \\
 Other &  &  &  &  &  &  \\
 1 & 2 & 3 & 4 & 5 & 6 & 7 \\
 \hline
\phantom{PGC 00}"    & "\phantom{00} & "\phantom{00} & "\phantom{0} & "\phantom{0} & \phantom{00000}" & bluish galaxy  \\                                                                                         
 J09263342+4800415  & 2308766 &  49788 & 0.042     &    1.0 & SB$_a$($\underline{\rm r}$s)a                                                          & excellent case; very strong  \\                            
\phantom{PGC 00}"    & "\phantom{00} & "\phantom{00} & "\phantom{0} & "\phantom{0} & \phantom{00000}" & (r) dark-spacer  \\                                                                                       
IC 2472             & 26779   & 235733 & 0.026     &    4.0 & (R$^{\prime}$:)SB($\underline{\rm r}$s)bc                                              & very unusual; huge blue  \\                                
\phantom{PGC 00}"    & "\phantom{00} & "\phantom{00} & "\phantom{0} & "\phantom{0} & \phantom{00000}" & ``frame" around \urs; yellow bar;  \\
\hline
 \end{tabular}
 \end{table*}
 \clearpage
 \begin{table*}
 \centering
 \setcounter{table}{1}
 \caption{(cont.) CVRHS Classifications for the         Galaxy Zoo 2 Ring Sample.}
 \label{tab:catalog}
 \begin{tabular}{lrrcrll}
 \hline
 2MASX/SDSS & PGC & file & $z$ & $T$ &                   CVRHS Type & Notes \\
 Other &  &  &  &  &  &  \\
 1 & 2 & 3 & 4 & 5 & 6 & 7 \\
 \hline
\phantom{PGC 00}"    & "\phantom{00} & "\phantom{00} & "\phantom{0} & "\phantom{0} & \phantom{00000}" & excellent, dimpled \RoneP  \\                                                                             
CGCG 122-010        & 27005   & 148704 & 0.024     &    3.0 & SAB(rs,l)b                                                                             & very good case  \\                                         
SBS 0927+493        & 27014   &  49936 & 0.034     &    0.0 & (L,R)SB$_a$0/a                                                                         & excellent case; strong bar  \\                             
\phantom{PGC 00}"    & "\phantom{00} & "\phantom{00} & "\phantom{0} & "\phantom{0} & \phantom{00000}" & ansae  \\                                                                                                 
 J09310699+6414418  & 3743207 & 107094 & 0.119     &    3.0 & SA(r)b:                                                                                & odd case  \\
\hline
 \end{tabular}
 \end{table*}
 \clearpage
 \begin{table*}
 \centering
 \setcounter{table}{1}
 \caption{(cont.) CVRHS Classifications for the         Galaxy Zoo 2 Ring Sample.}
 \label{tab:catalog}
 \begin{tabular}{lrrcrll}
 \hline
 2MASX/SDSS & PGC & file & $z$ & $T$ &                   CVRHS Type & Notes \\
 Other &  &  &  &  &  &  \\
 1 & 2 & 3 & 4 & 5 & 6 & 7 \\
 \hline
 J09370169-0213004  & 1098404 &   5834 & .....     &    2.0 & (R$^{\prime}$)SABab:                                                                   & superposed star  \\                                        
 J09372076+3853353  & 2138999 &  72968 & 0.061     &    2.0 & (R$_1$R$_2^{\prime}$)SAB$_a$(l)ab                                                      & very good \Rone\ plus one  \\                              
\phantom{PGC 00}"    & "\phantom{00} & "\phantom{00} & "\phantom{0} & "\phantom{0} & \phantom{00000}" & bright ansa; poorly resolved;  \\                                                                         
\phantom{PGC 00}"    & "\phantom{00} & "\phantom{00} & "\phantom{0} & "\phantom{0} & \phantom{00000}" & \Lff\ dark-spacer  \\                                                                                     
 J09372948+1627330  & 1509268 & 162207 & 0.073     &    3.0 & (R$_2^{\prime}$)SAB(s)b:                                                               & small and poorly resolved,  \\
\hline
 \end{tabular}
 \end{table*}
 \clearpage
 \begin{table*}
 \centering
 \setcounter{table}{1}
 \caption{(cont.) CVRHS Classifications for the         Galaxy Zoo 2 Ring Sample.}
 \label{tab:catalog}
 \begin{tabular}{lrrcrll}
 \hline
 2MASX/SDSS & PGC & file & $z$ & $T$ &                   CVRHS Type & Notes \\
 Other &  &  &  &  &  &  \\
 1 & 2 & 3 & 4 & 5 & 6 & 7 \\
 \hline
 J09445232+0825296  & 3460256 &  62835 & 0.084     &    0.0 & (R)SA(l)0/a                                                                            & poorly resolved  \\                                        
IC 2506             & 27940   & 134167 & 0.033     &    1.0 & (R$_1^{\prime}$)SAB$_a$($\underline{\rm r}$s,bl)a                                      & good case  \\                                              
CGCG 122-070        & 27976   & 151822 & 0.022     &    4.0 & SB(rs)bc                                                                               & .......... \\                                              
WAS 02              & ....... & 235556 & 0.034     & $-$1.0 & (R)SAB(l)0$^+$                                                                         & !excellent case; face-on;  \\                              
\phantom{PGC 00}"    & "\phantom{00} & "\phantom{00} & "\phantom{0} & "\phantom{0} & \phantom{00000}" & \Lff\ dark-spacer  \\
\hline
 \end{tabular}
 \end{table*}
 \clearpage
 \begin{table*}
 \centering
 \setcounter{table}{1}
 \caption{(cont.) CVRHS Classifications for the         Galaxy Zoo 2 Ring Sample.}
 \label{tab:catalog}
 \begin{tabular}{lrrcrll}
 \hline
 2MASX/SDSS & PGC & file & $z$ & $T$ &                   CVRHS Type & Notes \\
 Other &  &  &  &  &  &  \\
 1 & 2 & 3 & 4 & 5 & 6 & 7 \\
 \hline
\phantom{PGC 00}"    & "\phantom{00} & "\phantom{00} & "\phantom{0} & "\phantom{0} & \phantom{00000}" & attached one end  \\                                                                                      
 J09530860+2417570  & 1704155 & 151827 & 0.034     &    1.0 & (R$_1^{\prime}$)SB(r$^{\prime}$l)a:                                                    & very good, white  \\                                       
 J09535267+1733508  & 1535805 & 166102 & 0.071     & $-$1.0 & (R)S$\underline{\rm A}$B(l)0$^+$                                                       & excellent case; poorly  \\                                 
\phantom{PGC 00}"    & "\phantom{00} & "\phantom{00} & "\phantom{0} & "\phantom{0} & \phantom{00000}" & resolved; \Lff\ dark-spacer  \\                                                                           
 J09535907+2334349  & 1689795 & 157242 & 0.066     & $-$1.0 & (L)SA(r)0$^+$:                                                                         & poorly resolved  \\
\hline
 \end{tabular}
 \end{table*}
 \clearpage
 \begin{table*}
 \centering
 \setcounter{table}{1}
 \caption{(cont.) CVRHS Classifications for the         Galaxy Zoo 2 Ring Sample.}
 \label{tab:catalog}
 \begin{tabular}{lrrcrll}
 \hline
 2MASX/SDSS & PGC & file & $z$ & $T$ &                   CVRHS Type & Notes \\
 Other &  &  &  &  &  &  \\
 1 & 2 & 3 & 4 & 5 & 6 & 7 \\
 \hline
 J09590860+0814099  & 3462291 &  62352 & 0.097     &    0.0 & (R$_1^{\prime}$)SAB$_a$(l)0/a                                                          & very good but poorly resolved  \\                          
 J09592499+3404493  & 3749771 & 226596 & 0.095     &    3.0 & (R$_2^{\prime}$)S$\underline{\rm A}$B(l)b + IR?                                        & large circular, blue ring;  \\                             
\phantom{PGC 00}"    & "\phantom{00} & "\phantom{00} & "\phantom{0} & "\phantom{0} & \phantom{00000}" & \Lff\ dark-spacer; very faint  \\                                                                         
\phantom{PGC 00}"    & "\phantom{00} & "\phantom{00} & "\phantom{0} & "\phantom{0} & \phantom{00000}" & outer structure is unusual  \\                                                                            
 J09592781-0207077  & 1101256 &  29336 & .....     & $-$1.0 & (R)SAB(l)0$^+$                                                                         & poorly resolved  \\
\hline
 \end{tabular}
 \end{table*}
 \clearpage
 \begin{table*}
 \centering
 \setcounter{table}{1}
 \caption{(cont.) CVRHS Classifications for the         Galaxy Zoo 2 Ring Sample.}
 \label{tab:catalog}
 \begin{tabular}{lrrcrll}
 \hline
 2MASX/SDSS & PGC & file & $z$ & $T$ &                   CVRHS Type & Notes \\
 Other &  &  &  &  &  &  \\
 1 & 2 & 3 & 4 & 5 & 6 & 7 \\
 \hline
\phantom{PGC 00}"    & "\phantom{00} & "\phantom{00} & "\phantom{0} & "\phantom{0} & \phantom{00000}" & dark-spacer zones  \\                                                                                     
CGCG 240-005        & 29244   &  54789 & 0.023     &    0.0 & (R)SB(r,bl)0/a                                                                         & !excellent bar and round (r)  \\                           
J100457.50+123031.6 & 3533255 & 110631 & 0.030     &    7.0 & (R$^{\prime}$)SABd: pec                                                                & very odd; large blue knots  \\                             
 J10050435+0244220  & 1236100 &  22761 & 0.081     &    1.0 & (R)S$\underline{\rm A}$B(r)a                                                           & excellent case; very knotty  \\                            
\phantom{PGC 00}"    & "\phantom{00} & "\phantom{00} & "\phantom{0} & "\phantom{0} & \phantom{00000}" & (r)  \\
\hline
 \end{tabular}
 \end{table*}
 \clearpage
 \begin{table*}
 \centering
 \setcounter{table}{1}
 \caption{(cont.) CVRHS Classifications for the         Galaxy Zoo 2 Ring Sample.}
 \label{tab:catalog}
 \begin{tabular}{lrrcrll}
 \hline
 2MASX/SDSS & PGC & file & $z$ & $T$ &                   CVRHS Type & Notes \\
 Other &  &  &  &  &  &  \\
 1 & 2 & 3 & 4 & 5 & 6 & 7 \\
 \hline
CGCG 064-076        & 29508   & 110638 & 0.030     &    1.0 & (R$^{\prime}$)SAB(r,bl)a                                                               & .......... \\                                              
 J10084513+2321395  & 1686047 & 235389 & 0.111     &    0.0 & (RL)SA0/a                                                                              & very poorly resolved  \\                                   
 J10085396+6224078  & 2636018 &   8625 & 0.053     &    2.0 & SA($\underline{\rm r}$s)ab:                                                            & near two bright stars  \\                                  
 J10090223+4503483  & 3407419 &  54794 & 0.114     & $-$1.0 & (RL)SA0$^+$                                                                            & poorly resolved  \\                                        
 J10091123+2253303  & 1677848 & 235182 & 0.053     &    1.0 & (R$_1$R$_2^{\prime}$)SA(rl)a                                                           & strong banana \Lff\  \\
\hline
 \end{tabular}
 \end{table*}
 \clearpage
 \begin{table*}
 \centering
 \setcounter{table}{1}
 \caption{(cont.) CVRHS Classifications for the         Galaxy Zoo 2 Ring Sample.}
 \label{tab:catalog}
 \begin{tabular}{lrrcrll}
 \hline
 2MASX/SDSS & PGC & file & $z$ & $T$ &                   CVRHS Type & Notes \\
 Other &  &  &  &  &  &  \\
 1 & 2 & 3 & 4 & 5 & 6 & 7 \\
 \hline
\phantom{PGC 00}"    & "\phantom{00} & "\phantom{00} & "\phantom{0} & "\phantom{0} & \phantom{00000}" & whitish; \RoneL\ is yellowish  \\                                                                         
NGC 3151            & 29796   &  84046 & 0.023     & $-$1.0 & (R)SB$_a$(r$^{\prime}$l)0$^+$                                                          & excellent case; very diffuse  \\                           
\phantom{PGC 00}"    & "\phantom{00} & "\phantom{00} & "\phantom{0} & "\phantom{0} & \phantom{00000}" & features; strong ansae  \\                                                                                
CGCG 008-073        & 29801   &  29011 & 0.044     &    0.0 & (R$_1$)SA$\underline{\rm B}$(rl)0/a                                                    & !excellent, large case; very  \\                           
\phantom{PGC 00}"    & "\phantom{00} & "\phantom{00} & "\phantom{0} & "\phantom{0} & \phantom{00000}" & oval (rl); very faint \RoneP  \\
\hline
 \end{tabular}
 \end{table*}
 \clearpage
 \begin{table*}
 \centering
 \setcounter{table}{1}
 \caption{(cont.) CVRHS Classifications for the         Galaxy Zoo 2 Ring Sample.}
 \label{tab:catalog}
 \begin{tabular}{lrrcrll}
 \hline
 2MASX/SDSS & PGC & file & $z$ & $T$ &                   CVRHS Type & Notes \\
 Other &  &  &  &  &  &  \\
 1 & 2 & 3 & 4 & 5 & 6 & 7 \\
 \hline
 J10193427+6013346  & 2599504 & 193684 & 0.156     &    2.0 & SA(r)ab: pec                                                                           & a nearly round blue ring has  \\                           
\phantom{PGC 00}"    & "\phantom{00} & "\phantom{00} & "\phantom{0} & "\phantom{0} & \phantom{00000}" & asymmetric structure outside  \\                                                                          
 J10194223+5147523  & 2400298 &  46806 & 0.065     &    0.0 & (RL)SA(l)0/a:                                                                          & poorly resolved  \\                                        
 J10194382+4950323  & 2355120 &  63640 & 0.140     &    2.0 & (RL)S$\underline{\rm A}$B(l)ab                                                         & good but poorly resolved  \\                               
 J10194857+3806069  & 2117864 &  83664 & 0.053     &    0.0 & (R:)SAB(r)0/a                                                                          & excellent (r) is underfilled  \\
\hline
 \end{tabular}
 \end{table*}
 \clearpage
 \begin{table*}
 \centering
 \setcounter{table}{1}
 \caption{(cont.) CVRHS Classifications for the         Galaxy Zoo 2 Ring Sample.}
 \label{tab:catalog}
 \begin{tabular}{lrrcrll}
 \hline
 2MASX/SDSS & PGC & file & $z$ & $T$ &                   CVRHS Type & Notes \\
 Other &  &  &  &  &  &  \\
 1 & 2 & 3 & 4 & 5 & 6 & 7 \\
 \hline
\phantom{PGC 00}"    & "\phantom{00} & "\phantom{00} & "\phantom{0} & "\phantom{0} & \phantom{00000}" & still very clear; strong \Lff\  \\                                                                        
\phantom{PGC 00}"    & "\phantom{00} & "\phantom{00} & "\phantom{0} & "\phantom{0} & \phantom{00000}" & dark-spacer  \\                                                                                           
 J10231846+0812304  & 1341711 &  74615 & .....     &    1.0 & (R$_2^{\prime}$)SA(l)a:                                                                & poorly resolved  \\                                        
 J10232074+1204385  & 1403382 &  79930 & 0.107     &    2.0 & (R$^{\prime}$)SA(rl)ab pec                                                             & asymmetric outer structure;  \\                            
\phantom{PGC 00}"    & "\phantom{00} & "\phantom{00} & "\phantom{0} & "\phantom{0} & \phantom{00000}" & \urs\ very good  \\
\hline
 \end{tabular}
 \end{table*}
 \clearpage
 \begin{table*}
 \centering
 \setcounter{table}{1}
 \caption{(cont.) CVRHS Classifications for the         Galaxy Zoo 2 Ring Sample.}
 \label{tab:catalog}
 \begin{tabular}{lrrcrll}
 \hline
 2MASX/SDSS & PGC & file & $z$ & $T$ &                   CVRHS Type & Notes \\
 Other &  &  &  &  &  &  \\
 1 & 2 & 3 & 4 & 5 & 6 & 7 \\
 \hline
\phantom{PGC 00}"    & "\phantom{00} & "\phantom{00} & "\phantom{0} & "\phantom{0} & \phantom{00000}" & asymmetric; \Lff\ dark-spacer  \\                                                                         
 J10274339+2022428  & 3755947 & 164238 & 0.111     &    1.0 & (R$_1^{\prime}$)S$\underline{\rm A}$B(l)a:                                             & very poorly resolved  \\                                   
J102756.16+042456.0 & ....... & 196012 & 0.154     &    3.0 & (R$_2^{\prime}$)SAB(l)b:                                                               & poorly resolved, but likely  \\                            
\phantom{PGC 00}"    & "\phantom{00} & "\phantom{00} & "\phantom{0} & "\phantom{0} & \phantom{00000}" & has an \RtwoP  \\                                                                                         
MCG +07-22-008      & 30794   & 207853 & .....     &    3.0 & SB(r)b pec                                                                             & (r) is very good and has  \\
\hline
 \end{tabular}
 \end{table*}
 \clearpage
 \begin{table*}
 \centering
 \setcounter{table}{1}
 \caption{(cont.) CVRHS Classifications for the         Galaxy Zoo 2 Ring Sample.}
 \label{tab:catalog}
 \begin{tabular}{lrrcrll}
 \hline
 2MASX/SDSS & PGC & file & $z$ & $T$ &                   CVRHS Type & Notes \\
 Other &  &  &  &  &  &  \\
 1 & 2 & 3 & 4 & 5 & 6 & 7 \\
 \hline
\phantom{PGC 00}"    & "\phantom{00} & "\phantom{00} & "\phantom{0} & "\phantom{0} & \phantom{00000}" & bluish-tinted (r)  \\                                                                                     
J103229.86+113215.6 & 1395109 & 217730 & 0.095     &    2.0 & (R$_1^{\prime}$)SAB(l)ab:                                                              & very poorly resolved; (l) has  \\                          
\phantom{PGC 00}"    & "\phantom{00} & "\phantom{00} & "\phantom{0} & "\phantom{0} & \phantom{00000}" & conspicuous ``handles"; \Lff\  \\                                                                         
\phantom{PGC 00}"    & "\phantom{00} & "\phantom{00} & "\phantom{0} & "\phantom{0} & \phantom{00000}" & dark-spacer  \\                                                                                           
 J10323391+3333298  & 2034532 & 113794 & 0.043     &    1.0 & (R$^{\prime}$,R)SA$\underline{\rm B}$(r)a                                              & very good  \\
\hline
 \end{tabular}
 \end{table*}
 \clearpage
 \begin{table*}
 \centering
 \setcounter{table}{1}
 \caption{(cont.) CVRHS Classifications for the         Galaxy Zoo 2 Ring Sample.}
 \label{tab:catalog}
 \begin{tabular}{lrrcrll}
 \hline
 2MASX/SDSS & PGC & file & $z$ & $T$ &                   CVRHS Type & Notes \\
 Other &  &  &  &  &  &  \\
 1 & 2 & 3 & 4 & 5 & 6 & 7 \\
 \hline
\phantom{PGC 00}"    & "\phantom{00} & "\phantom{00} & "\phantom{0} & "\phantom{0} & \phantom{00000}" & spiral arcs in (r)  \\                                                                                    
 J10364449+4956324  & 2358103 &  51862 & 0.043     & $-$1.0 & (L)SB($\underline{\rm r}$s)0$^+$                                                       & (r) dark-spacer  \\                                        
 J10364875+0531091  & 1284744 &  26383 & 0.030     &    0.0 & S$\underline{\rm A}$B(r,bl)0/a                                                         & very good case  \\                                         
 J10365013-0138526  & 31489   &   6311 & 0.028     &    0.0 & (R$_1^{\prime}$L)SA$\underline{\rm B}$$_a$(l)0/a                                       & strong bluish ansae; broad  \\                             
\phantom{PGC 00}"    & "\phantom{00} & "\phantom{00} & "\phantom{0} & "\phantom{0} & \phantom{00000}" & (l) region  \\
\hline
 \end{tabular}
 \end{table*}
 \clearpage
 \begin{table*}
 \centering
 \setcounter{table}{1}
 \caption{(cont.) CVRHS Classifications for the         Galaxy Zoo 2 Ring Sample.}
 \label{tab:catalog}
 \begin{tabular}{lrrcrll}
 \hline
 2MASX/SDSS & PGC & file & $z$ & $T$ &                   CVRHS Type & Notes \\
 Other &  &  &  &  &  &  \\
 1 & 2 & 3 & 4 & 5 & 6 & 7 \\
 \hline
 J10420253+1544147  & 1492143 & 181506 & 0.051     &    0.0 & (RL)SA(rs)0/a                                                                          & excellent case; (RL) very  \\                              
\phantom{PGC 00}"    & "\phantom{00} & "\phantom{00} & "\phantom{0} & "\phantom{0} & \phantom{00000}" & broad; white galaxy  \\                                                                                   
 J10420255+0746237  & 3091596 &  57799 & 0.078     &    0.0 & (R$^{\prime}$)SAB(r$^{\prime}$l)0/a                                                    & very good case; \Lff\  \\                                  
\phantom{PGC 00}"    & "\phantom{00} & "\phantom{00} & "\phantom{0} & "\phantom{0} & \phantom{00000}" & dark-spacer  \\                                                                                           
 J10421046+0823309  & 1345931 &  61467 & 0.158     &    3.0 & (R$_2^{\prime}$)SA$\underline{\rm B}$(s,nb)b                                           & very good face-on case, but  \\
\hline
 \end{tabular}
 \end{table*}
 \clearpage
 \begin{table*}
 \centering
 \setcounter{table}{1}
 \caption{(cont.) CVRHS Classifications for the         Galaxy Zoo 2 Ring Sample.}
 \label{tab:catalog}
 \begin{tabular}{lrrcrll}
 \hline
 2MASX/SDSS & PGC & file & $z$ & $T$ &                   CVRHS Type & Notes \\
 Other &  &  &  &  &  &  \\
 1 & 2 & 3 & 4 & 5 & 6 & 7 \\
 \hline
\phantom{PGC 00}"    & "\phantom{00} & "\phantom{00} & "\phantom{0} & "\phantom{0} & \phantom{00000}" & case; \urs\ is very bright and  \\                                                                        
\phantom{PGC 00}"    & "\phantom{00} & "\phantom{00} & "\phantom{0} & "\phantom{0} & \phantom{00000}" & circular; (R) highly elongated  \\                                                                        
\phantom{PGC 00}"    & "\phantom{00} & "\phantom{00} & "\phantom{0} & "\phantom{0} & \phantom{00000}" & as in NGC 7187  \\                                                                                        
 J10453300+0809218  & 3760109 &  74535 & 0.050     &    2.0 & (R$_1^{\prime}$)SAB(p,r$^{\prime}$l)ab                                                 & very good but poorly  \\                                   
\phantom{PGC 00}"    & "\phantom{00} & "\phantom{00} & "\phantom{0} & "\phantom{0} & \phantom{00000}" & resolved; short plumes  \\
\hline
 \end{tabular}
 \end{table*}
 \clearpage
 \begin{table*}
 \centering
 \setcounter{table}{1}
 \caption{(cont.) CVRHS Classifications for the         Galaxy Zoo 2 Ring Sample.}
 \label{tab:catalog}
 \begin{tabular}{lrrcrll}
 \hline
 2MASX/SDSS & PGC & file & $z$ & $T$ &                   CVRHS Type & Notes \\
 Other &  &  &  &  &  &  \\
 1 & 2 & 3 & 4 & 5 & 6 & 7 \\
 \hline
\phantom{PGC 00}"    & "\phantom{00} & "\phantom{00} & "\phantom{0} & "\phantom{0} & \phantom{00000}" & variety; \RP\ is bluish; bar is a  \\                                                                     
\phantom{PGC 00}"    & "\phantom{00} & "\phantom{00} & "\phantom{0} & "\phantom{0} & \phantom{00000}" & small oval and is yellow  \\                                                                              
 J10483113+2312320  & 1683364 & 161493 & 0.060     &    0.0 & (R$_1$)SAB$_a$(l)0/a                                                                   & (l) handles; \Lff\  \\                                     
\phantom{PGC 00}"    & "\phantom{00} & "\phantom{00} & "\phantom{0} & "\phantom{0} & \phantom{00000}" & dark-spacer  \\                                                                                           
 J10483391+1823030  & 3760887 & 173532 & 0.098     &    0.0 & (R)SAB0/a                                                                              & very interesting face-on pure  \\
\hline
 \end{tabular}
 \end{table*}
 \clearpage
 \begin{table*}
 \centering
 \setcounter{table}{1}
 \caption{(cont.) CVRHS Classifications for the         Galaxy Zoo 2 Ring Sample.}
 \label{tab:catalog}
 \begin{tabular}{lrrcrll}
 \hline
 2MASX/SDSS & PGC & file & $z$ & $T$ &                   CVRHS Type & Notes \\
 Other &  &  &  &  &  &  \\
 1 & 2 & 3 & 4 & 5 & 6 & 7 \\
 \hline
NGC 3398            & 32564   &  49023 & 0.010     &    2.0 & (R$_1$L)SAB($\underline{\rm r}$s)ab                                                    & interesting case with  \\                                  
\phantom{PGC 00}"    & "\phantom{00} & "\phantom{00} & "\phantom{0} & "\phantom{0} & \phantom{00000}" & possible (x$_1$r)  \\                                                                                     
CGCG 038-033        & 32571   &  25976 & 0.052     &    2.0 & (R$^{\prime}$)SB(p,$\underline{\rm r}$s,bl)ab                                          & .......... \\                                              
J105141.38+090825.7 & ....... & 189316 & 0.027     &    2.0 & SAB(r$_w$)ab:                                                                          & face-on with very broad (r)  \\                            
\phantom{PGC 00}"    & "\phantom{00} & "\phantom{00} & "\phantom{0} & "\phantom{0} & \phantom{00000}" & zone; (r) dark-spacer  \\
\hline
 \end{tabular}
 \end{table*}
 \clearpage
 \begin{table*}
 \centering
 \setcounter{table}{1}
 \caption{(cont.) CVRHS Classifications for the         Galaxy Zoo 2 Ring Sample.}
 \label{tab:catalog}
 \begin{tabular}{lrrcrll}
 \hline
 2MASX/SDSS & PGC & file & $z$ & $T$ &                   CVRHS Type & Notes \\
 Other &  &  &  &  &  &  \\
 1 & 2 & 3 & 4 & 5 & 6 & 7 \\
 \hline
\phantom{PGC 00}"    & "\phantom{00} & "\phantom{00} & "\phantom{0} & "\phantom{0} & \phantom{00000}" & similar to 58735  \\                                                                                      
 J10560931+2940139  & 1873962 & 131154 & 0.051     &    1.0 & (R$_1^{\prime}$)SA$\underline{\rm B}$(r)a                                              & (r) blue; \RoneP\ very \\                           
\phantom{PGC 00}"    & "\phantom{00} & "\phantom{00} & "\phantom{0} & "\phantom{0} & \phantom{00000}" & faint; \Lff\ dark-spacer  \\                                                                         
 J10564145+4843068  & 2324007 & 238533 & 0.087     & $-$1.0 & (RL)SA0$^+$                                                                            & poorly resolved but still  \\                              
\phantom{PGC 00}"    & "\phantom{00} & "\phantom{00} & "\phantom{0} & "\phantom{0} & \phantom{00000}" & clear; \Lff\ dark-spacer  \\
\hline
 \end{tabular}
 \end{table*}
 \clearpage
 \begin{table*}
 \centering
 \setcounter{table}{1}
 \caption{(cont.) CVRHS Classifications for the         Galaxy Zoo 2 Ring Sample.}
 \label{tab:catalog}
 \begin{tabular}{lrrcrll}
 \hline
 2MASX/SDSS & PGC & file & $z$ & $T$ &                   CVRHS Type & Notes \\
 Other &  &  &  &  &  &  \\
 1 & 2 & 3 & 4 & 5 & 6 & 7 \\
 \hline
\phantom{PGC 00}"    & "\phantom{00} & "\phantom{00} & "\phantom{0} & "\phantom{0} & \phantom{00000}" & handles on (rl), one side  \\                                                                             
\phantom{PGC 00}"    & "\phantom{00} & "\phantom{00} & "\phantom{0} & "\phantom{0} & \phantom{00000}" & especially  \\                                                                                            
 J10595064+2403283  & 1698889 & 161670 & 0.044     & $-$1.0 & SA(r)0$^+$:                                                                            & in glare of very bright star  \\                           
 J10595465+1005302  & 33129   &  75857 & 0.036     &    0.0 & (R$_1$)SAB(r)0/a                                                                       & very good case; round dark  \\                             
\phantom{PGC 00}"    & "\phantom{00} & "\phantom{00} & "\phantom{0} & "\phantom{0} & \phantom{00000}" & spaces in \Rone; possible X  \\
\hline
 \end{tabular}
 \end{table*}
 \clearpage
 \begin{table*}
 \centering
 \setcounter{table}{1}
 \caption{(cont.) CVRHS Classifications for the         Galaxy Zoo 2 Ring Sample.}
 \label{tab:catalog}
 \begin{tabular}{lrrcrll}
 \hline
 2MASX/SDSS & PGC & file & $z$ & $T$ &                   CVRHS Type & Notes \\
 Other &  &  &  &  &  &  \\
 1 & 2 & 3 & 4 & 5 & 6 & 7 \\
 \hline
\phantom{PGC 00}"    & "\phantom{00} & "\phantom{00} & "\phantom{0} & "\phantom{0} & \phantom{00000}" & \Lff\ dark-spacer; excellent  \\                                                                          
\phantom{PGC 00}"    & "\phantom{00} & "\phantom{00} & "\phantom{0} & "\phantom{0} & \phantom{00000}" & \RoneP  \\                                                                                                
CGCG 038-106        & 33407   &  19899 & 0.053     &    0.0 & (L:)SA(r$^{\prime}$l,r)0/a                                                             & excellent (r)  \\                                          
 J11033617+0046111  & 1174978 & 244803 & 0.096     &    2.0 & (R$^{\prime}$)SA(s)ab:                                                                 & very poorly resolved  \\                                   
UGC 06125           & 33404   & 207680 & 0.021     &    3.0 & (R$_2^{\prime}$)SA(r)b                                                                 & !!excellent case; very bright  \\
\hline
 \end{tabular}%
 \end{table*}

\end{document}